\documentclass[aps,prd,twocolumn,preprintnumbers,showpacs,superscriptaddress,nofootinbib,floatfix,10pt]{revtex4-2}
\pdfoutput=1
\usepackage{textcomp}
\usepackage[english]{babel}
\usepackage{amsmath,amssymb,amsbsy,booktabs}
\usepackage{bm}
\usepackage{color}
\usepackage{array}
\usepackage{amstext}
\usepackage{graphicx}
\usepackage{amsfonts}
\usepackage{dcolumn}
\usepackage{rotating}
\usepackage{epstopdf}
\usepackage{esint}
\usepackage{url}
\usepackage[colorlinks=true,
linkcolor=blue,
filecolor=blue,
anchorcolor=blue,
urlcolor=blue,
citecolor=blue
]{hyperref}

\usepackage{array}
\usepackage{booktabs}
\usepackage{multirow}

\usepackage[utf8]{inputenc}

\usepackage{xcolor,colortbl}

\usepackage[T1]{fontenc} 

\usepackage{cancel}
\usepackage{slashed}
\usepackage{hhline}
\usepackage{subfigure} 
\usepackage{enumerate}
\usepackage{times}
\usepackage[version=4]{mhchem}

\DeclareMathOperator{\tr}{Tr}

\allowdisplaybreaks

\begin{document}

\title{\Large \bf \boldmath Probing new physics with polarized $\tau$ and $\Lambda_c$ in quasielastic $\nu_{\tau}\!+\!n\!\to\! \tau^-\!+\!\Lambda_c$ scattering process}

\author{Ya-Ru Kong}
\email{yarukong@mails.ccnu.edu.cn}
\affiliation{Institute of Particle Physics and Key Laboratory of Quark and Lepton Physics~(MOE),\\
Central China Normal University, Wuhan, Hubei 430079, China}

\author{Li-Fen Lai}
\email{lailifen@mails.ccnu.edu.cn}
\affiliation{School of Physics and Electronic Information, Shangrao Normal University, Shangrao 334001, China}

\author{Xin-Qiang Li}
\email{xqli@mail.ccnu.edu.cn }
\affiliation{Institute of Particle Physics and Key Laboratory of Quark and Lepton Physics~(MOE),\\
Central China Normal University, Wuhan, Hubei 430079, China}
\affiliation{Center for High Energy Physics, Peking University, Beijing 100871, China}

\author{Xin-Shuai Yan}
\email{yanxinshuai@htu.edu.cn (Corresponding author)}
\affiliation{Institute of Particle and Nuclear Physics, Henan Normal University, Xinxiang, Henan 453007, China}

\author{Ya-Dong Yang}
\email{yangyd@mail.ccnu.edu.cn}
\affiliation{Institute of Particle Physics and Key Laboratory of Quark and Lepton Physics~(MOE),\\
Central China Normal University, Wuhan, Hubei 430079, China}
\affiliation{Institute of Particle and Nuclear Physics, Henan Normal University, Xinxiang, Henan 453007, China}

\author{Dong-Hui Zheng}
\email{zhengdh@mails.ccnu.edu.cn}
\affiliation{Institute of Particle Physics and Key Laboratory of Quark and Lepton Physics~(MOE),\\
Central China Normal University, Wuhan, Hubei 430079, China}
		
\begin{abstract}
The absence of semitauonic decays of charmed hadrons makes the decay processes mediated by the quark-level $c\to d \tau^+ \nu_{\tau}$ transition inadequate for probing a generic new physics (NP) with all kinds of Dirac structures. To fill in this gap, we consider in this paper the quasielastic neutrino scattering process $\nu_{\tau}+n\to \tau^-+\Lambda_c$, and propose searching for NP through the polarizations of the $\tau$ lepton and the $\Lambda_c$ baryon. In the framework of a general low-energy effective Lagrangian, we perform a comprehensive analysis of the (differential) cross sections and polarization vectors of the process both within the Standard Model and in various NP scenarios, and scrutinize possible NP signals. We also explore the influence on our findings due to the uncertainties and the different parametrizations of the $\Lambda_c \to N$ transition form factors, and show that they have become one of the major challenges to further constrain possible NP through the quasielastic scattering process.
\end{abstract}

\pacs{}

\maketitle

\section{Introduction} 
\label{sec:intro}

Over the past few years, several intriguing anomalies have been observed in the processes mediated by the quark-level $b\to c l \bar{\nu}_{l}$ transitions, particularly in the ratios $R_{D^{(\ast)}}$~\cite{BaBar:2012obs,BaBar:2013mob,Belle:2015qfa,LHCb:2015gmp,Belle:2016dyj,Belle:2017ilt,LHCb:2017smo,LHCb:2017rln,Belle:2019rba,LHCb:2023zxo,LHCb:2023cjr},
\begin{align}
	R_{D^{(\ast)}}\equiv \frac{\mathcal{B}(B\to D^{(*)}\tau^-\nu_{\tau})}{\mathcal{B}(B\to D^{(*)}\ell^-\nu_{\ell})},
\end{align}
with $\ell=e,\mu$. These anomalies continuously challenge the lepton flavor universality, a central feature of the Standard Model (SM) of particle physics, and arouse a surge of phenomenological studies of new physics (NP) beyond the SM in $B$ physics (for recent reviews, see, e.g., Refs.~\cite{Bifani:2018zmi,Bernlochner:2021vlv,Albrecht:2021tul,London:2021lfn}).
In view of the potential violation of the lepton flavor universality in $B$-meson decays, it is 
also natural to investigate if such phenomena also emerge in the charm sector. 

Among the various processes used to probe the phenomena, the ones mediated by the quark-level $c\to d \tau^+ \nu_{\tau}$ transition attract certain attention~\cite{CLEO:2006jxt,BESIII:2019vhn,Fleischer:2019wlx,Becirevic:2020rzi}. In particular, a ratio $R_{\tau/\mu}$, somewhat similar to $R_{D^{(*)}}$, 
can be defined as  
\begin{align}
	R_{\tau/\mu}=\frac{\Gamma(D^+\to \tau^+\nu_{\tau})}{\Gamma(D^+\to \mu^+\nu_{\mu})}\,,
\end{align}
and serve as an important avenue to test the SM in the charm sector~\cite{CLEO:2006jxt,BESIII:2019vhn}. Interestingly enough, the ratio $R_{\tau/\mu}$ is constructed from the purely leptonic $D$-meson decays rather than from the semileptonic ones, which is in contrast to the ratios $R_{D^{(*)}}$. The underlying reason for this is that the largest accessible phase space for semileptonic $D$-meson decays is 
given by $m_{D^+}-m_{\pi^0}\simeq 1.735$~GeV, which is smaller than the $\tau$-lepton mass, rendering the semitauonic $D$-meson decays kinematically forbidden. The same conclusion also holds for the charmed-baryon decays.

The absence of semitauonic decays of charmed hadrons makes, therefore, the decay processes mediated by the $c\to d \tau^+ \nu_{\tau}$ transition suitable for probing NP with only a subset of Dirac structures. For example, the purely leptonic $D$-meson decays are known to be only sensitive to the axial and pseudo-scalar four-fermion operators of a general low-energy effective Lagrangian (denoted by $\mathcal{L}_{\text{eff}}$ as introduced in Eq.~\eqref{eq:Leff}), making the tauonic vector, scalar, and tensor operators seemingly inaccessible at low-energy regime~\cite{Fleischer:2019wlx,Becirevic:2020rzi,Leng:2020fei,Colangelo:2021dnv}. Although these operators can be probed through the high-$p_T$ dilepton invariant mass tails at high-energy 
colliders under additional assumptions~\cite{Fuentes-Martin:2020lea,Allwicher:2022gkm}, other new processes and observables, particularly the 
low-energy ones, are still badly needed in order to pinpoint all the possible NP Dirac structures. In some cases, these low-energy processes and observables can also provide very complementary information about NP~\cite{Lai:2021sww,Lai:2022ekw}.

In this paper, we will consider the quasielastic (QE) neutrino scattering process $\nu_{\tau}+n\to \tau^-+\Lambda_c$ induced by the quark-level $\nu_{\tau} d \to \tau^- c$ transition. This process is free from the kinematic problem that the semitauonic charmed-baryon decays face 
and involves all the effective operators of $\mathcal{L}_{\text{eff}}$. 
However, even with the purely tauonic $D$-meson decays and the high-$p_T$ dilepton invariant mass analyses, it
still cannot provide enough observables to fully pinpoint all the NP Dirac structures and determine the corresponding complex 
Wilson coefficients (WCs). Thus, we will also propose searching for NP through the polarizations of the $\tau$ lepton and the $\Lambda_c$ baryon.\footnote{We note that the polarizations of the final lepton and the produced nucleon in a charged-current QE neutrino-nucleus scattering process induced by the quark-level $\nu_{\ell} d \to \ell^- u$ or $\bar{\nu}_{\ell} u \to \ell^+ d/s$ transition have also been discussed in Refs.~\cite{Graczyk:2004uy, Fatima:2018tzs, Graczyk:2019xwg}.} The polarization observables to be considered in this work 
involve all the effective operators of $\mathcal{L}_{\text{eff}}$, and can fill the gap (at least partially), though they  
are generally more difficult to measure than the cross sections.
Based on a combined constraint on the WCs of the effective operators 
set by the measured branching ratio of $D^+\to\tau^+\nu_{\tau}$ decay~\cite{BESIII:2019vhn} 
and the analysis of the high-$p_T$ dilepton invariant mass tails~\cite{Fuentes-Martin:2020lea}, 
we will perform a comprehensive analysis of all the observables involved both within the SM and in various NP scenarios, and scrutinize possible NP signals.

The hadronic matrix elements of the scattering process will be parametrized by the $n\to \Lambda_c$ transition form factors, which are in turn related to the $\Lambda_c\to N$ (nucleon) form factors by complex conjugation. However, since a scattering process generally occupies a negative kinematic range ($q^2<0$) while a decay process happens at the positive one ($q^2>0$), an extrapolation of the $\Lambda_c\to N$ transition form factors from positive to negative $q^2$ becomes necessary. This requires that the form-factor parametrization must possess analyticity in the proper $q^2$ range~\cite{Sobczyk:2019uej,Lai:2021sww,Lai:2022ekw}. In this paper, we will consider three different models with three different form-factor parametrizations for the $\Lambda_c\to N$ transition form factors to compute 
the cross sections and polarization vectors in various NP scenarios. Our major results will be, however, based on the lattice QCD (LQCD) calculations~\cite{Meinel:2017ggx}, since they also provide the theoretical uncertainties, which we will propagate to all the observables considered. Nonetheless, a detailed comparison of all the observables calculated with different form-factor parametrizations will be provided as well.

The paper is organized as follows. In Sec.~\ref{sec:models}, we begin with a 
brief introduction of our theoretical framework, including the most general low-energy effective Lagrangian as well as the kinematics, the cross sections, and the various polarization vectors of the scattering process. In such a framework, we study in subsection~\ref{sec:totalcross} the total cross section and the averaged polarization vectors in various NP scenarios, and then in subsection~\ref{sec:diffcross} the differential cross sections and the $Q^2$-dependent polarization observables. In subsection~\ref{sec:lowglimit}, we revisit the scattering process together with the $Q^2$-dependent observables in the limit of small WCs (i.e., small-$g_i$). The subsequent two subsections contain our exploration of the influence on our findings due to the uncertainties and the different parametrizations of the $\Lambda_c \to N$ transition form factors. Finally, we collect our main conclusions in Sec.~\ref{sec:con}, and relegate further details on the form factors and explicit expressions of the various observables to the appendices.

\section{Theoretical framework}
\label{sec:models}

\subsection{Low-energy effective Lagrangian}

Without introducing the right-handed neutrinos, the most general low-energy effective 
Lagrangian responsible for the $\nu_{\tau} d \to \tau^- c$ transition can be written as   
\begin{align}\label{eq:Leff}
	\mathcal{L}_{\text{eff}}=&-\frac{4G_F}{\sqrt{2}}V_{cd}\Big [ (1+g^{L}_{V})\mathcal{O}^{L}_{V}+
	g^{R}_{V}\mathcal{O}^{R}_{V}+g^{L}_{S}\mathcal{O}^{L}_{S}\nonumber \\[0.12cm]
	&+g^{R}_{S}\mathcal{O}^{R}_{S}+g^{L}_{T}\mathcal{O}^{L}_{T} \Big]+\text{H.c.}\,,
\end{align}
with 
\begin{align}
	\mathcal{O}^{L,R}_{V}&=(\bar{c}\gamma^{\mu}P_{L,R} d)(\bar{\tau}\gamma_{\mu}P_{L}\nu_{\tau})\,, 
	\nonumber \\[0.12cm]
	\mathcal{O}^{L,R}_{S}&=(\bar{c}P_{L,R} d)(\bar{\tau}P_{L}\nu_{\tau})\,, 
	\nonumber \\[0.12cm]
    \mathcal{O}^{L}_{T}&=(\bar{c}\sigma^{\mu \nu} P_{L} d)(\bar{\tau}\sigma_{\mu \nu} P_{L} \nu_{\tau})\,, 
    \label{eq:operators}
\end{align}
where $P_{R,L}\!=\!(1\pm \gamma_5)/2$ are the right- and left-handed projectors, and $\sigma^{\mu \nu}\!= \! i[\gamma^{\mu},\gamma^{\nu}]/2$ the antisymmetric tensor. Note that the tensor operators with mixed quark and lepton chiralities vanish due to Lorentz invariance. The WCs $g_i$ in Eq.~\eqref{eq:Leff} parametrize possible deviations from the SM and are complex in general. Such a framework is only applicable up to an energy scale of $\mathcal{O}(m_b)$, with $m_b$ denoting the bottom-quark mass, above which new degrees of freedom would appear. 
 
It should be pointed out that the $\mathcal{L}_{\text{eff}}$ can also be presented in another operator basis, 
in which the majority of basis operators posses definite parity (see, e.g., Ref.~\cite{Becirevic:2020rzi}).
The WCs associated with this set of basis operators can be related to the $g_i$ in Eq.~\eqref{eq:Leff} through 
the following relations:
\begin{align}\label{eq:WC_relation}
	g_{V,A}=g^R_V\pm g^L_V,  \quad  g_{S,P}=g^R_S\pm g^L_S, \quad g_T=g^L_T\,.
\end{align}
And the former become very handy for discussing the $D$-meson leptonic decays, since these decays 
are only sensitive to $g_A$ and $g_P$, as shown in Eq.~\eqref{eq:D_decay}. 
However, we will focus on the operators listed in Eq.~\eqref{eq:operators}, since 
we will also take account of the constraints set through the analysis of the dilepton invariant mass tails in 
$pp\to \tau \nu_{\tau}$ processes at high $p_T$~\cite{Fuentes-Martin:2020lea}, 
which are based on the very same set of basis operators as in Eq.~\eqref{eq:operators} and much severer in general than the ones 
set by the $D^+\!\to\! \tau^+\nu_{\tau}$ decay (see the colored regions in Fig.~\ref{fig:constraints}).

\subsection{Cross section, form factors, and kinematics}
\label{subsec:Cross section}

The differential cross section of the QE scattering process $\nu_{\tau}(k)+n(p)\to \tau^-(k^\prime)+\Lambda_c(p^\prime)$, with $p=(m_n, 0)$, $p^\prime=(E_{\Lambda_c}, \pmb{p}^\prime)$, $k=(E, \pmb{k})$, and $k^\prime=(E^\prime, \pmb{k}^\prime)$, is given by 
\begin{align}\label{eq:diff_cross}
	d\sigma=&\frac{1}{4 p\cdot k}\frac{d^3\pmb{k}^\prime}{(2\pi)^3}\frac{1}{2E^\prime}\frac{d^3\pmb{p}^\prime}{(2\pi)^3}\frac{1}{2E_{\Lambda_c}}\overline{|\mathcal{M}|}^2\nonumber \\[0.12cm]
	&\times (2\pi)^4\delta^4(p+k-p^\prime-k^\prime),
\end{align}
where the amplitude $\mathcal{M}$ can be generically written as~\cite{Penalva:2020xup} 
\begin{align}\label{eq:amplitude}
	\mathcal{M}=\frac{4G_F}{\sqrt{2}}V_{cd}\left(J_H J^L
	+J^{\alpha}_H J_{\alpha}^L
	+J^{\alpha\beta}_H J_{\alpha\beta}^L\right),
\end{align} 
when all the effective operators in Eq.~\eqref{eq:Leff} are taken into account.  
The lepton currents in Eq.~\eqref{eq:amplitude} are defined as 
\begin{align}
	J_{(\alpha\beta)}^L=\bar{u}_{\tau}(k^\prime, r^\prime) \Gamma_{(\alpha\beta)}P_Lu_{\nu_{\tau}}(k,r),  
\end{align}
with $\Gamma_{(\alpha\beta)}=(1, \gamma_{\alpha},\sigma_{\alpha\beta})$, 
while the hadron currents as 
\begin{align}\label{eq:hadron current}
	J^{(\alpha\beta)}_H&=\langle \Lambda_c(p^{\prime},s^{\prime}) |\bar{c}O^{(\alpha\beta)}_H d|n(p,s)\rangle \,, 
\end{align}
with
\begin{align}
	O_H&=\frac{1}{2}\left(g_S+g_P\gamma_5\right)\,, \nonumber \\[0.12cm]
	O^{\alpha}_H&=\frac{1}{2}\gamma^{\alpha}\left(g^{\prime}_V-g^{\prime}_A\gamma_5\right) \,, \nonumber \\[0.12cm]
    O^{\alpha\beta}_H&=g^L_T\sigma^{\alpha\beta}P_L\,,
\end{align}
where $g^{\prime}_{V,A}=(1+g^{L}_{V}\pm g^{R}_{V})$, $g_{S,P}$ are given by Eq.~\eqref{eq:WC_relation}, 
and $r$ and $s$ ($r^\prime$ and $s^\prime$) denote the spins of initial (final) lepton and baryon, respectively. The amplitude squared $|\mathcal{M}|^2$ is obtained by summing up the initial- and final-state spins; more details are elaborated in Appendix~\ref{app:Amplitude}.

The hadronic matrix elements $\langle\Lambda_c|\bar{c}O^{(\alpha\beta)}_H d|n\rangle$ in Eq.~\eqref{eq:hadron current} are identical to the complex conjugate of $\langle n|(\bar{c}O^{(\alpha\beta)}_H d)^{\dagger}|\Lambda_c\rangle$, which are further parametrized by the $\Lambda_c \to N$ transition form factors~\cite{Feldmann:2011xf,Meinel:2017ggx,Das:2018sms}. Since a scattering process generally occupies a different kinematic range ($q^2<0$) from that of a decay ($q^2>0$), theoretical analyses of the scattering process require an extrapolation of the form factors to negative $q^2$. Thus, the form-factor parametrizations suitable for our purpose must be analytic in the proper $q^2$ range. 

Interestingly, there exist already several schemes that meet our selection criterion and have been utilized to parametrize the $\Lambda_c \to N$ form factors by various models. For instance, a dipole parametrization scheme has been employed within the MIT bag model (MBM)~\cite{Chodos:1974je,Chodos:1974pn} and the nonrelativistic quark model (NRQM)~\cite{Kokkedee1969}, and a double-pole one in the relativistic constituent quark model (RCQM)~\cite{Ivanov:1996fj,Branz:2009cd}. Although the form-factor parametrizations in each scheme do not result in pathological behaviors in the $q^2<0$ range, only the form factors associated with the matrix element $\langle N|\bar{d}\gamma^{\mu}P_L c|\Lambda_c\rangle$ were calculated in these models. The primary scheme we consider was initially proposed to parametrize the $B\to \pi$ vector form factor~\cite{Bourrely:2008za}, and has been recently utilized in the LQCD calculations of the $\Lambda_c\to N$ transition form factors~\cite{Meinel:2017ggx}. In contrast to other model evaluations, the LQCD calculation not only takes care of all the form factors, but also provides an error estimation. Thus, we will adopt the latest LQCD results~\cite{Meinel:2017ggx} throughout this work. Meanwhile, given that the model calculations of the $N\to \Lambda_c$ form factors can significantly affect the predictions of $\Lambda_c$ weak production in neutrino QE processes~\cite{DeLellis:2004ovi,Sobczyk:2019uej}, we will also analyze the QE scattering process $\nu_{\tau}+n\to \tau^-+\Lambda_c$ in terms of the form factors calculated within the models MBM, NRQM, and RCQM in various NP scenarios; for more details about the form factors in these different models, we refer the readers to Appendix~\ref{appendix:form factor}.

\begin{figure}[t]
	\centering
	\includegraphics[width=0.48\textwidth]{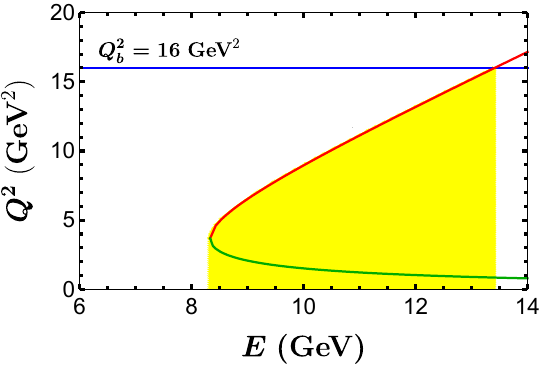}
	\caption{Criteria for selecting the neutrino beam energy $E$, where the red (green) curve denotes the $E$-$Q^2_{\max(\min)}$ relation given by Eq.~\eqref{eq:LFV_Q2_range}, and the blue line represents the condition $Q^2\leq 16\,\text{GeV}^2$ required by our theoretical framework. The yellow range indicates the eligible $E$.} 
	\label{fig:Eselection} 
\end{figure}

The kinematics of the QE scattering process is bounded by~\cite{Lai:2021sww} 
\begin{align} 
	\frac{\alpha-E\sqrt{\lambda}}{m_n+2E} \leq
	q^2 \leq \frac{\alpha +E\sqrt{\lambda}}{m_n+2E}\,,\label{eq:LFV_Q2_range}
\end{align}
where
\begin{align}
	\alpha &\!\equiv \! E(m_{\Lambda_c}^2\!-\! m_n^2\!+\!m^2_{\tau}\!-\! 2m_nE)\!+\!m_n m^2_{\tau}\,, \nonumber \\[0.2cm]
	\lambda &\!\equiv\! m_{\Lambda_c}^4\!+\!(m_n^2\!+\!2m_nE\!-\!m^2_{\tau})^2\!-\!2m_{\Lambda_c}^2(m_n^2\!+\!2m_nE\!+\!m^2_{\tau})\,. \nonumber 
\end{align}
This condition indicates that the neutrino beam energy $E$ determines the maximal and minimal values of $Q^2$ ($Q^2=-q^2$), which, in turn, implies that any constraints on $Q^2_{\max}$ and $Q^2_{\min}$ restrict the $E$ selection. An explicit example is that a minimal requirement for $E$ ($E\gtrsim 8.33$~GeV) of the scattering process can be obtained by using the condition $Q^2_{\max}=Q^2_{\min}$; this can also be visualized in Fig.~\ref{fig:Eselection} by noting the intersection point of the red and green curves that represent the $E$-$Q^2_{\max}$ and $E$-$Q^2_{\min}$ relations, respectively. Besides the kinematic constraint on $Q^2_{\max}$, 
we also consider the limit from our theoretical framework. As our analyses are carried out in the framework of $\mathcal{L}_{\text{eff}}$ given by Eq.~\eqref{eq:Leff}, to ensure the validity of our results, we require $Q^2_{\max}$ to not exceed $Q^2_{b}=16\,\text{GeV}^2\approx m_b^2$. Such a requirement, depicted by the blue line in Fig.~\ref{fig:Eselection}, indicates an upper bound $E\lesssim 13.41\,\text{GeV}$, provided that the observables one is interested in, such as the total cross section, involve $Q^2_{\max}$. Otherwise, $E$ is not bounded from above, since one can always concentrate on the lower $Q^2$ range, even though a high $Q^2_{\max}$ is available due to a high $E$. 

It is interesting to note that the $\tau$-optimized $\nu_\tau$ flux at the Deep Underground Neutrino Experiment (DUNE) drops below $10^8\,\text{m}^{-2} \text{year}^{-1}$ at $E_{\nu_{\tau}}\gtrsim 14$~GeV~\cite{DUNE2020,Machado:2020yxl}, which is close to the upper bound of $E$ shown in Fig.~\ref{fig:Eselection}. If the proposed QE scattering process were measured at the DUNE, one could then explore all the observables considered in this work within the whole, available $Q^2$ range, while maintaining a relatively high $\nu_\tau$ beam flux. 
It should be pointed out that the neutrino oscillation experiments in the few-GeV range at DUNE use detectors constructed of liquid argon (see, e.g., 
Ref.~\cite{DUNE:2015lol}), where nuclear effects are significant. But the knowledge of those effects remains imperfect, 
which induces important uncertainties for the experiments of neutrino oscillation as well as the proposed QE scattering process at DUNE.

\subsection{Polarization vectors of the final lepton and baryon}
\label{subsec:polarization}

The polarization four-vector $\mathcal{P}_l^{\mu}$ of the 
$\tau$ lepton produced in the scattering process $\nu_{\tau}+n\to \tau^-+\Lambda_c$ can be conveniently obtained by using the density matrix formalism as~\cite{Athar:2020kqn}
\begin{align}\label{eq:eta}
	\mathcal{P}^{\mu}_{l}=\frac{\tr[\rho_{l}(k^\prime)\gamma^{\mu}\gamma_5]}{\tr[\rho_{l}(k^\prime)]}\,,
\end{align}
where the spin density matrix $\rho_{l}(k^\prime)$ of the $\tau$ lepton is given by 
\begin{align}\label{eq:rho_tau}
	\rho_{l}(k^\prime)=\mathcal{J}^{(\alpha\beta,\alpha^\prime\beta^\prime)}\left[\Lambda(k^\prime)
	\Gamma_{(\alpha\beta)}P_L\Lambda(k)P_R\widetilde{\Gamma}_{(\alpha^\prime\beta^\prime)}\Lambda(k^\prime)\right].
\end{align}
Now a clarification of the various symbols in Eq.~\eqref{eq:rho_tau} is in order. Firstly, the hadronic tensor $\mathcal{J}^{(\alpha\beta,\alpha^\prime\beta^\prime)}$ is given by 
\begin{align}
	\mathcal{J}^{(\alpha\beta,\alpha^\prime\beta^\prime)}&=
	\frac{1}{2}\sum_{ss^\prime}J_H^{(\alpha\beta)}J_H^{(\alpha^\prime\beta^\prime)\dagger}\nonumber \\[0.12cm]
	&=\frac{1}{2}\tr\left[\Lambda(p^\prime)
	\mathcal{M}^{(\alpha\beta)}\Lambda(p)\widetilde{\mathcal{M}}^{(\alpha^\prime\beta^\prime)}\right],
\end{align}
where $\mathcal{M}_{(\alpha\beta)}$ denotes the Dirac $\gamma$ structure of the hadronic matrix element 
$\langle\Lambda_c|\bar{c}O^{(\alpha\beta)}_H d|n\rangle$ in Eq.~\eqref{eq:hadron current}. Clearly, $\mathcal{M}_{(\alpha\beta)}$ involves not only the WCs $g_i$ but also the form factors. The prefactor $1/2$ accounts for the spin average over the neutron spin. Secondly, $\widetilde{\mathcal{M}}^{(\alpha^\prime\beta^\prime)}\!=\!\gamma^0\mathcal{M}^{(\alpha^\prime\beta^\prime)\dagger}\gamma^0$, 
$\widetilde{\Gamma}_{(\alpha^\prime\beta^\prime)}\!=\!\gamma^0\Gamma^{\dagger}_{(\alpha^\prime\beta^\prime)}\gamma^0$, 
and $\Lambda(k)\!=\!(\slashed{k}+m_k)$ is the spin projection operator for 
a spin $1/2$ fermion with momentum $k$ and mass $m_k$.    

The polarization four-vector $\mathcal{P}_{h}^{\mu}$ of the produced 
$\Lambda_c$ baryon can be obtained in a similar way, with the spin density matrix $\rho_{h}(p^\prime)$ given by  
\begin{align}\label{eq:rho_lambda}
	\rho_{h}(p^\prime)\!=\!\mathcal{L}_{(\alpha\beta,\alpha^\prime\beta^\prime)}\!\left[\Lambda(p^\prime)
	\mathcal{M}^{(\alpha\beta)}\Lambda(p)\widetilde{\mathcal{M}}^{(\alpha^\prime\beta^\prime)}\Lambda(p^\prime)\!\right],
\end{align}
where the leptonic tensor $\mathcal{L}_{(\alpha\beta,\alpha^\prime\beta^\prime)}$ can be written as 
\begin{align}
	\mathcal{L}_{(\alpha\beta,\alpha^\prime\beta^\prime)}&=\frac{1}{2}\sum_{rr^\prime}J^L_{(\alpha\beta)}J^{L\dagger}_{(\alpha^\prime\beta^\prime)}\nonumber \\[0.12cm]
	&=\frac{1}{2}\tr[\Lambda(k^\prime)\Gamma_{(\alpha\beta)}P_L\Lambda(k)P_R\widetilde{\Gamma}_{(\alpha^\prime\beta^\prime)}]\,. 
\end{align}

The polarization vectors $\mathcal{P}^{\mu}_{l,h}$ of the outgoing lepton and  baryon can be decomposed as
\begin{align}\label{eq:polvect}
	\mathcal{P}^{\mu}_{l,h}\!=\!P^{l,h}_L (N^{l,h}_L)^{\mu}\!+\!P^{l,h}_P (N^{l,h}_P)^{\mu}\!+\!P^{l,h}_T(N^{l,h}_T)^{\mu},
\end{align}
where the two sets of four-vectors $N^{l,h}_L$, $N^{l,h}_T$, and $N^{l,h}_P$ are defined, respectively, as 
\begin{equation}\label{eq:basis_lepton}
\begin{aligned}
	\left(N^{l}_L\right)^{\mu}&=\left(\frac{|\pmb{k}^\prime|}{m_{\tau}},\frac{k^{\prime0}\pmb{k}^\prime}{m_{\tau}|\pmb{k}^\prime|}\right), \\
	\left(N^{l}_T\right)^{\mu}&=\left(0, \frac{\pmb{k}\times \pmb{k}^\prime}{|\pmb{k}\times \pmb{k}^\prime|}\right), \\
	\left(N^{l}_P\right)^{\mu}&=\left(0, \frac{\pmb{k}^\prime\times(\pmb{k}\times \pmb{k}^\prime)}{|\pmb{k}^\prime\times(\pmb{k}\times \pmb{k}^\prime)|}\right), 
\end{aligned}
\end{equation}
and 
\begin{equation}\label{eq:basis_hadron}
\begin{aligned}
	\left(N^{h}_L\right)^{\mu}&=\left(\frac{|\pmb{p}^\prime|}{m_{\Lambda_c}},\frac{p^{\prime0}\pmb{p}^\prime}{m_{\Lambda_c}|\pmb{p}^\prime|}\right), \\
	\left(N^{h}_T\right)^{\mu}&=\left(0, \frac{\pmb{p}^\prime\times \pmb{k}}{|\pmb{p}^\prime\times \pmb{k}|}\right), \\
	\left(N^{h}_P\right)^{\mu}&=\left(0, \frac{\pmb{p}^\prime\times(\pmb{p}^\prime\times \pmb{k})}{|\pmb{p}^\prime\times(\pmb{p}^\prime\times \pmb{k})|}\right),
\end{aligned}
\end{equation}
indicating the longitudinal ($L$), transverse ($T$), and perpendicular ($P$) directions of the final $\tau$ lepton and $\Lambda_c$ baryon in their reaction planes accordingly. It is then fairly straightforward to obtain the components of $\mathcal{P}^{\mu}$ in Eq.~\eqref{eq:polvect} through
\begin{align}\label{eq:polcomp}
	P^{l,h}_a=-(\mathcal{P}\cdot N^{l,h}_a), \quad a=L,\, P,\,T.
\end{align}

In order to study the dependence of these polarization vectors on the neutrino energy $E$, one often introduces the average polarizations $\langle P^{l,h}_{a}\rangle$, which are defined as~\cite{Graczyk:2004uy,Fatima:2020pvv} 
\begin{align}
	\langle P^{l,h}_a\rangle=\frac{\int_{Q^2_{\text{min}}}^{Q^2_{\text{max}}}P^{l,h}_a(Q^2)\frac{d\sigma}{dQ^2}dQ^2}{\int_{Q^2_{\text{min}}}^{Q^2_{\text{max}}}\frac{d\sigma}{dQ^2}dQ^2}\,.
\end{align}
To characterize the overall degree of polarization of the outgoing particles,  
one can also define the overall average polarization $\langle P^{l,h}\rangle$ as
\begin{align}
	\langle P^{l,h} \rangle=\sqrt{\langle P^{l,h}_L\rangle^2+\langle P^{l,h}_P\rangle^2+\langle P^{l,h}_T\rangle^2}\,.
\end{align}

\subsection{Constraints on the WCs of $\mathcal{L}_{\text{eff}}$}
\label{sec:constraints}

Here we discuss briefly the most relevant and stringent constraints on the WCs $g_i$ from the charmed-hadron weak decays and the high-$p_T$ dilepton invariant mass tails. 
  
Given that the semitauonic decays of charmed hadrons are kinematically forbidden, the $D$-meson tauonic decays become the only decay processes that can be used to constrain the WCs $g_i$ in Eq.~\eqref{eq:Leff}. Here we   
consider the $D^+\!\to\! \tau^+\nu_{\tau}$ decay with its branching ratio given by~\cite{Fleischer:2019wlx,Becirevic:2020rzi,Buras:2020xsm}  
\begin{align}\label{eq:D_decay}
	\mathcal{B}(D^+\!\to\! \tau^+\nu_{\tau})&=\frac{G^2_F |V_{cd}|^2 f^2_{D^+} m_{D^+}m^2_{\tau}}{8\pi}
	\left(1-\frac{m^2_{\tau}}{m^2_{D^+}}\right)^2 \nonumber \\[0.12cm]
	&\times \left|1-g_A+g_P\frac{m^2_{D^+}}{m_{\tau}(m_c+m_d)}\right|^2 \tau_{D^+}\,,
\end{align}
where $g_{A}$ and $g_{P}$ are introduced in Eq.~\eqref{eq:WC_relation}.
With the inputs listed in Table~\ref{table:input}, $|V_{cd}|=0.22438\pm 0.00044$ from the global fit~\cite{Workman:2022ynf}, 
and $f_{D^+}=212.0\pm 0.7$~MeV from an average of the LQCD simulations~\cite{FlavourLatticeAveragingGroupFLAG:2021npn,Carrasco:2014poa,Bazavov:2017lyh}, we can obtain the parameter space of the WCs $g_i$ allowed by the measured branching fraction $\mathcal{B}(D^+\!\to\!\tau^+\nu_{\tau})
=(1.20\pm0.24_{\text{stat}}\pm0.12_{\text{syst}})\times 10^{-3}$~\cite{BESIII:2019vhn}; similar works have also been conducted in Refs.~\cite{Fleischer:2019wlx,Becirevic:2020rzi}. At the same time, constraints on these WCs can also be set through the analysis of the dilepton invariant mass tails in 
$pp\to \tau \nu_{\tau}$ processes at high $p_T$~\cite{Fuentes-Martin:2020lea}. 

 \begin{table}[t]
	\renewcommand\arraystretch{1.5} 
	\tabcolsep=0.92cm
	\centering
	\caption{Values of the input parameters relevant for Eq.~\eqref{eq:D_decay}, which are all from Ref.~\cite{Workman:2022ynf}.}
	\label{table:input} 
	\begin{tabular}[t]{cc}
		\hline \hline
		Parameter &Value \\ \hline
		$m_{\tau}$ & $1.77686$~GeV \\
		$m_{D^+}$ & $1.86965$~GeV \\
		$\tau_{D^+}$ & $1.04$~ps \\
		$G_F$ & $1.1663787\times 10^{-5}~\mathrm{GeV}^{-2}$ \\
		$m_c$ & $1.27$~GeV \\
		$m_d$ & $0$~MeV \\
		\hline \hline
	\end{tabular}
\end{table} 

\begin{figure*}[t]
	\centering
	\includegraphics[height=3.4cm,width=3.4cm]{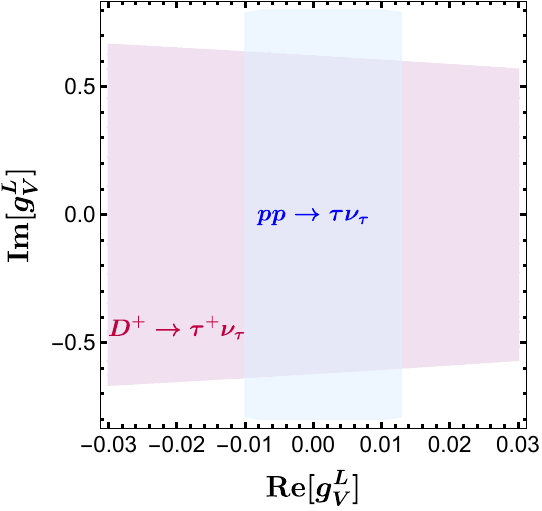}\; 
	\includegraphics[height=3.4cm,width=3.4cm]{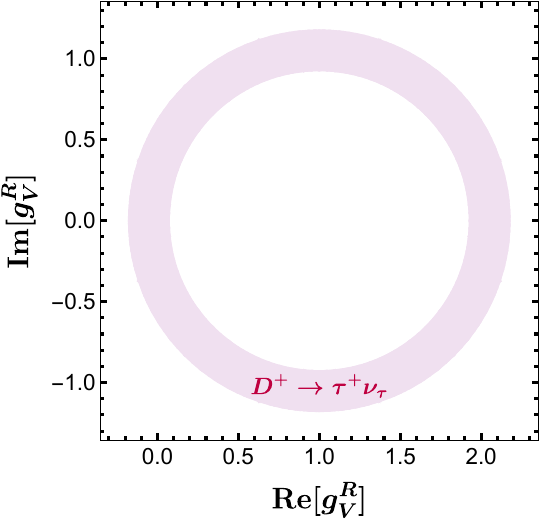}\; 
	\includegraphics[height=3.4cm,width=3.4cm]{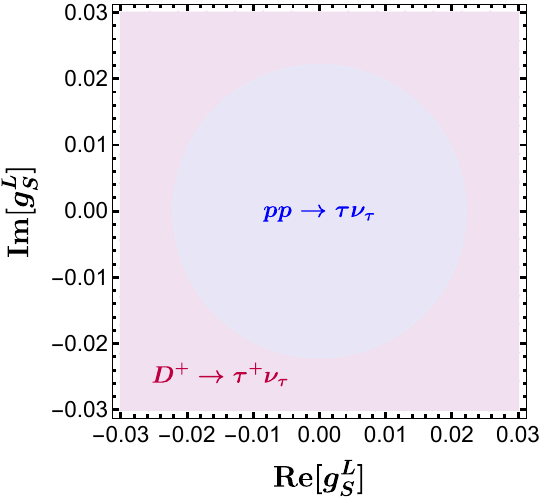}\;
	\includegraphics[height=3.4cm,width=3.4cm]{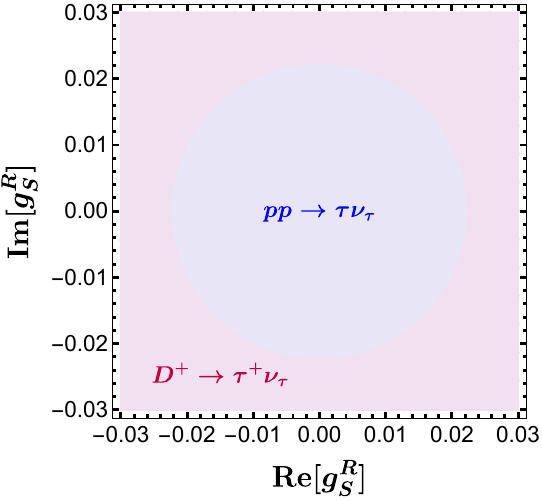}\; 
	\includegraphics[height=3.4cm,width=3.52cm]{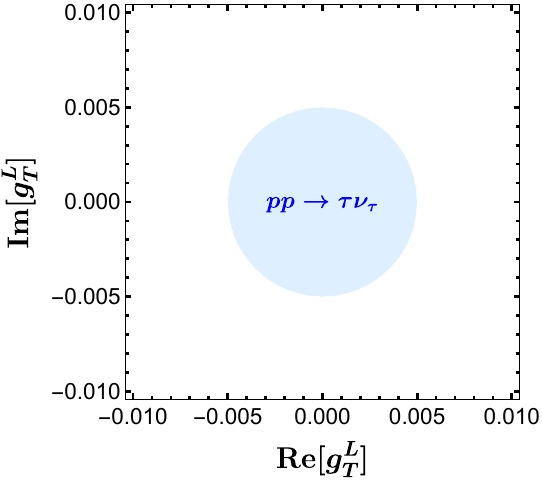}
	\caption{Constraints on the WCs $g_i$ within $1\sigma$ level. The region colored in pink is set by the measured branching fraction of $D^+\!\to\!\tau^+\nu_{\tau}$ decay~\cite{BESIII:2019vhn}, while the region colored in light blue is allowed by the high-$p_T$ dilepton invariant mass tails in $pp\to \tau \nu_{\tau}$ processes~\cite{Fuentes-Martin:2020lea}.}
	\label{fig:constraints}
\end{figure*}

\begin{figure*}[t]
	\centering
	\includegraphics[height=3.4cm,width=3.4cm]{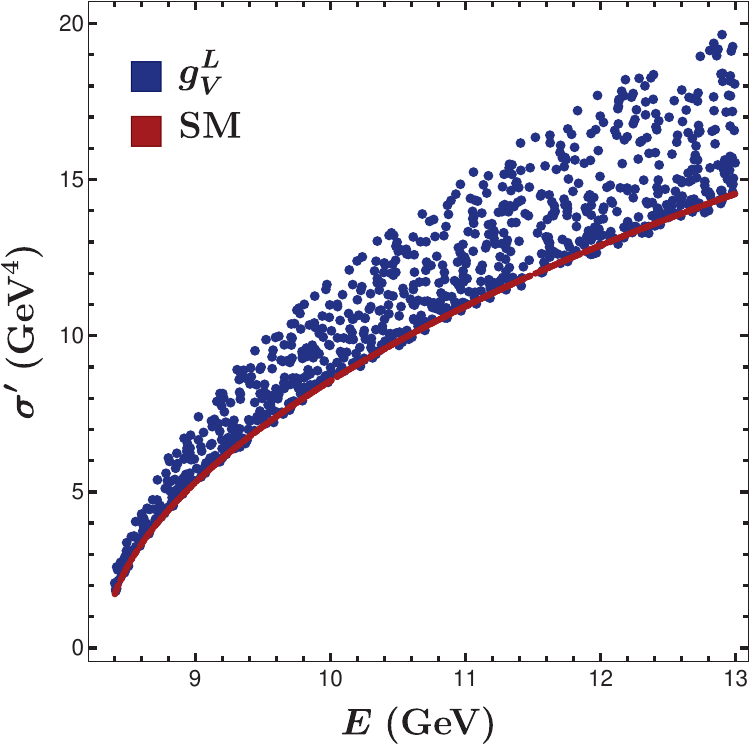}\; 
	\includegraphics[height=3.4cm,width=3.4cm]{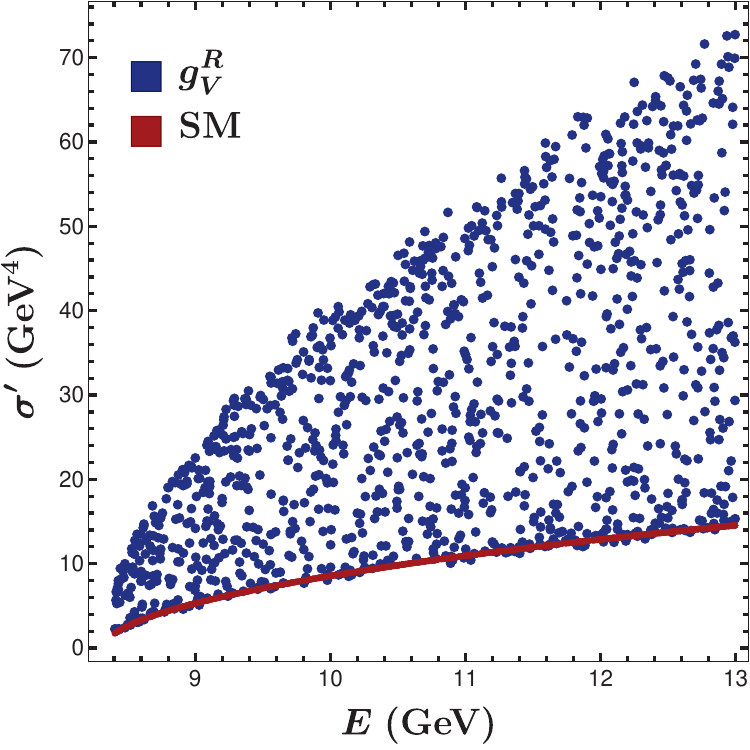}\;
	\includegraphics[height=3.4cm,width=3.4cm]{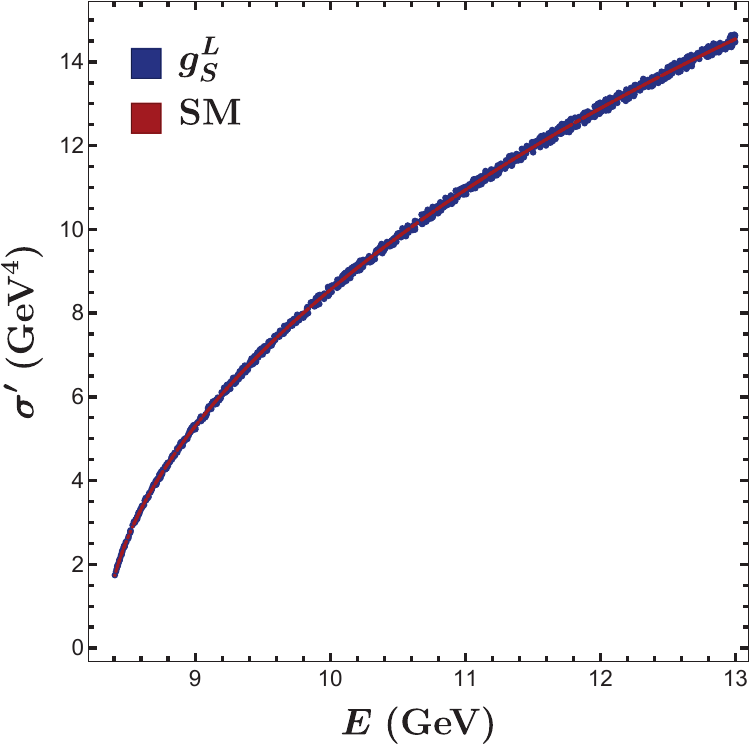}\;
	\includegraphics[height=3.4cm,width=3.4cm]{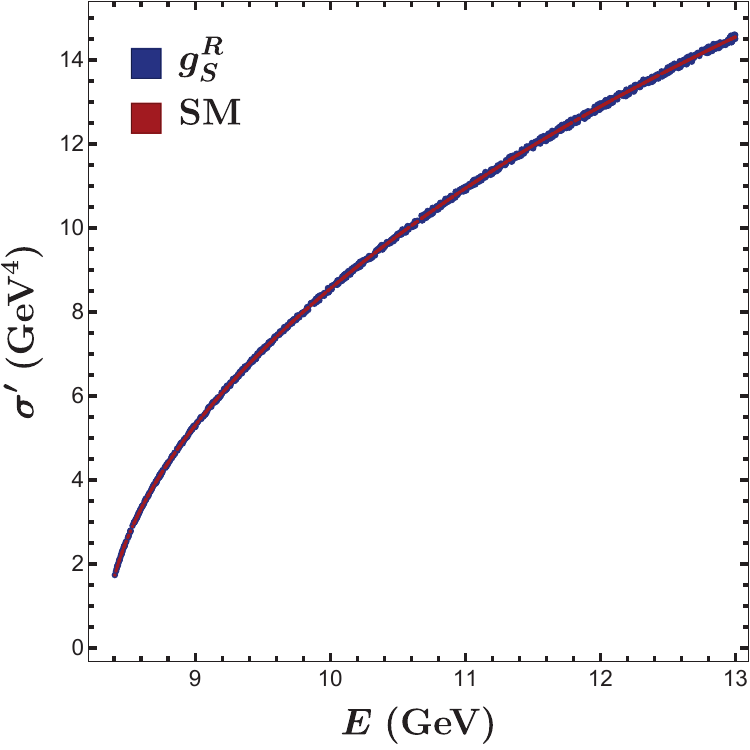}\;
	\includegraphics[height=3.4cm,width=3.4cm]{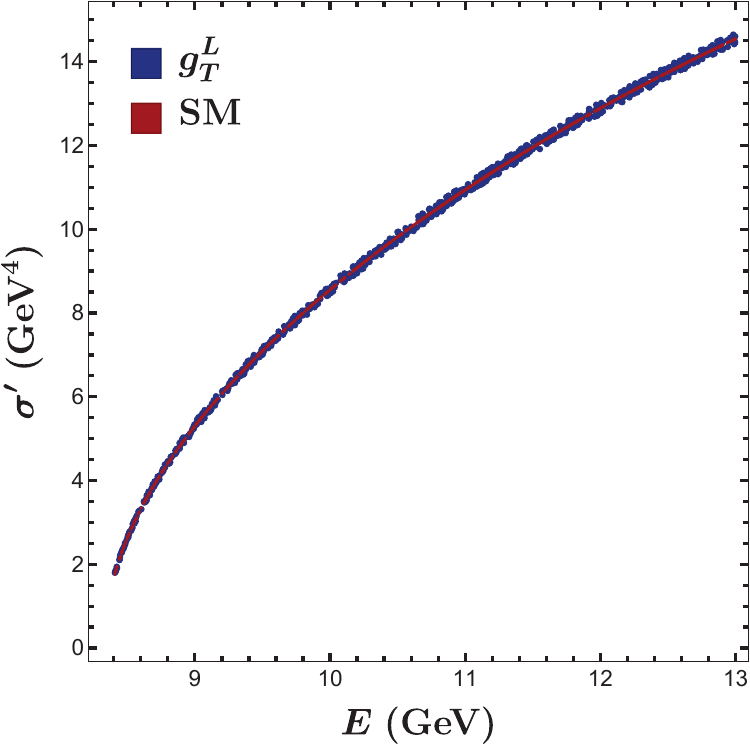}
	\caption{The total cross section $\sigma^{\prime}$, with $\sigma^{\prime}=8\pi m^2_n\sigma/(G^2_F|V_{cd}|^2)$, of the scattering process $\nu_{\tau}+n\to \tau^-+\Lambda_c$ as a function of the neutrino energy $E$. The dark red curve denotes the SM contribution, while the dark blue points represent the total contributions from both the SM and the NP in the presence of a single $g_i$, whose values are varied randomly within the overlapped regions in color shown in Fig.~\ref{fig:constraints}.}
	\label{fig:TotalCross}
\end{figure*} 

We combine in Fig.~\ref{fig:constraints} the aforementioned constraints at the $1\sigma$ level. It can be seen that the most stringent constraints on $g^L_S$, $g^R_S$, and $g^L_T$ are set by the high-$p_T$ dilepton invariant mass tails, whereas the bound on $g^R_V$ is entirely dominated by the measured branching fraction of $D^+\!\to\! \tau^+\nu_{\tau}$ decay. Meanwhile, although the boundary of the real part of $g^L_V$ is set by the high-$p_T$ dilepton invariant mass tails, the imaginary part is bounded by the $D^+\!\to\! \tau^+\nu_{\tau}$ decay, as indicated by the overlapped region in color. 
It should be pointed out that all the constraints denoted by the colored regions in Fig.~\ref{fig:constraints} are obtained by setting the rest of WCs to zero. In order to fully constrain the NP operators in Eq.~\eqref{eq:Leff}, more processes and observables are clearly needed. 

Our proposed QE scattering process together with the polarization vectors, as will be shown in the next section, is exactly what one is looking for. 
Before delving into detailed numerical analyses to justify this statement, let us take the $g^R_V$ case (i.e., except for $g^R_V\neq0$ all the other WCs vanish) for a simple illustration.  
From Fig.~\ref{fig:constraints} we have observed that the WC $g^R_V$ is solely constrained by the $D^+\!\to\! \tau^+\nu_{\tau}$ decay, 
as denoted by the pink ring area. For simplicity, let us drop the errors of the constraint for the moment, 
so that the ring now becomes a circle (see Eq.~\eqref{eq:D_decay}). 
Meanwhile, the (differential) cross section of our proposed scattering process 
can also provide a constraint, which will be denoted by another circle (see Eq.~\eqref{eq:amplitude2}). 
Assuming these two circles intersect at two points---as it happens quite often---one then obtains two sets of possible values 
for the real and imaginary parts of $g^R_V$. To further identify the correct one, 
one must invoke another observable that involves $g^R_V$. 
Clearly, the detailed formulae of $P^{l,h}_a$ in Appendix~\ref{app:numerator} 
indicate that those polarization observables can fill the gap. 
Nevertheless, it should be pointed out that compared with the cross sections of the scattering process, 
the polarization observables are generally more difficult to measure, and thus it will be experimentally more demanding to obtain the same accuracy of those observables as of the cross sections.     

\section{Numerical Results and Discussions}
\label{sec:Numresults}

\subsection{Total cross section and average polarizations}
\label{sec:totalcross}

\begin{figure*}[t]
	\centering
	\includegraphics[width=0.24\textwidth]{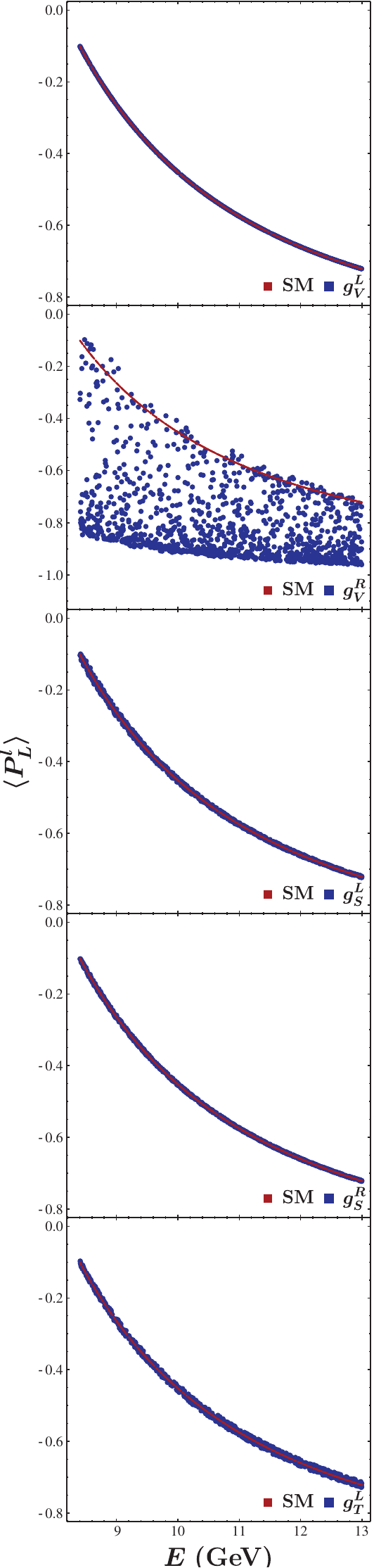}\;
	\includegraphics[width=0.24\textwidth]{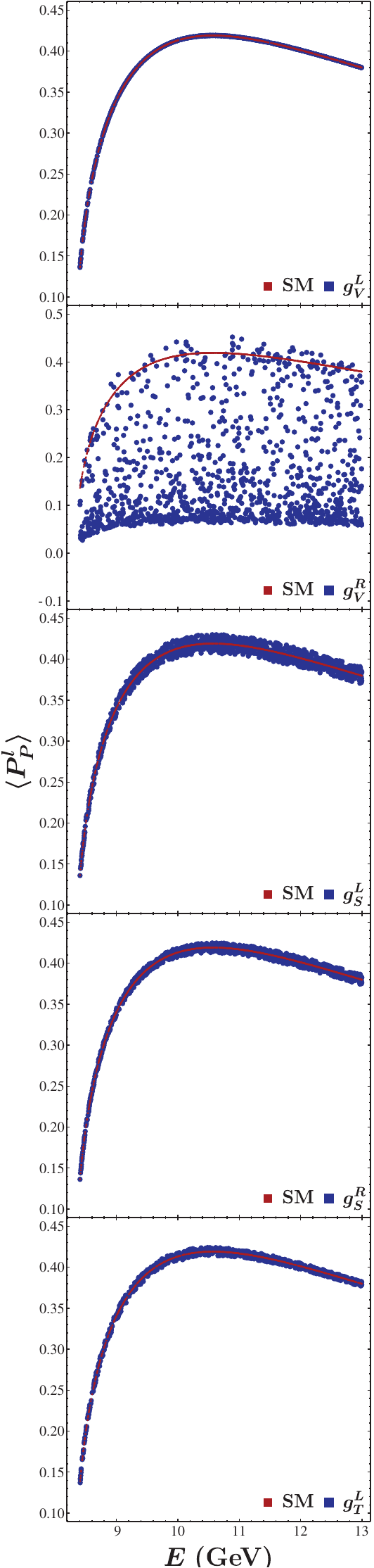}\;
	\includegraphics[width=0.245\textwidth]{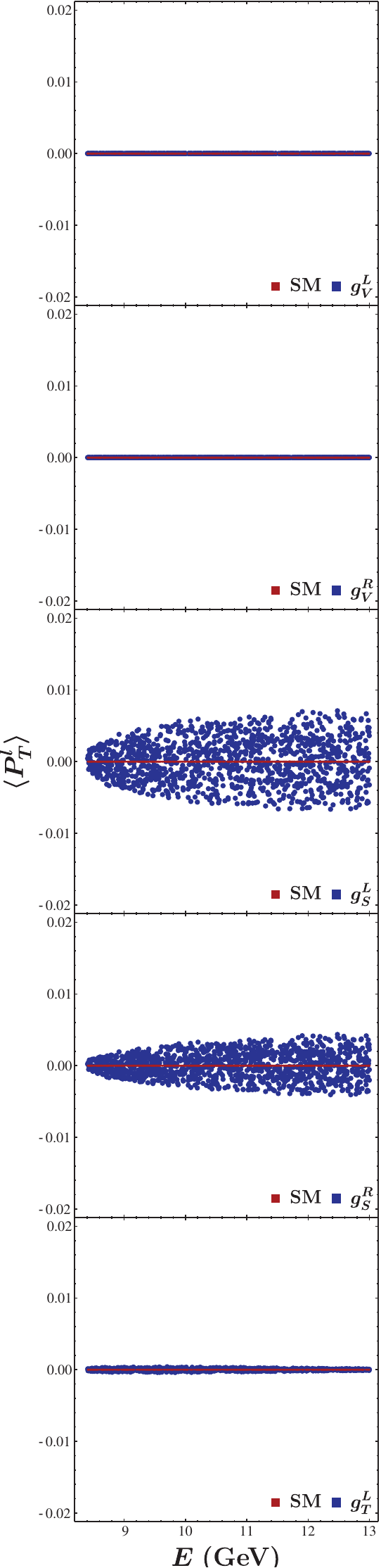}\;
	\includegraphics[width=0.235\textwidth]{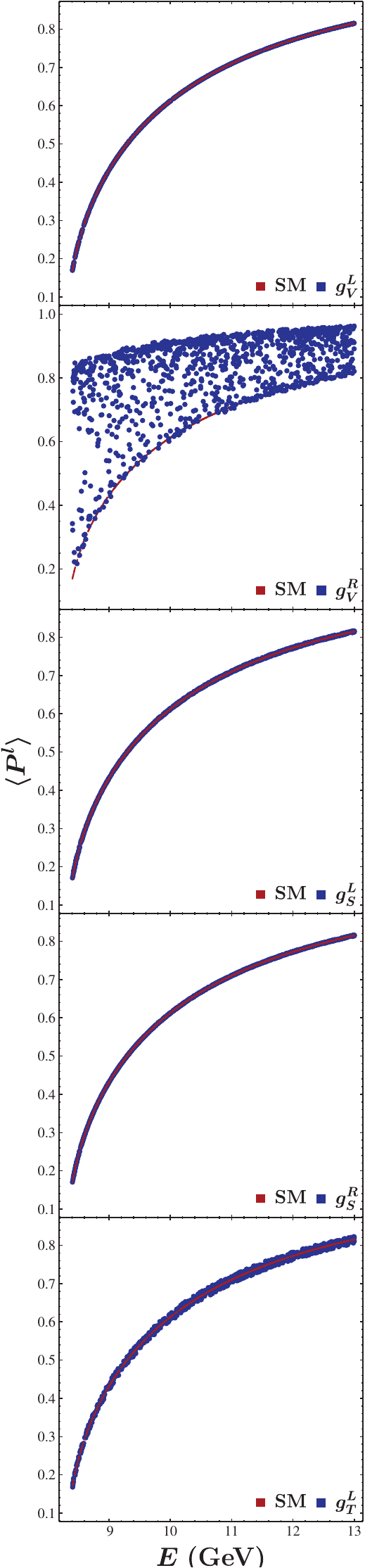}
	\caption{The average polarizations $\langle P^{l}_L\rangle$, $\langle P^{l}_P\rangle$, $\langle P^{l}_T\rangle$, and $\langle P^{l}\rangle$ for the scattering process $\nu_{\tau}+n\to \tau^-+\Lambda_c$ as a function of the neutrino energy $E$. The color captions are the same as in Fig.~\ref{fig:TotalCross}.}
	\label{fig:leptonpolarization}
	\vspace{-0.3cm}
\end{figure*}

\begin{figure*}[t]
	\centering
	\includegraphics[width=0.2408\textwidth]{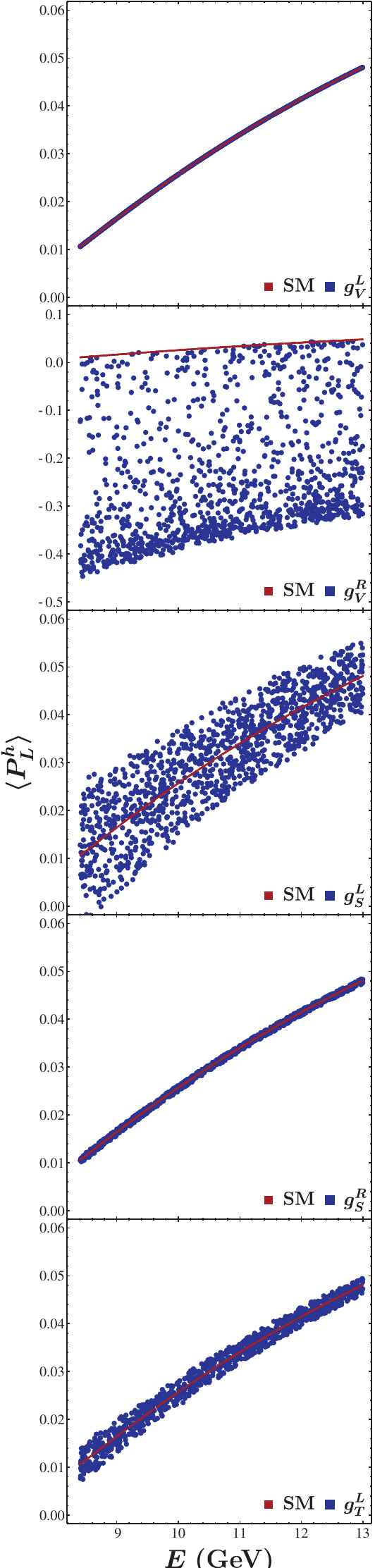}\;
	\includegraphics[width=0.2348\textwidth]{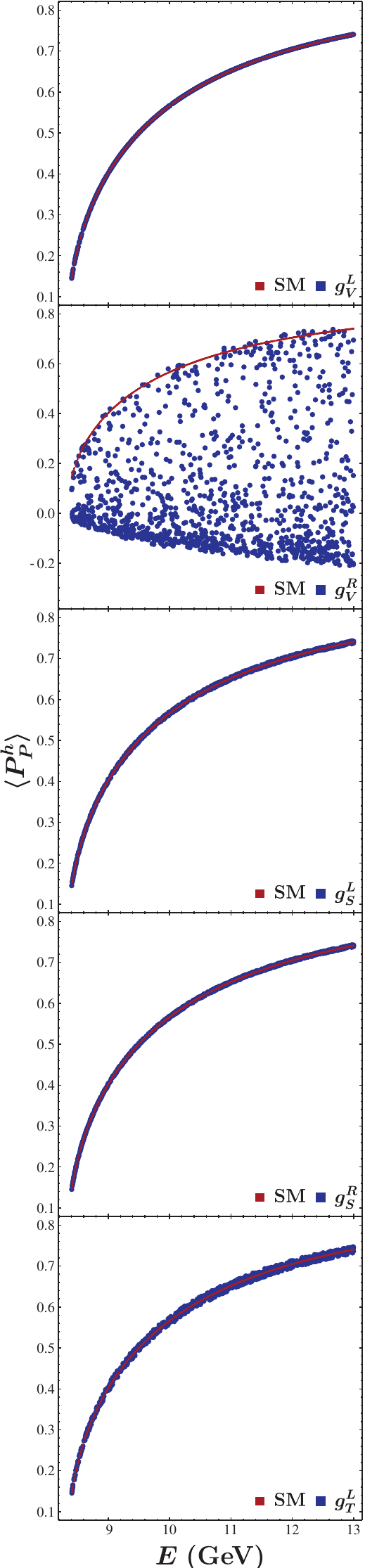}\;
	\includegraphics[width=0.245\textwidth]{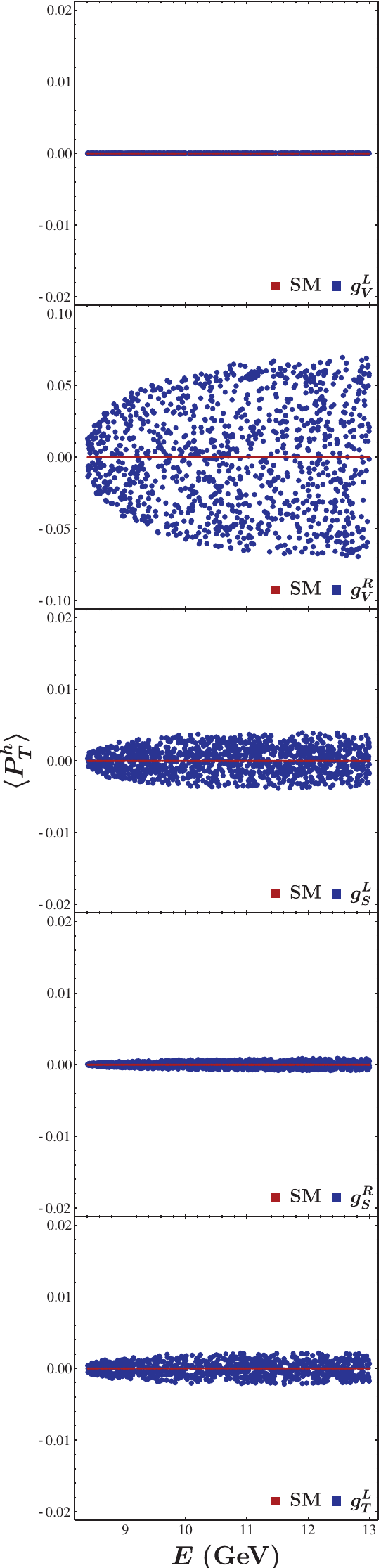}\;
	\includegraphics[width=0.234\textwidth]{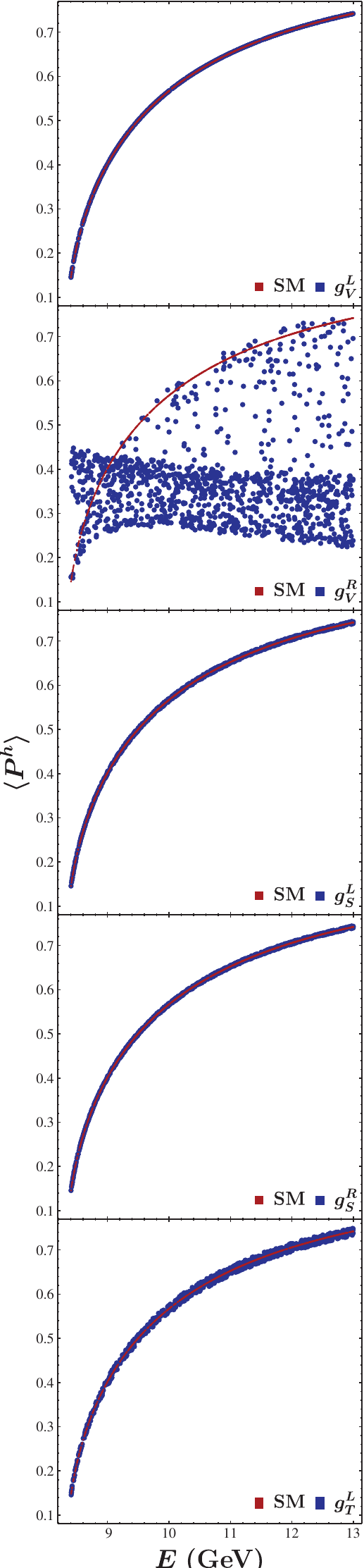}
	\caption{The average polarizations $\langle P^{h}_L\rangle$, $\langle P^{h}_P\rangle$, $\langle P^{h}_T\rangle$, and $\langle P^{h}\rangle$ for the scattering process $\nu_{\tau}+n\to \tau^-+\Lambda_c$ as a function of the neutrino energy $E$. The color captions are the same as in Fig.~\ref{fig:TotalCross}.}
	\label{fig:hadronpolarization}
	\vspace{-0.3cm}
\end{figure*}

We start with studying the dependence of the total cross section $\sigma^{\prime}$, with $\sigma^{\prime}=8\pi m^2_n\sigma/(G^2_F|V_{cd}|^2)$, and the average polarizations $\langle P^{l,h}_{a}\rangle$ on the neutrino energy $E$. To this end, by considering the range $E\in [8.33,13]$~GeV and varying randomly the WCs $g_i$ within the overlapped regions in color shown in Fig.~\ref{fig:constraints}, we plot in Fig.~\ref{fig:TotalCross} the total cross section $\sigma$ of the scattering process $\nu_{\tau}+n\to \tau^-+\Lambda_c$ as a function of $E$, both within the SM and in various NP scenarios.\footnote{For simplicity, we will neglect the possible nuclear effects~\cite{Kuzmin:2004yb,Hagiwara:2003di,Graczyk:2004uy,Sobczyk:2019urm,Nieves:2017lij} when discussing all the observables, which induce additional important uncertainties besides the experimental ones and the ones to be discussed in the subsections~\ref{sec:Obswitherror} and \ref{sec:obsdiffform}.} 
It can be seen that a few interesting features already emerge. Firstly, a higher beam energy clearly favors a larger total cross section. Secondly, the cross section can be significantly affected by the allowed parameter space of $g^R_V$ and $g^L_V$ shown in Fig.~\ref{fig:TotalCross}, especially by the former. This in turn indicates larger opportunity for improving the limits on $g^{L,R}_V$ through the proposed QE scattering process. On the other hand, for $g^L_S$, $g^R_S$, and $g^T_S$, stringent constraints from the high-$p_T$ dilepton invariant mass tails do not leave much room for possible deviations from the SM predication. Thus, to further improve the constraints on these $g_i$, demanding experimental setup for the scattering process is certainly necessary. Finally, although the allowed parameter spaces for $g^L_S$ and $g^R_S$ are identical to each other (see Eq.~\eqref{eq:D_decay} and Fig.~\ref{fig:constraints}), their imprints on the total cross section are slightly different, especially 
at the high-$E$ range, as shown vaguely in Fig.~\ref{fig:TotalCross}. 
Such a small difference in fact results from the different interference between $\mathcal{O}^L_V$ and $\mathcal{O}^{L,R}_S$; more details could be found in Appendix~\ref{app:Amplitude}.

In Fig.~\ref{fig:leptonpolarization}, we show the average polarizations $\langle P^{l}_L\rangle$, $\langle P^{l}_P\rangle$, 
$\langle P^{l}_T\rangle$, and $\langle P^{l}\rangle$ of the $\tau$ lepton as a function of the neutrino beam energy $E$ in various scenarios. Let us start with the SM case. As depicted by the red curves in Fig.~\ref{fig:leptonpolarization}, both the absolute values of $\langle P^{l}_L\rangle$ and $\langle P^{l}\rangle$ increase along with the increase of $E$, which is not surprising, since the $\tau$ lepton produced through the scattering process $\nu_{\tau}+n\to \tau^-+\Lambda_c$ is left-handed in the SM. On the other hand, $\langle P^{l}_P\rangle$ reaches its peak around $E=10$~GeV, while $\langle P^{l}_T\rangle=0$ irrespective of $E$ because $P^{l}$ in this case misses the terms containing $\varepsilon_{\{k\}\{k^\prime\}\{N_a\}\{p\}}$,\footnote{Note that $P^{l,h}_T$ will also vanish if all the WCs $g_i$ are real, since  $\varepsilon_{\{k\}\{k^\prime\}\{N_a\}\{p\}}$ is always accompanied by the imaginary unit $i$, as shown in Appendix~\ref{app:numerator}.}
which essentially characterize the $T$-component of $P^{l}$; see Appendix~\ref{app:numerator} for more details. 
Note that $\langle P^{l}_T\rangle=0$ in the SM qualifies itself as a null test observable. Measuring a tiny but nonzero $\langle P^{l}_T\rangle$ induced by NP effects could be, however, challenging, as indicated by the plots in the third column of Fig.~\ref{fig:leptonpolarization}. 

We now move on to the NP scenarios. From the four figures on the top panel in Fig.~\ref{fig:leptonpolarization}, we observe that contributions to the average polarization $\langle P^{l}_{a}\rangle$ from the SM and the WC $g^L_{V}$ are indistinguishable, because they share the same effective operator $\mathcal{O}_V^L$ (see Eq.~\eqref{eq:Leff}). For the WC $g^R_{V}$, on the other hand, large deviations of $\langle P^{l}_{L,P}\rangle$ from their SM predictions are possible due to the sizable allowed parameter space of $g^R_{V}$, while $\langle P^{l}_T\rangle$ still remains zero in this case due to the same reason as in the SM. Similar to the case of total cross section, possible deviations of all $\langle P^{l}_a\rangle$ from their SM predictions are relatively small for the WCs $g^L_{S}$, $g^R_{S}$, and $g^L_{T}$ due to the stringent constraints on them from the current data, as shown in Fig.~\ref{fig:constraints}. 

Similar to the SM case, we can make the following observations in the NP scenarios. Firstly, there exist small differences between $\langle P^{l}_{a}\rangle$ associated with the WCs $g^L_{S}$ and $g^R_{S}$ due to their different operator structures. One can see that the overall blue bands from $g^L_{S}$ are slightly broader than from $g^R_{S}$ in the $\langle P^{l}_{a}\rangle$-$E$ planes. Secondly, the fuzzy blue bands in the $\langle P^{l}_{L}\rangle$-$E$ plane from $g^{L,R}_{S}$ imply that a relatively low $E$ is more favored to further constrain these two WCs, whereas a relatively high $E$ would be more advantaged for further limiting $g^{L}_{T}$ through $\langle P^{l}_{L}\rangle$. The situation is, however, totally opposite in probing
$g^L_{S}$, $g^R_{S}$, and $g^L_{T}$ through $\langle P^{l}_{T}\rangle$. Finally, only a relatively high $E$ is favored for probing $g^L_{S}$, $g^R_{S}$, and $g^L_{T}$ through $\langle P^{l}_{P}\rangle$.    

We also show in Fig.~\ref{fig:hadronpolarization} the average polarizations $\langle P^{h}_L\rangle$, $\langle P^{h}_P\rangle$, 
$\langle P^{h}_T\rangle$, and $\langle P^{h}\rangle$ of the $\Lambda_c$ baryon as a function of $E$. Contrary to the $\tau$-lepton case, the predominant polarization mode of the $\Lambda_c$ baryon produced through the QE scattering
process is perpendicular in the SM. Although $\langle P^{h}_L\rangle$ increases along with the increase of $E$, its overall polarization degree is only of $\mathcal{O}(10^{-2})$. Meanwhile, $\langle P^{h}_T\rangle$ is always zero irrespective of $E$ for a similar reason as $\langle P^{l}_T\rangle$ in the $\tau$-lepton case.

\begin{figure*}[ht]
	\centering
	\includegraphics[width=0.235\textwidth]{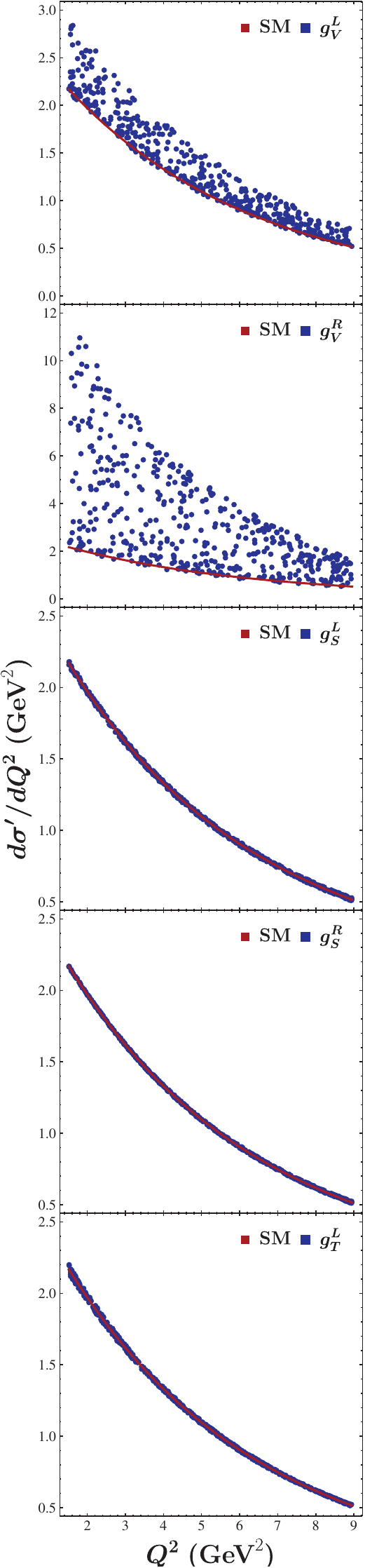}\;
	\includegraphics[width=0.237\textwidth]{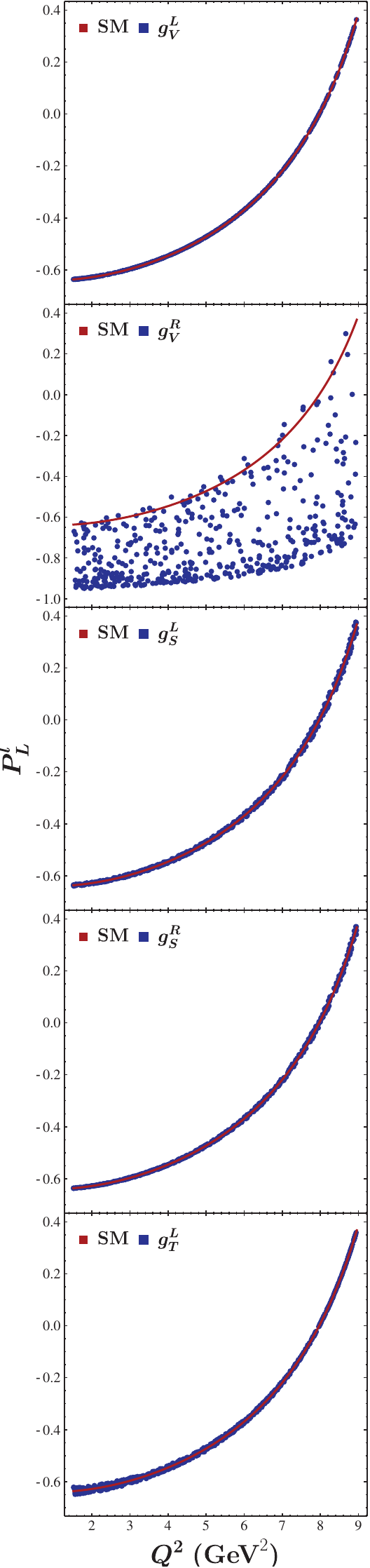}\;
	\includegraphics[width=0.233\textwidth]{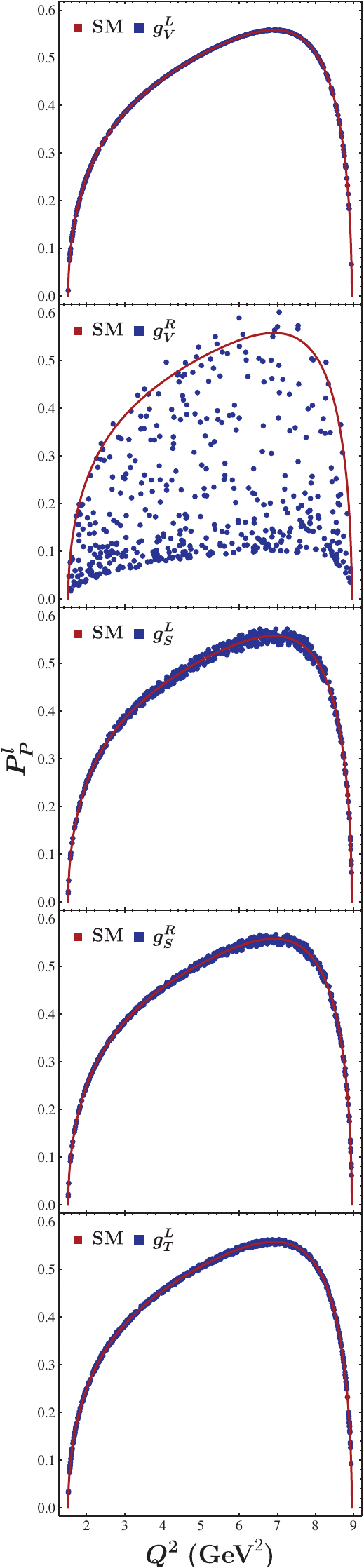}\;
	\includegraphics[width=0.243\textwidth]{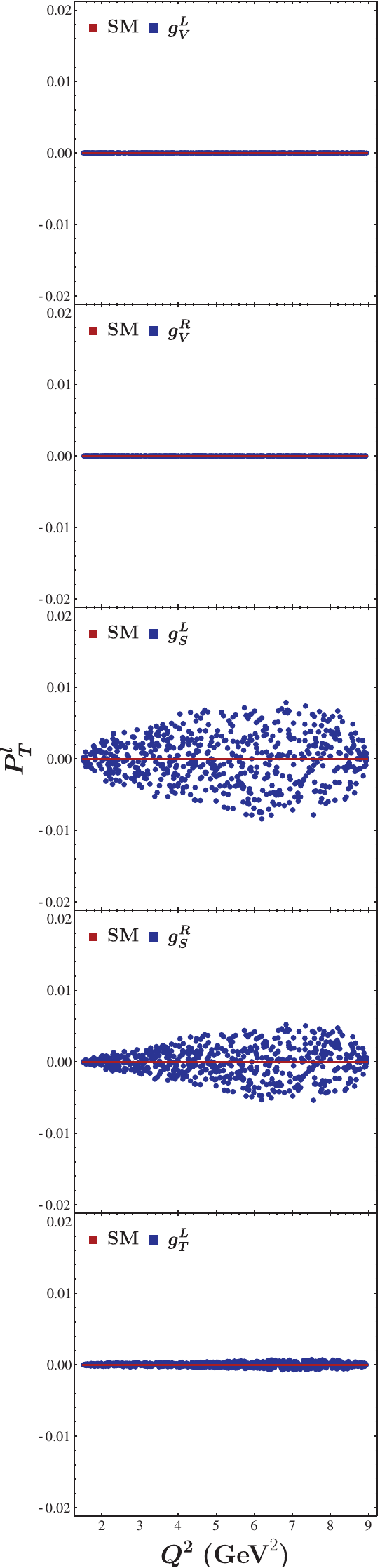}
	\caption{Variations of the differential cross section as well as the polarizations $P^{l}_L$, $P^{l}_P$, and $P^{l}_T$ with respect to $Q^2$, where we have set the neutrino beam energy at $E=10$~GeV, after taking into account the interesting behavior of $\langle P^{l}_P\rangle$ shown in Fig.~\ref{fig:leptonpolarization} and the neutrino beam flux at the DUNE~\cite{DUNE2020,Machado:2020yxl}. The color captions are the same as in Fig.~\ref{fig:TotalCross}.}
	\label{fig:leptonpolarizationQ2}
\end{figure*}

\begin{figure*}[ht]
	\centering
	\includegraphics[width=0.243\textwidth]{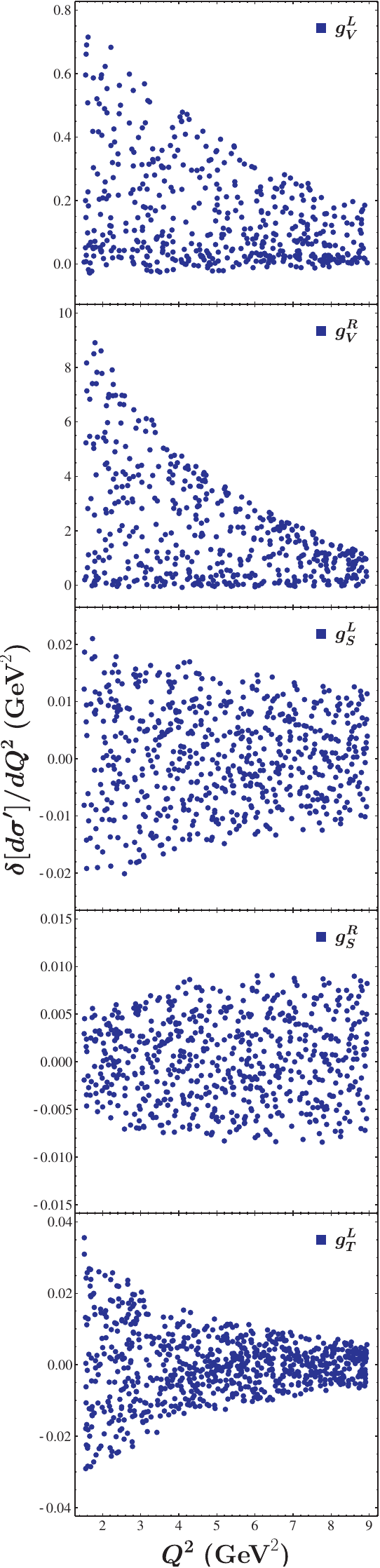}\;
	\includegraphics[width=0.241\textwidth]{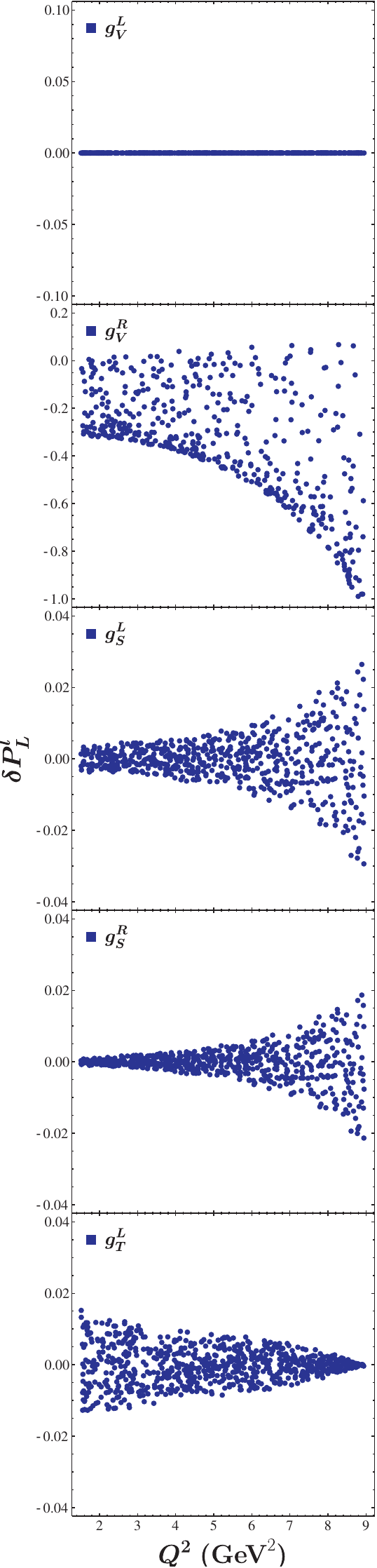}\;
	\includegraphics[width=0.241\textwidth]{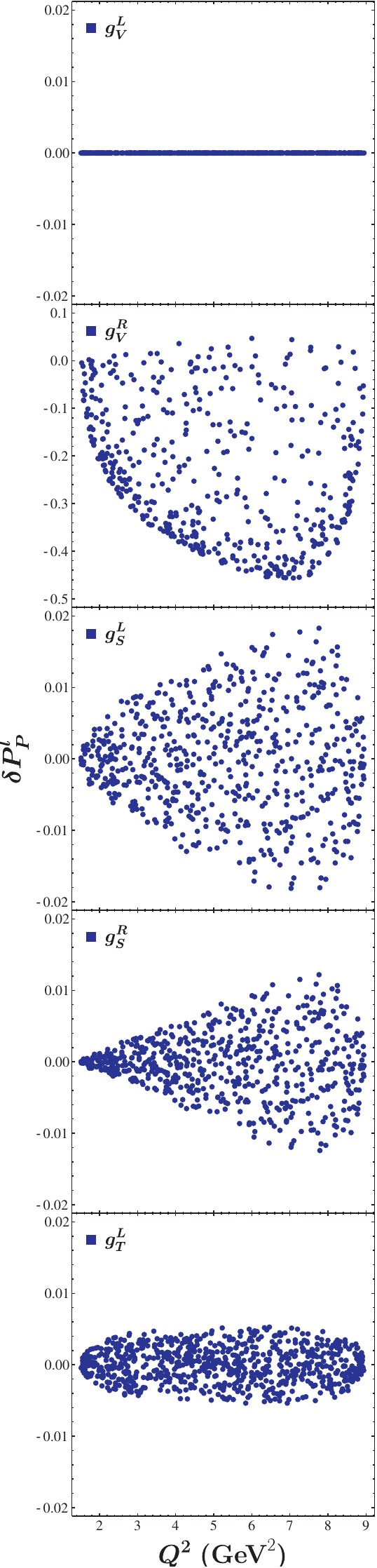}\;
	\includegraphics[width=0.243\textwidth]{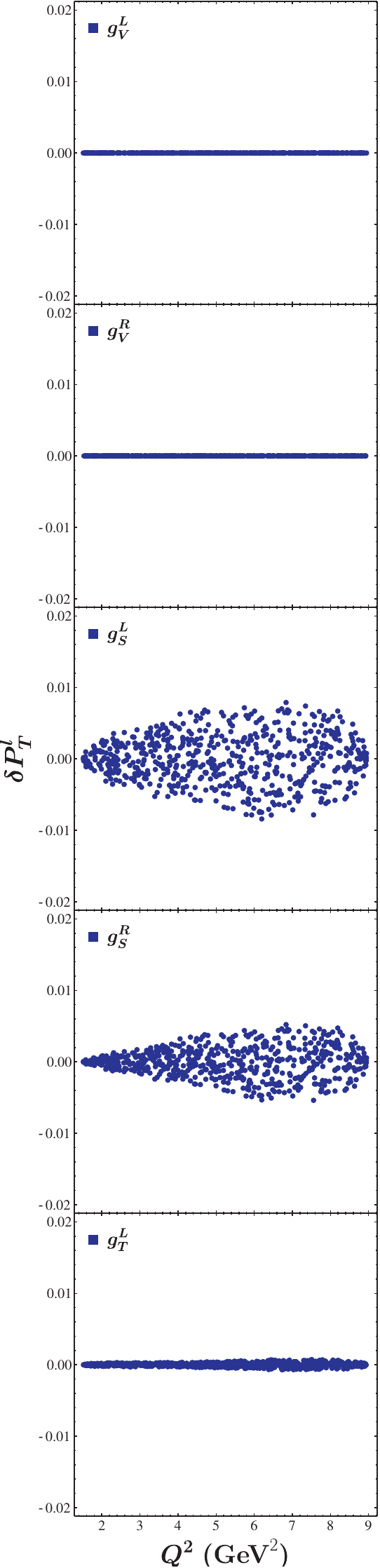}
	\caption{Deviations from the SM predictions for the differential cross section and the polarizations $P^{l}_L$, $P^{l}_P$, and $P^{l}_T$ in different NP scenarios.}
	\label{fig:diffleptonpolarizationQ2}
\end{figure*}

\begin{figure*}[htbp]
	\centering
	\includegraphics[width=0.2895\textwidth]{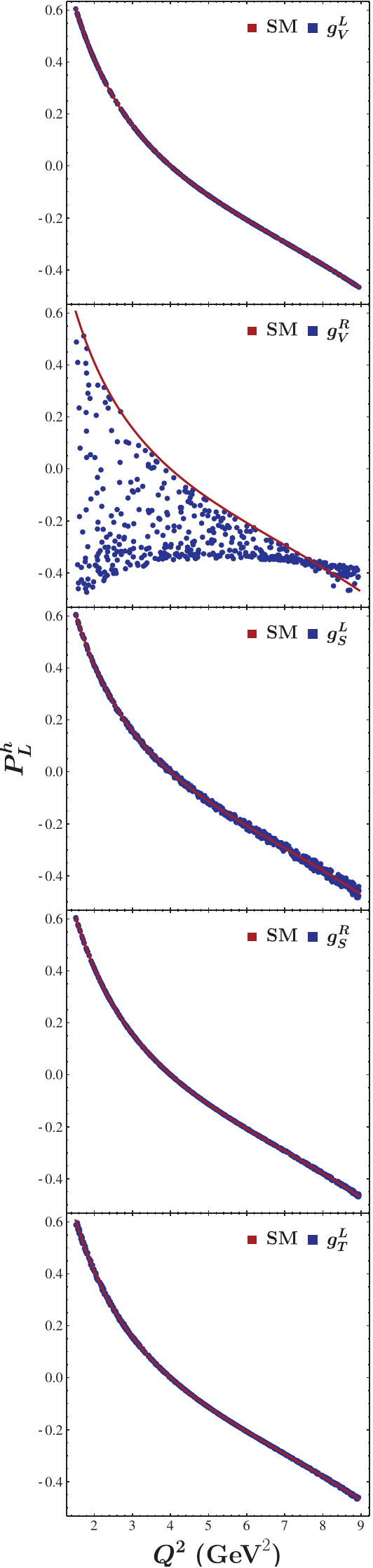}\quad
	\includegraphics[width=0.2825\textwidth]{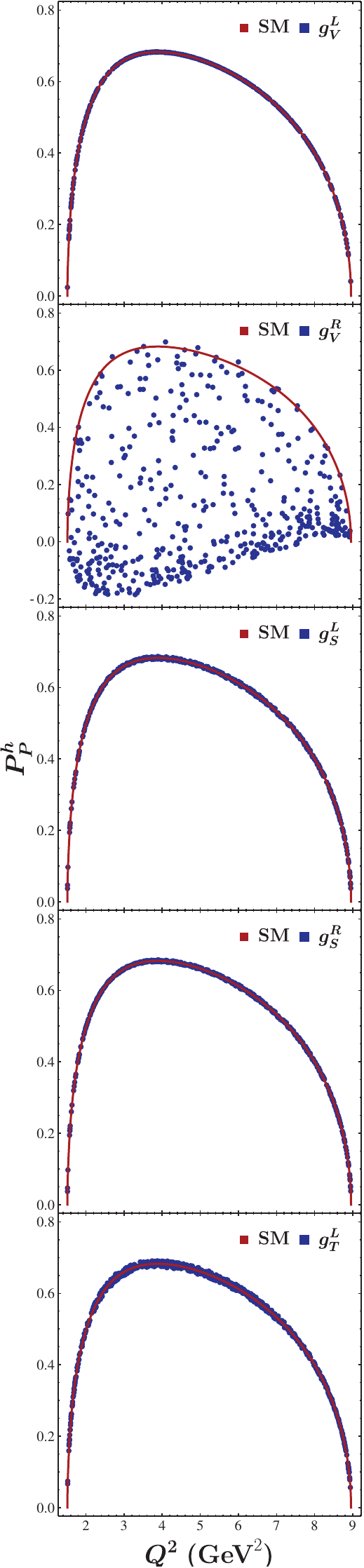}\quad
	\includegraphics[width=0.295\textwidth]{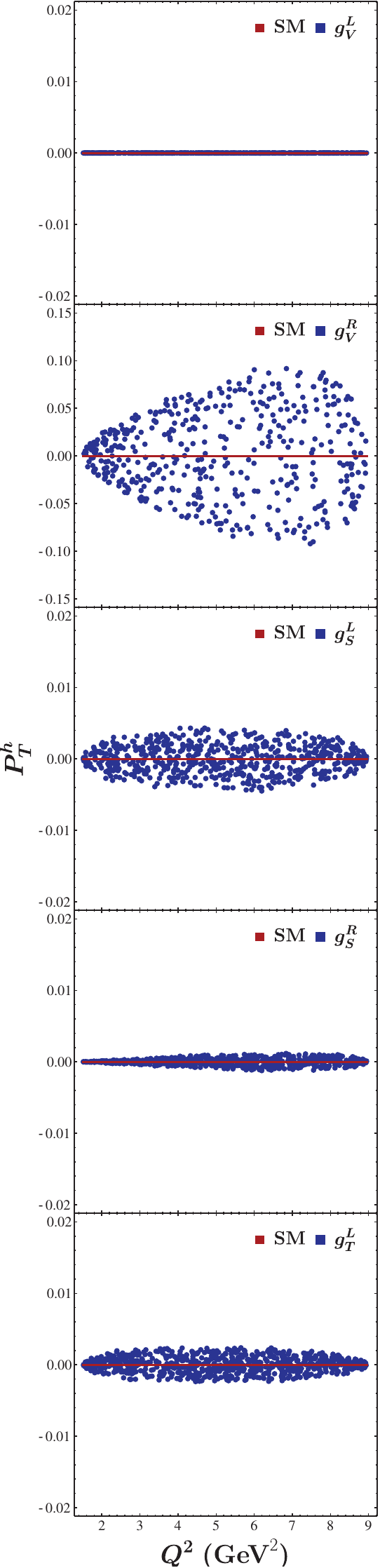}
	\caption{Variations of the polarizations $P^{h}_L$, $P^{h}_P$, and $P^{h}_T$ with respect to $Q^2$, where the neutrino beam energy has also been set at $E=10$~GeV for consistency. The color captions are the same as in Fig.~\ref{fig:TotalCross}.}
	\label{fig:hadronpolarizationQ2}
\end{figure*}

\begin{figure*}[htbp]
	\centering
	\includegraphics[width=0.29\textwidth]{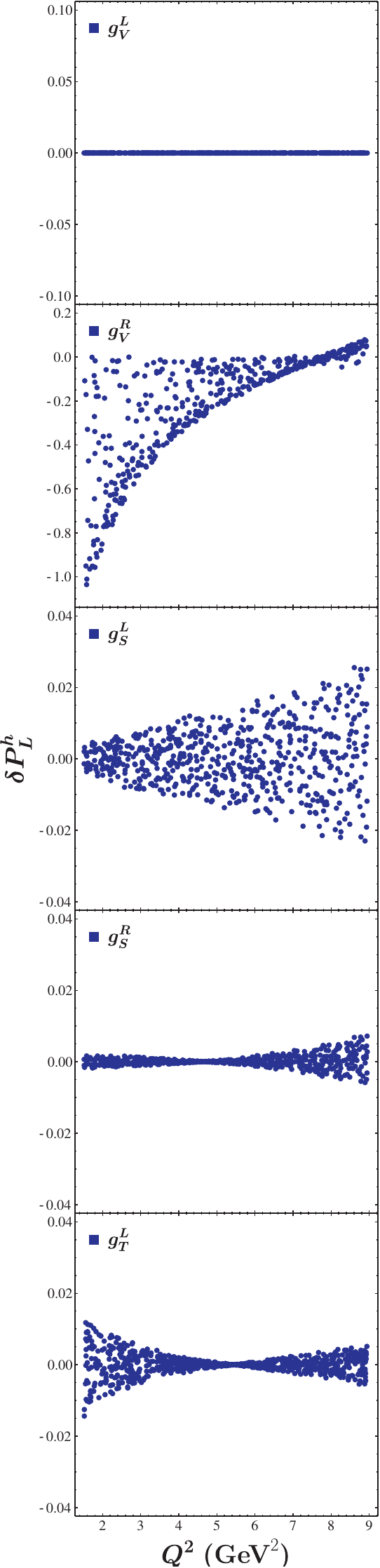}\;
	\includegraphics[width=0.29\textwidth]{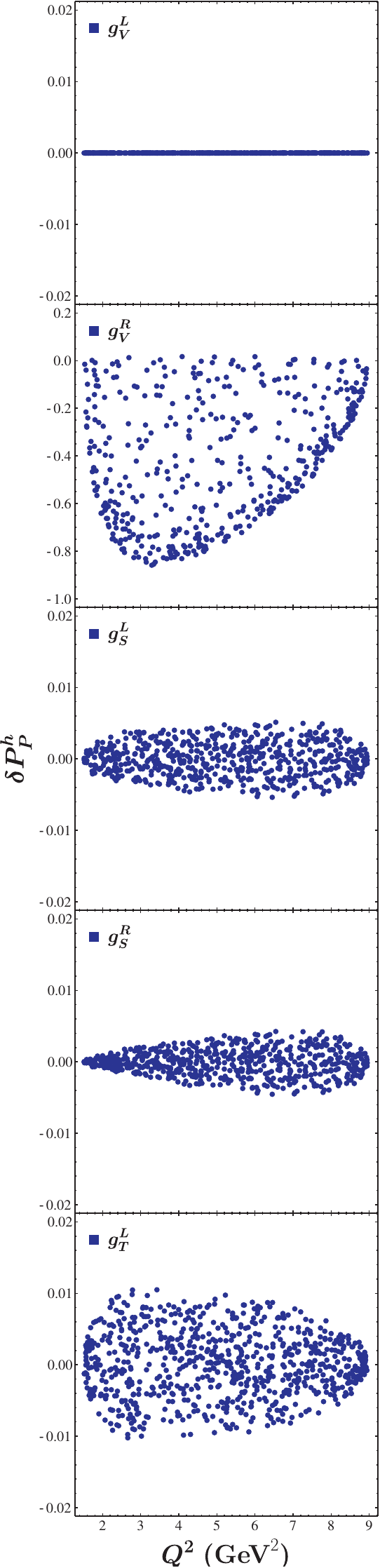}\;
	\includegraphics[width=0.29\textwidth]{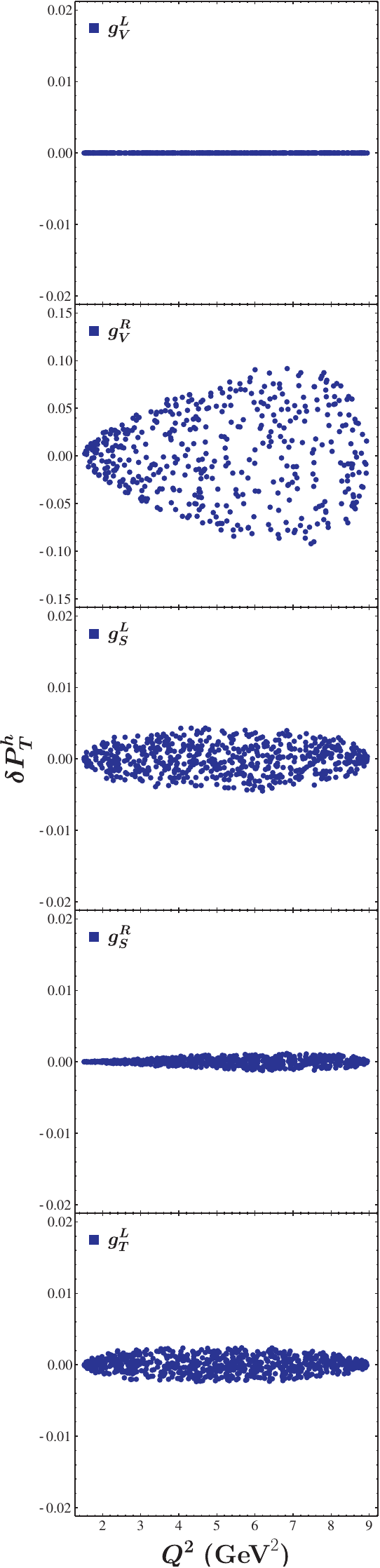}
	\caption{Deviations from the corresponding SM predictions for the polarizations $P^{h}_L$, $P^{h}_P$, and $P^{h}_T$ in different NP scenarios.}
	\label{fig:dhadronpolarizationQ2}
\end{figure*} 

For the NP scenarios in this case, we observe some similar features too. Firstly, the average polarizations $\langle P^{h}_{a}\rangle$ induced by   $g^L_{V}$ are also indistinguishable from the SM case, as shown by the first four plots on the top panel in Fig.~\ref{fig:hadronpolarization}, due to the same reason as mentioned in the $\tau$-lepton case. Secondly, a large opportunity exists clearly for improving the limit on $g^{R}_V$ through the measurements of these polarization vectors of the $\Lambda_c$ baryon. Note that, contrary to $\langle P^{l}_T\rangle$, $\langle P^{h}_T\rangle$ would be nonzero in the presence of the very same NP scenario. Finally, all $\langle P^{h}_a\rangle$ induced by $g^L_{S}$, $g^R_{S}$, and $g^L_{T}$ are small due to the stringent constraints on these WCs. However, given the small value of $\langle P^{h}_L\rangle$ predicted in the SM, possible deviations induced by these NP effects, especially by $g^L_{S}$, could still reach more than $100\%$ at the low-$E$ range. 

\subsection{Differential cross section and $Q^2$-dependent polarizations}
\label{sec:diffcross}

Taking into account the interesting behavior of $\langle P^{l}_P\rangle$ shown in Fig.~\ref{fig:leptonpolarization} and the neutrino beam flux at the DUNE~\cite{DUNE2020,Machado:2020yxl}, we will set $E=10$~GeV as our benchmark beam energy and explore how the differential cross section 
and the polarizations $P^{l,h}_{a}$ vary with respect to $Q^2$. To this end, by letting the WCs $g_i$ vary randomly within the overlapped regions in color shown in Fig.~\ref{fig:constraints}, we plot in Fig.~\ref{fig:leptonpolarizationQ2} the resulting differential cross sections and polarizations $P^{l}_{a}$ as a function of $Q^2$ in various NP scenarios, together with the SM predictions. Let us scrutinize the SM case first. As indicated by the red curves in Fig.~\ref{fig:leptonpolarizationQ2}, the differential cross section of the scattering process clearly prefers the low-$Q^2$ range in the SM. A similar conclusion also holds for the polarization $P^{l}_{L}$, even though it experiences a crossover at $Q^2\simeq 8~\mathrm{GeV}^2$. 
$P^{l}_{P}$ peaks roughly at $Q^2\simeq 8~\mathrm{GeV}^2$, while unsurprisingly $P^{l}_{T}$ remains zero irrespective of $Q^2$.  

We now move on to discuss the NP scenarios shown in Fig.~\ref{fig:leptonpolarizationQ2}, from which an overall pattern similar to that found in the previous subsection is observed. Firstly, large deviations from the SM prediction for the differential cross section are only possible for $g^{L,R}_V$, while large deviations for the polarizations $P^{l}_{L,P}$ can be expected only for $g^{R}_V$. Secondly, due to the stringent experimental constraints on $g^L_{S}$, $g^R_{S}$, and $g^L_{T}$, deviations from the SM predictions for the differential cross section and the polarizations $P^{l}_{a}$ in these three NP scenarios become much smaller. 

To have a clearer view of these deviations from the corresponding SM predictions, let us define $\delta[d\sigma^\prime]/dQ^2=d\sigma^\prime/dQ^2|_\mathrm{NP} -d\sigma^\prime/dQ^2|_\mathrm{SM}$ and $\delta P^{l,h}_a=P^{l,h}_a|_\mathrm{NP}-P^{l,h}_a|_\mathrm{SM}$, and
plot them explicitly in Fig.~\ref{fig:diffleptonpolarizationQ2}.
It can be seen that the deviations $\delta P^{l}_{a}$ remain zero for the  $g^{L}_V$ scenario, making the (differential) cross section the only avenue to probe $g^{L}_V$ through the scattering process. For $g^{R}_V$, a relatively high $Q^2$ is certainly preferred to observe the potentially maximum deviations $\delta P^{l}_{L,P}$ but at the expense of observing the maximum deviation of the differential cross section, whereas $\delta P^{l}_{T}=0$ in the whole $Q^2$ range. In the case of $g^{L}_S$ and $g^{R}_S$, the overall deviation patterns are similar for the three polarizations $P^{l}_{a}$, but opposite for the differential cross section. Nonetheless, a relatively high $Q^2$, e.g., $Q^2\simeq 7.5~\mathrm{GeV}^2$, can be of benefit for probing $g^{L}_S$ and $g^{R}_S$ through these observables. In the presence of $g^{L}_T$, on the other hand, the situation is a little complicated. From the four plots on the bottom panel, we observe that the low-$Q^2$ range clearly favors the deviations of the differential cross section and the polarization $P^{l}_{L}$, whereas the slightly high-$Q^2$ range favors the deviations $\delta P^{l}_{P,T}$. Overall, the maximum $\delta P^{l}_{L}$ and $\delta P^{l}_{P}$ could reach $1$ and $0.45$ in the $g^{R}_V$ scenario, respectively. However, the maximum $\delta P^{l}_{L}$ for the $g^{L}_S$, $g^{R}_S$, and $g^{L}_T$ scenarios could only amount to $0.02$ at most, and the situation is even more challenging for $\delta P^{l}_{P,T}$.

Similar analyses can be applied to the polarizations $P^{h}_L$, $P^{h}_P$, and 
$P^{h}_T$ of the $\Lambda_c$ baryon. In Fig.~\ref{fig:hadronpolarizationQ2}, we show the variations of these observables with respect to $Q^2$ both within the SM and in the various NP scenarios. It is found that $P^{h}_a$ exhibit similar characteristics as of $P^{l}_a$ shown in Fig.~\ref{fig:leptonpolarizationQ2}. For instance, both $P^{l}_{T}$ and $P^{h}_{T}$ remain zero irrespective of the kinematics $Q^2$. In addition, both $P^{l}_{L}$ and $P^{h}_{L}$ experience a crossover and peak at the low-$Q^2$ range. Finally, both $P^{l}_{P}$ and $P^{h}_{P}$ drop down to zero at $Q^2_{\text{min}}$ and $Q^2_{\text{max}}$. Nevertheless, distinct differences between these two sets of observables are also observed. An obvious example is that $P^{l}_{P}$ and $P^{h}_{P}$ peak at different $Q^2$, $Q^2\simeq 7~\mathrm{GeV}^2$ for the former whereas $Q^2\simeq 4~\mathrm{GeV}^2$ for the later. In addition, the crossover positions of $P^{l}_{L}$ and $P^{h}_{L}$ lie at different $Q^2$, $Q^2\simeq 8~\mathrm{GeV}^2$ for the former whereas $Q^2\simeq 4~\mathrm{GeV}^2$ for the later. 
 
With regard to $\delta P^{h}_a$, the deviations from the corresponding SM predictions for the polarizations $P^{h}_a$, our results are shown in Fig.~\ref{fig:dhadronpolarizationQ2}. Compared to the deviations $\delta P^{l}_a$ shown in Fig.~\ref{fig:diffleptonpolarizationQ2}, $\delta P^{h}_a$ are characterized by some new features. Firstly, for the $g^R_V$ scenario, in contrast to $\delta P^{l}_L$ and $\delta P^{l}_P$, $\delta P^{h}_L$ and $\delta P^{h}_P$ prefer a relatively low $Q^2$, which is also favored by the deviation of the differential cross section shown in Fig.~\ref{fig:diffleptonpolarizationQ2}. In addition, contrary to $\delta P^{l}_T$, $\delta P^{h}_T$ is not equal to zero in this scenario. Secondly, the overall sizes of $\delta P^{h}_a$ in the presence of $g^L_S$, $g^R_S$, and 
$g^L_T$ are smaller than that of $\delta P^{l}_a$, especially of $\delta P^{l}_{P,T}$. Finally, for the $g^R_S$ and $g^L_T$ scenarios, the minima of 
$\delta P^{h}_L$ arise both at the medium-$Q^2$ range, whereas the minima of $\delta P^{l}_L$ arise at the $Q^2_{\text{min}}$ and $Q^2_{\text{max}}$, respectively.   

Thus far, we have explored in detail the behaviors of the differential cross section and the polarizations $P^{l,h}_{a}$ with respect to $Q^2$ and pointed out the possible $Q^2$ regions, in which these observables reach their maxima in various scenarios. However, we have not provided any explanations of these observed behaviors. We will postpone it to the next subsection, where it will be worked out in the small-$g_i$ limit.

\subsection{Polarization observables in the small-$g_i$ limit}
\label{sec:lowglimit}

\begin{figure*}[ht]
	\centering
	\includegraphics[width=5.6cm]{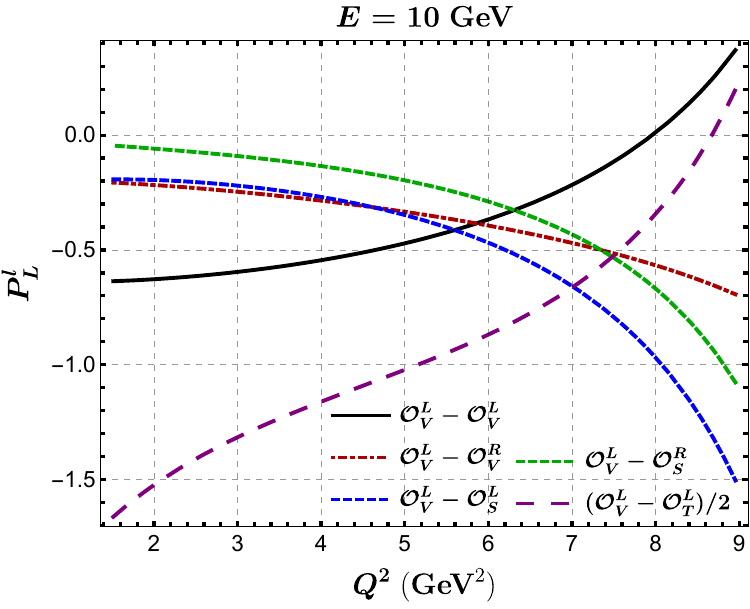}\;
	\includegraphics[width=5.6cm]{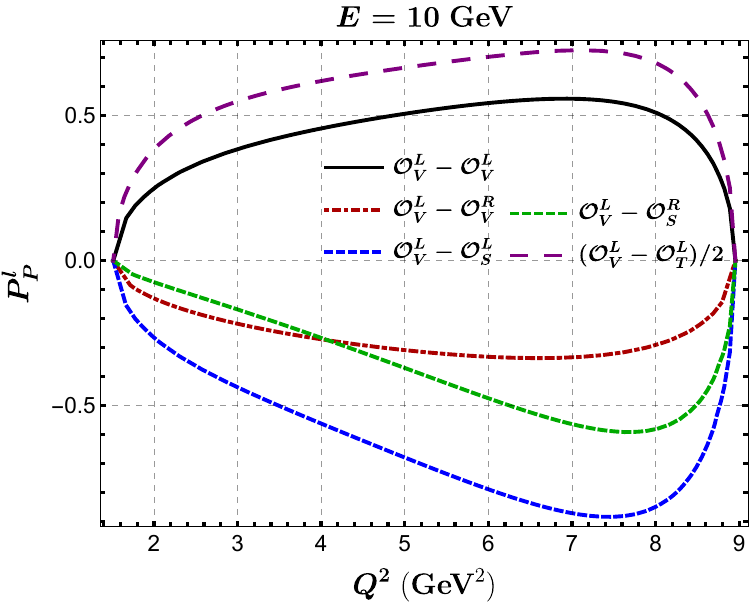}\;
	\includegraphics[width=5.51cm]{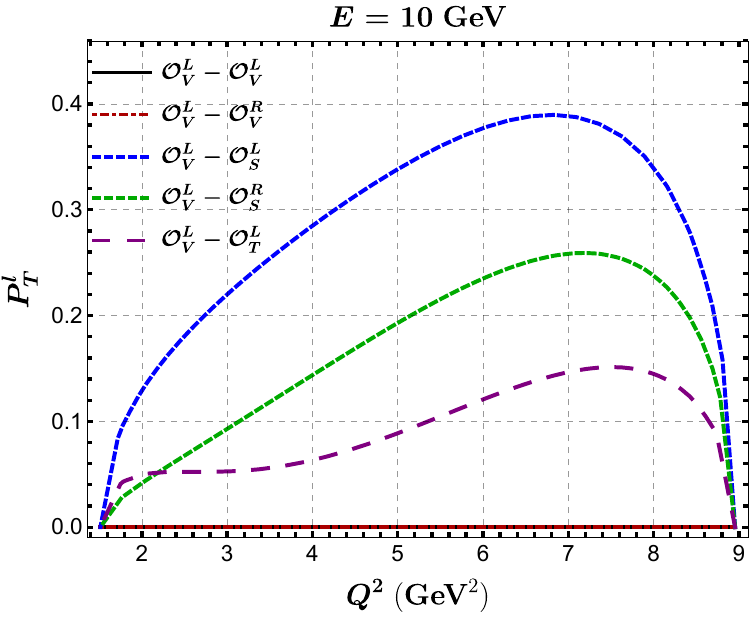}\\
	\vspace{0.2cm}
	\includegraphics[width=5.6cm]{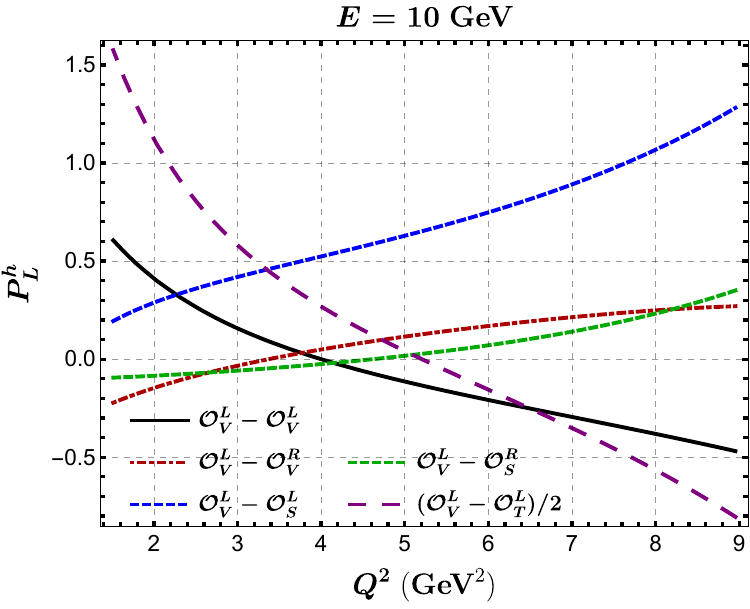}\;
	\includegraphics[width=5.6cm]{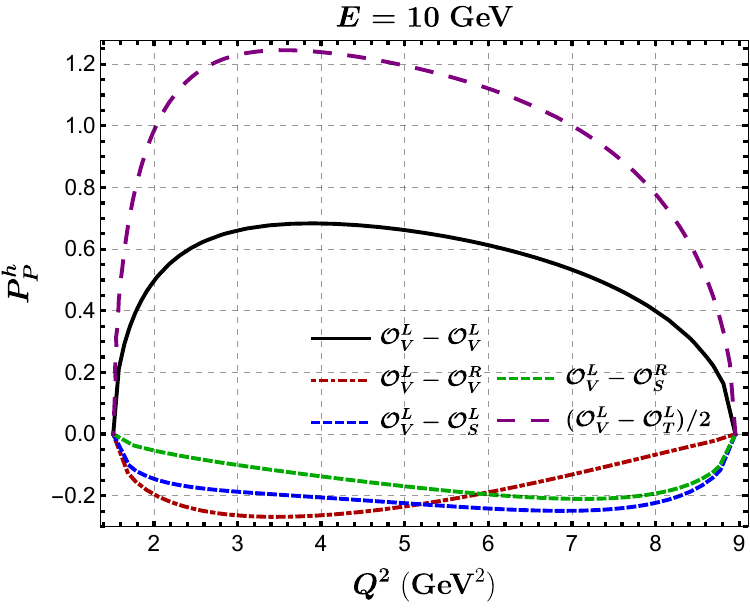}\;
	\includegraphics[width=5.545cm]{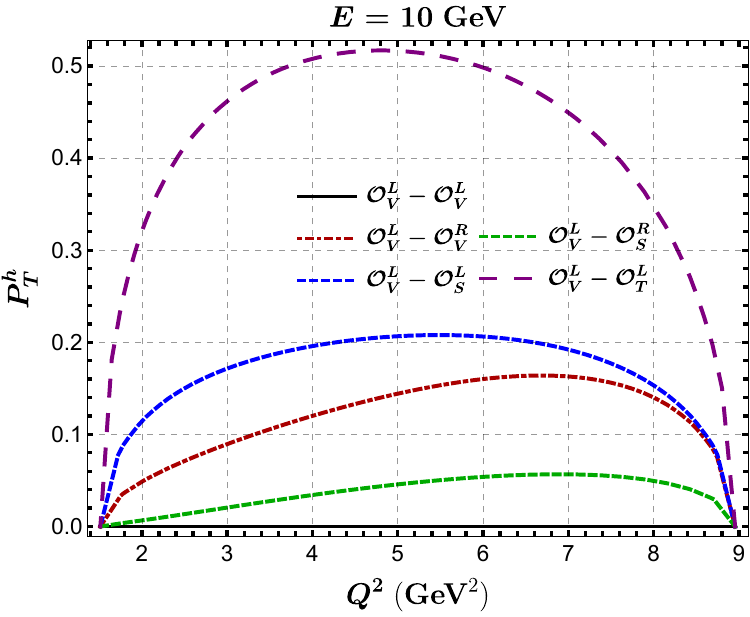}
	\caption{Variations of $(P_{\text{Int}})^l_a$ (top panel) and $(P_{\text{Int}})^h_a$ (bottom panel) with respect to $Q^2$ in different NP scenarios. Note that the mixing $\mathcal{O}^L_V-\mathcal{O}^L_V$ denoted by the solid dark curve represents in fact $(P_{\text{SM}})^{l,h}_a$, and the mixing $(\mathcal{O}^L_V-\mathcal{O}^L_T)/2$ indicates that only half of $(P_{\text{Int}})^{l,h}_a$ is depicted in this scenario.}
	\label{fig:PolaQ}
	\vspace{-0.3cm}
\end{figure*}

In the previous subsections, we have let the WCs $g_i$ vary randomly within the overlapped regions in color shown in Fig.~\ref{fig:constraints}, which are set by the measured branching fraction of $D^+\!\to\!\tau^+\nu_{\tau}$ decay~\cite{BESIII:2019vhn} and the high-$p_T$ dilepton invariant mass tails in $pp\to \tau \nu_{\tau}$ processes~\cite{Fuentes-Martin:2020lea}. However, the stringent experimental constraints on $g^L_{S}$, $g^R_{S}$, and $g^L_{T}$, together with the overall small deviations $\delta P^{l,h}_a$ shown in Figs.~\ref{fig:diffleptonpolarizationQ2} and \ref{fig:dhadronpolarizationQ2}, 
strongly motivate us to focus on the small-$g_i$ regions. In this case, we can expand the polarizations $P^{l,h}_a$ in terms of $g_i$ and keep only the terms up to $\mathcal{O}(g_i)$. As will be shown in the following, examining  $P^{l,h}_a$ in such a limit can shed light on the interesting behaviors of the deviations $\delta P^{l,h}_a$ shown in Figs.~\ref{fig:diffleptonpolarizationQ2} and \ref{fig:dhadronpolarizationQ2}.  

Given that only a single nonzero $g_i$ is activated at a time, the two traces in the polarization four-vector 
$\mathcal{P}^{\mu}$ (see, e.g., Eq.~\eqref{eq:eta}) can be written, respectively, as  
\begin{align}
	\tr[\rho]&= D_{\text{SM}}+(g_i)^* D_{VL,i}+(g_i)D^*_{VL,i}+\mathcal{O}(|g_i|^2)\nonumber \\[0.02cm]
	&=D_{\text{SM}}+2\text{Re}[g^*_{i}D_{VL,i}]+\mathcal{O}(|g_i|^2)\,, 
\end{align}
and 
\begin{align}
		\tr[\rho\gamma^{\mu}\gamma^5]=\mathcal{N}^{\mu}_{\text{SM}}+2\text{Re}[g^*_i\mathcal{N}^{\mu}_{VL,i} ]+\mathcal{O}(|g_i|^2)\,,
\end{align}
where $D_{\text{SM}}$ and $\mathcal{N}^{\mu}_{\text{SM}}$ stand for the SM contributions to the two traces $\tr[\rho]$ and $\tr[\rho\gamma^{\mu}\gamma^5]$ respectively, while $D_{VL,i}$ and $\mathcal{N}^{\mu}_{VL,i}$ denote the contributions to these two traces from the interference between the SM and the NP operator associated with $g_i$; explicit expressions of the various terms in the two traces can be found in Appendices~\ref{app:Amplitude} and \ref{app:numerator}. Clearly, the pure NP contributions are of $\mathcal{O}(|g_i|^2)$ and can be, therefore, neglected in the small-$g_i$ regions.

The polarization four-vector can now be approximated as  
\begin{align}\label{eq:4polarization_vector}
	\mathcal{P}^{\mu}&\simeq\frac{\mathcal{N}^{\mu}_{\text{SM}}+2\text{Re}[g^*_i\mathcal{N}^{\mu}_{VL,i} ]}{D_{\text{SM}}+2\text{Re}[g^*_{i}D_{VL,i}]} \nonumber \\[0.02cm]
	&\simeq \frac{\mathcal{N}^{\mu}_{\text{SM}}}{D_{\text{SM}}}+\frac{2\text{Re}[g^*_i\mathcal{N}^{\mu}_{VL,i} ]}
	{D_{\text{SM}}}-\frac{2\text{Re}[g^*_{i}D_{VL,i}]}{D_{\text{SM}}}\frac{\mathcal{N}^{\mu}_{\text{SM}}}{D_{\text{SM}}} \nonumber \\[0.02cm]
	&=\mathcal{P}^{\mu}_{\text{SM}}+\frac{2\text{Re}[g^*_i\mathcal{N}^{\mu}_{VL,i} ]}
	{D_{\text{SM}}}-\frac{2\text{Re}[g^*_{i}D_{VL,i}]}{D_{\text{SM}}}\mathcal{P}^{\mu}_{\text{SM}}\nonumber \\[0.02cm]
	&=\mathcal{P}^{\mu}_{\text{SM}}+\mathcal{P}^{\mu}_{\text{Int}}\,,
\end{align}
where we have ignored all the contributions from the higher-order terms of $g_i$, and introduced the new polarization four-vector $\mathcal{P}^{\mu}_{\text{Int}}$, with 
\begin{align}\label{eq:eta_int}
	\mathcal{P}^{\mu}_{\text{Int}}\equiv \frac{2\text{Re}[g^*_i\mathcal{N}^{\mu}_{VL,i} ]}
	{D_{\text{SM}}}-\frac{2\text{Re}[g^*_{i}D_{VL,i}]}{D_{\text{SM}}}\mathcal{P}^{\mu}_{\text{SM}}\,,
\end{align} 
which is induced by the interference between the SM and the NP operator associated with $g_i$. Projecting $\mathcal{P}^{\mu}_{\text{SM}}$ and $\mathcal{P}^{\mu}_{\text{Int}}$ onto the orthogonal bases (see Eqs.~\eqref{eq:basis_lepton} and \eqref{eq:basis_hadron}), we eventually obtain
\begin{align}\label{eq:low_g_P}
	P^{l,h}_{L,P}&=(P_{\text{SM}})^{l,h}_{L,P}+\text{Re}[g_i](P_{\text{Int}})^{l,h}_{L,P}\,, \\[0.02cm]
	P^{l,h}_{T}&=\text{Im}[g_i](P_{\text{Int}})^{l,h}_{T}\,,
\end{align}
where $(P_{\text{SM}})^{l,h}_{T}=0$ has been used. 

From the definition of $\mathcal{P}^{\mu}_{\text{Int}}$ in Eq.~\eqref{eq:eta_int}, one can already see that $\mathcal{N}^{\mu}_{VL,VL}=\mathcal{N}^{\mu}_{\text{SM}}$ and 
$D_{VL,VL}=D_{\text{SM}}$ for the $g_V^L$ scenario. Since both $\mathcal{N}^{\mu}_{\text{SM}}$ and $D_{\text{SM}}$ are real, $\mathcal{P}^{\mu}_{\text{Int}}$ vanishes, which in turn 
leads to $P_{\text{Int}}=0$. In other words, it is impossible to distinguish the $g_V^L$ scenario from the SM through the polarization vectors, which has already been observed repetitively in the previous subsections.

We then show in Fig.~\ref{fig:PolaQ} the variations of $(P_{\text{SM}})^{l,h}_{a}$ and $(P_{\text{Int}})^{l,h}_{a}$ with respect to $Q^2$ in various NP scenarios, where, for simplicity, we have labeled them by $P^{l,h}_a$ uniformly. From the $P^l_L$-$Q^2$ plot (the left-top one in Fig.~\ref{fig:PolaQ}), one can see that $(P_{\text{Int}})^l_L$ behave in a very similar way for the $g^L_S$ and $g^R_S$ scenarios, which are denoted by the blue and green dashed curves, respectively. Together with another straightforward observation that the magnitude of $(P_{\text{Int}})^l_L$ at any $Q^2$ in the $g^L_S$ case is always larger than in the $g^R_S$ case, it is expected that the maximum deviation $\delta P^l_L$ for the $g^L_S$ and $g^R_S$ scenarios must have a similar shape but with the former broader than the latter. Such a behavior has already been observed explicitly in Fig.~\ref{fig:diffleptonpolarizationQ2}. From the dot-dashed red curve, one can see that, below $Q^2=5~\mathrm{GeV}^2$, $(P_{\text{Int}})^l_L$ for the $g^R_V$ scenario behaves just like that for $g^L_S$, indicating a similar shape of $\delta P^l_L$ within this $Q^2$ range. However, the shape of $\delta P^l_L$ will become narrower as $Q^2$ increases, even narrower than that for the $g^R_S$ scenario at the high-$Q^2$ range. Such an expectation is, unfortunately, buried by the vast shadow of the $\delta P^l_L$-$Q^2$ plot shown in Fig.~\ref{fig:diffleptonpolarizationQ2}, due to the large parameter space of $g^R_V$. Compared with $(P_{\text{SM}})^{l}_{L}$ denoted by the black curve, the absolute value of $(P_{\text{Int}})^l_L$ for the $g^L_T$ scenario (see the long-dashed purple curve) is always larger. However, their difference decreases as $Q^2$ increases, justifying that a low $Q^2$ is favored to observe a maximum deviation of $\delta P^{l}_{L}$ in the $g^L_T$ scenario, as shown in Fig.~\ref{fig:diffleptonpolarizationQ2}. 

We now turn to discuss the various curves in the $P^l_P$-$Q^2$ plot (the middle-top one in Fig.~\ref{fig:PolaQ}). It can be seen that the blue and green dashed curves behave in a similar way---both peak roughly at 
$Q^2=7.5~\mathrm{GeV}^2$---but with different magnitudes. Although the dashed purple curve also peaks at a similar $Q^2$, it behaves less dramatically within the range $Q^2\in [3,7]$~GeV$^2$. Nonetheless, all these three curves drop to zero at $Q^2_{\text{min}}$ and $Q^2_{\text{max}}$. Taking all these points into account, one can understand the interesting features of the deviation $\delta P^{l}_{P}$ observed in the $g^L_S$, $g^R_S$, and $g^L_T$ scenarios, as shown in Fig.~\ref{fig:diffleptonpolarizationQ2}. For the $g^R_V$ scenario, as indicated by the dot-dashed red curve, the deviation $\delta P^{l}_{P}$ shall behave similarly to that for the $g^R_S$ scenario but with a more flattened curvature at the high-$Q^2$ range. This is different from the behaviors of the deviation $\delta P^{l}_{L}$ in the same NP scenarios, as can be clearly seen from Fig.~\ref{fig:diffleptonpolarizationQ2}. 

Let us move on to the $P^h_L$-$Q^2$ plot (the left-bottom one in Fig.~\ref{fig:PolaQ}). A couple of observations can already be made. Firstly, all of 
the curves except the dashed blue one experience 
a crossover, indicating that the deviations $\delta P^h_L$ become zero at a certain $Q^2$ for the $g^R_V$, $g^R_S$, and $g^L_T$ scenarios, while in the $g^L_S$ case $\delta P^h_L$ increases along with the increase of $Q^2$. Secondly, both the green and purple dashed curves cross the $P^h_L=0$ line at $Q^2\simeq 5$~GeV$^2$, suggesting a similar behavior of $\delta P^h_L$ for the $g^R_S$ and $g^L_T$ scenarios. However, the pattern of small at the $Q_{\text{min}}^2$ while relatively large at the 
$Q^2_{\text{max}}$ region of $(P_{\text{Int}})^h_L$ reveals that the deviation $\delta P^h_L$ must be narrower at the $Q_{\text{min}}^2$ than  
at the $Q_{\text{max}}^2$ one for the $g^R_S$ scenario. This is contrary to the pattern of $\delta P^h_L$ observed for the $g^L_T$ scenario, as can be clearly seen from Fig.~\ref{fig:dhadronpolarizationQ2}. Finally, the similar behavior between the green and blue dashed curves indicates that the deviation $\delta P^h_L$ shall behave similarly for the $g^R_V$ and $g^R_S$ scenarios, provided they are both assumed at the small-$g_i$ limit.

With regard to the $P^h_P$-$Q^2$ plot (the middle-bottom one in Fig.~\ref{fig:PolaQ}), one can draw some similar observations as from the $P^l_P$-$Q^2$ plot. For instance, the similar behavior between the green and blue dashed curves predicts a close shape of $\delta P^h_P$ for the $g^L_S$ and $g^R_S$ scenarios. The small difference between the resulting values of $(P_{\text{Int}})^{h}_{P}$, however, suggests that the deviation $\delta P^h_P$ for the former must be broader than for the latter, as shown in Fig.~\ref{fig:dhadronpolarizationQ2}. Meanwhile, the blue and green dashed curves in the $P^h_P$-$Q^2$ and $P^l_P$-$Q^2$ plots indicate that both $\delta P^h_P$ and $\delta P^l_P$ in these two scenarios shall peak at $Q^2\simeq 7$~GeV$^2$. Another example is that the red and purple dashed curves reveal that the maximal $\delta P^h_P$ occurs at low $Q^2$, $Q^2\simeq 3.4$~GeV$^2$, contrary to its counterpart $\delta P^l_P$, for the $g^R_L$ and $g^L_T$ scenarios. 

We conclude this subsection by giving a brief discussion of the $P^l_T$-$Q^2$ and $P^h_T$-$Q^2$ plots in Fig.~\ref{fig:PolaQ}. Since the SM contribution to $P^{l,h}_T$ denoted by the dark line is zero, the shapes of other curves reveal not only the behaviors of the polarizations $P^{l,h}_T$ but also the deviations $\delta P^{l,h}_T$ directly. It can be seen that the blue, green, and purple dashed curves in the $P^l_T$-$Q^2$ plot behave similarly in general with only some small differences, indicating a similar pattern of the deviation $\delta P^{l}_T$ for the $g^L_S$, $g^R_S$, and $g^L_T$ scenarios. The blue, green, and purple dashed curves in the $P^h_T$-$Q^2$ plot, on the other hand, behave quite differently in both their curvatures and peak positions, justifying the distinct shapes of $\delta P^h_T$ for the $g^L_S$, $g^R_S$, and $g^L_T$ scenarios, as shown in Fig.~\ref{fig:dhadronpolarizationQ2}. Finally, the deviation $\delta P^h_T$ for the $g^R_V$ scenario in Fig.~\ref{fig:dhadronpolarizationQ2} behaves just like the dashed red curve in Fig.~\ref{fig:PolaQ}, even though the latter works only in the small-$g_i$ 
limit. 

\subsection{Observables with uncertainties due to the form factors}
\label{sec:Obswitherror}

\begin{figure*}[t]
	\centering
	\includegraphics[width=4.25cm]{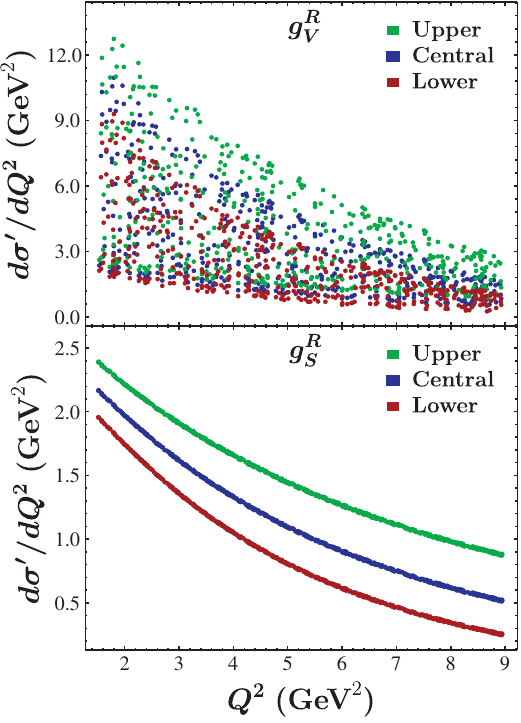} \;
	\includegraphics[width=4.25cm]{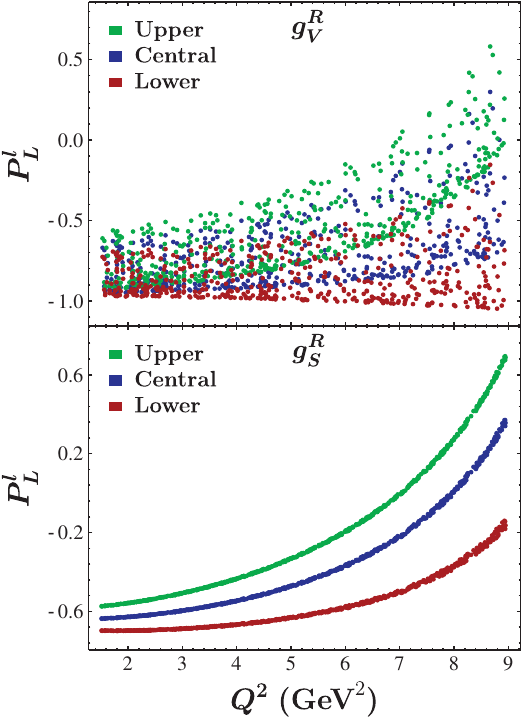} \;
	\includegraphics[width=4.125cm]{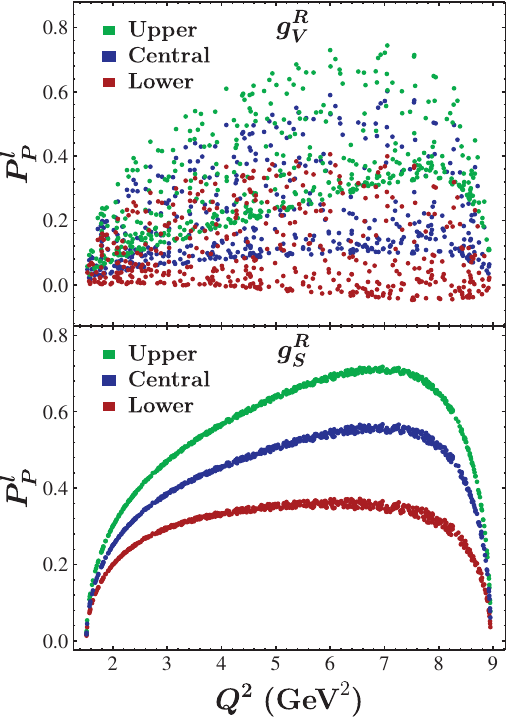} \;
	\includegraphics[width=4.4cm]{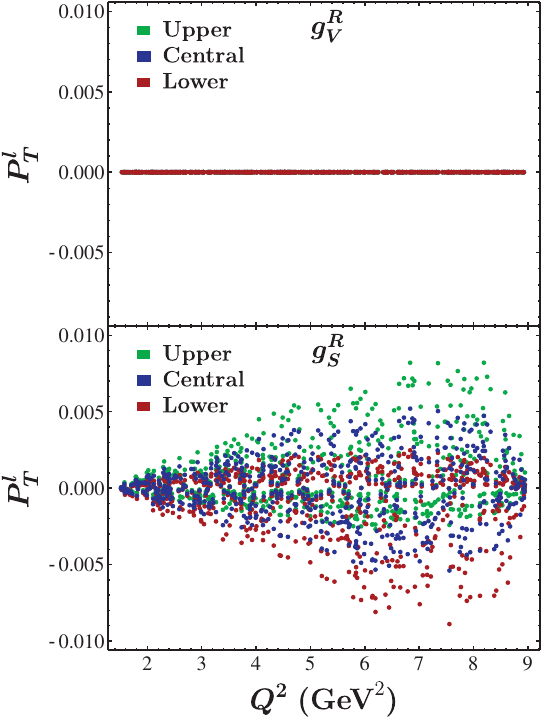}
	\caption{Uncertainties of the differential cross section as well as the polarizations $P^l_L$, $P^l_P$, and $P^l_T$ due to the $\Lambda_c\to N$ transition form factors, in the $g^R_V$ (top panel) and $g^R_S$ (bottom panel) scenarios. The blue points denote the resulting observables calculated with the central values, whereas the green (red) points the observables computed with the upper (lower) values at $1\sigma$ level of these form factors. Note that the neutrino beam energy has been fixed at $E=10$~GeV for consistency.}
	\label{fig:ErrorL}
\end{figure*}

\begin{figure*}[t]
	\centering
	\includegraphics[width=4.25cm]{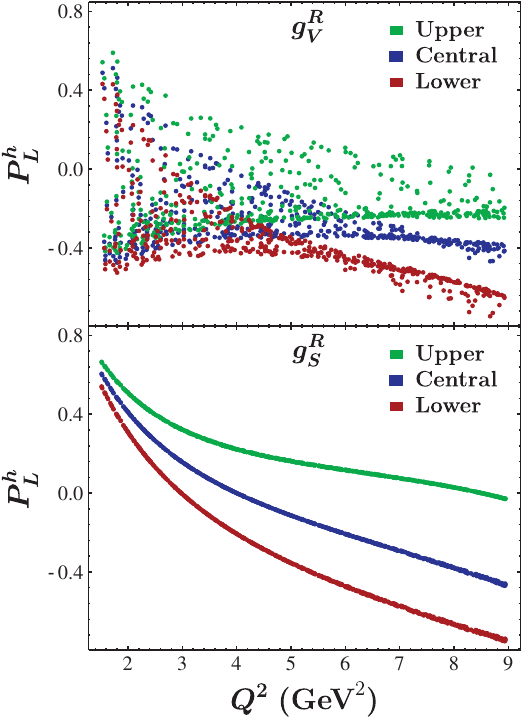} \quad
	\includegraphics[width=4.13cm]{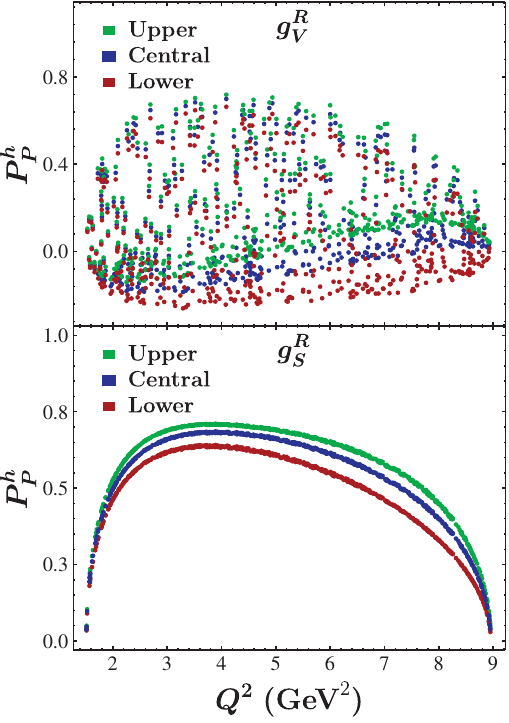} \quad
	\includegraphics[width=4.4cm]{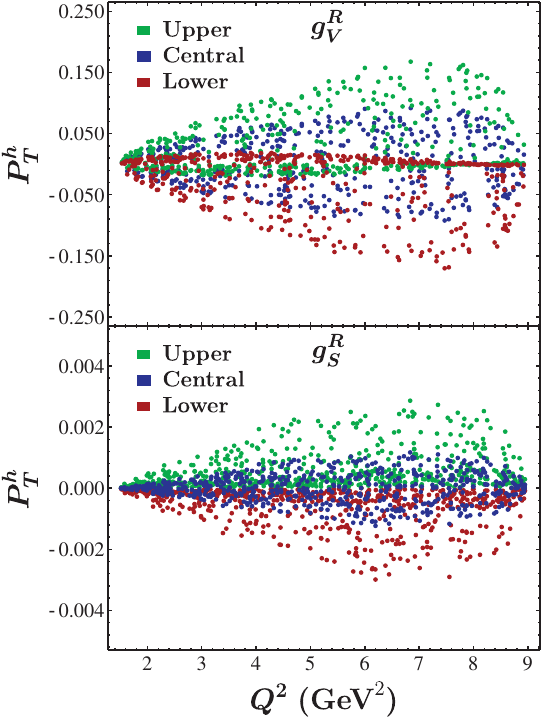}
	\caption{Uncertainties of the polarizations $P^h_L$, $P^h_P$, and $P^h_T$ due to the $\Lambda_c\to N$ transition form factors, in the $g^R_V$ (top panel) and $g^R_S$ (bottom panel) scenarios. The other captions are the same as in Fig.~\ref{fig:ErrorL}.}
	\label{fig:ErrorH}
\end{figure*}

\begin{figure*}[t]
	\centering
	\includegraphics[width=4.17cm]{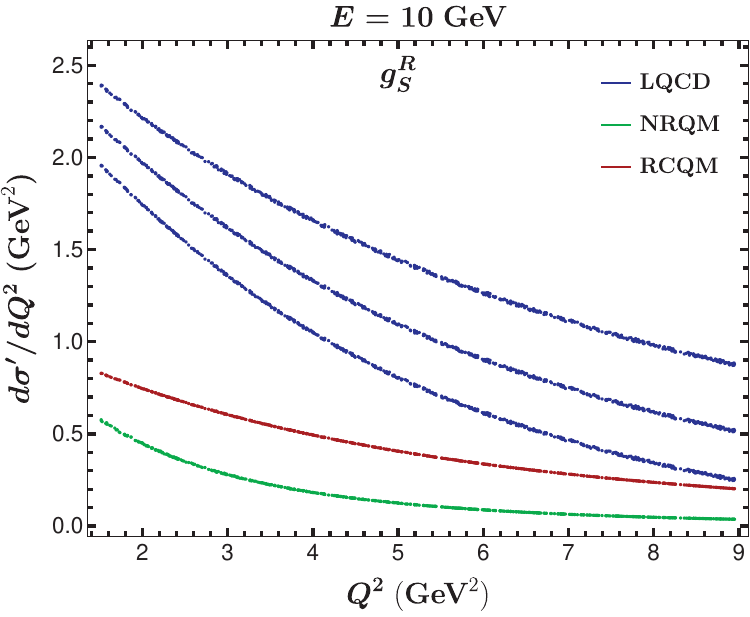} \;
	\includegraphics[width=4.2cm]{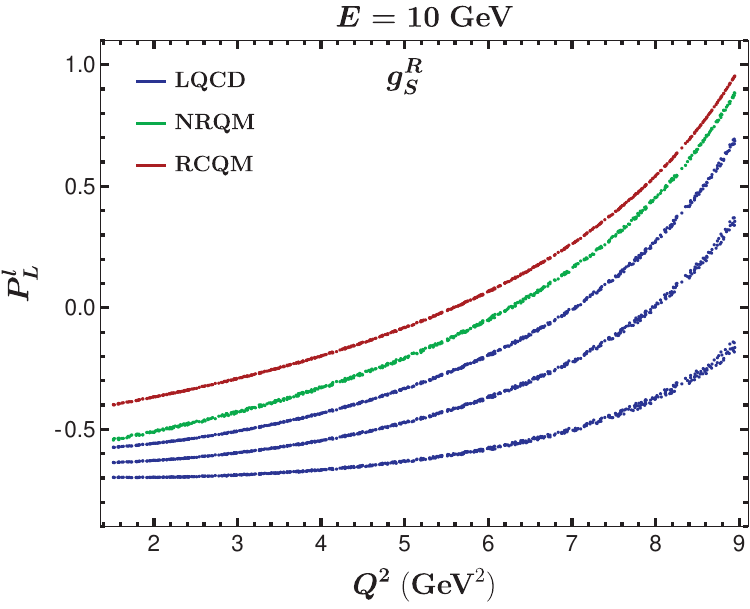} \;
	\includegraphics[width=4.17cm]{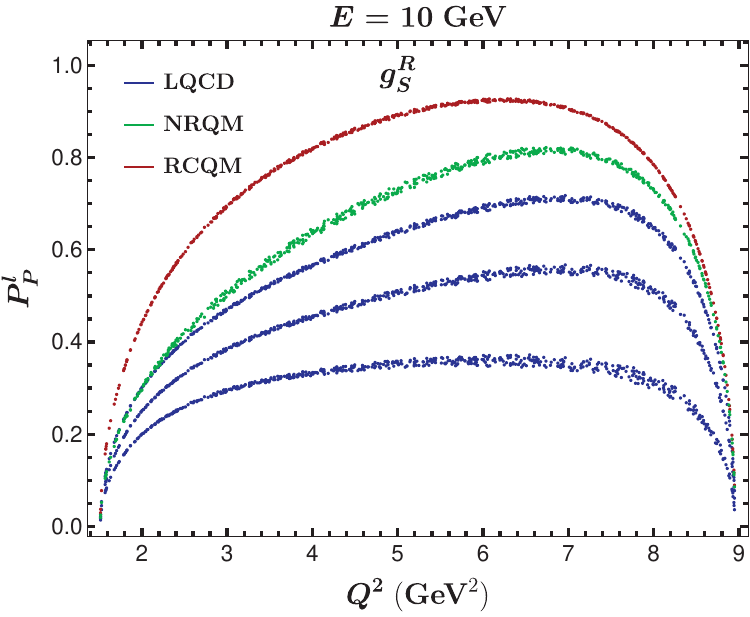} \;
	\includegraphics[width=4.37cm]{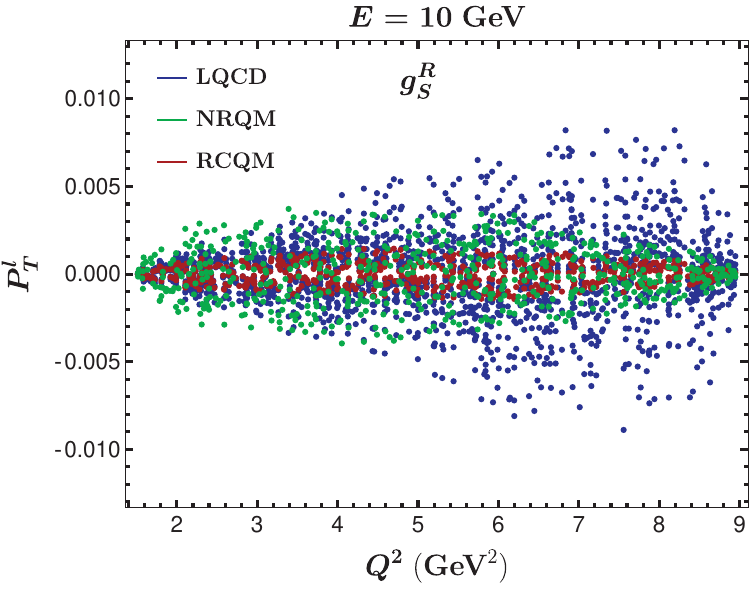} \\
	\vspace{0.2cm}
	\includegraphics[width=4.2cm]{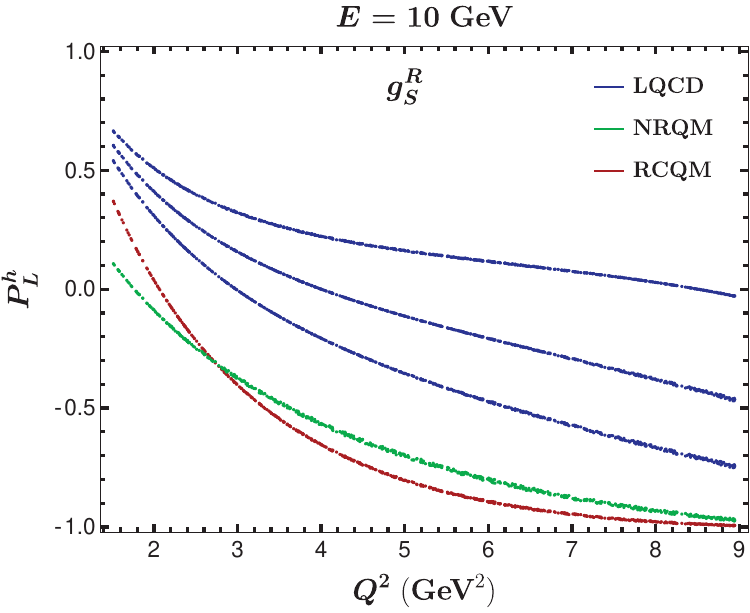} \;
	\includegraphics[width=4.17cm]{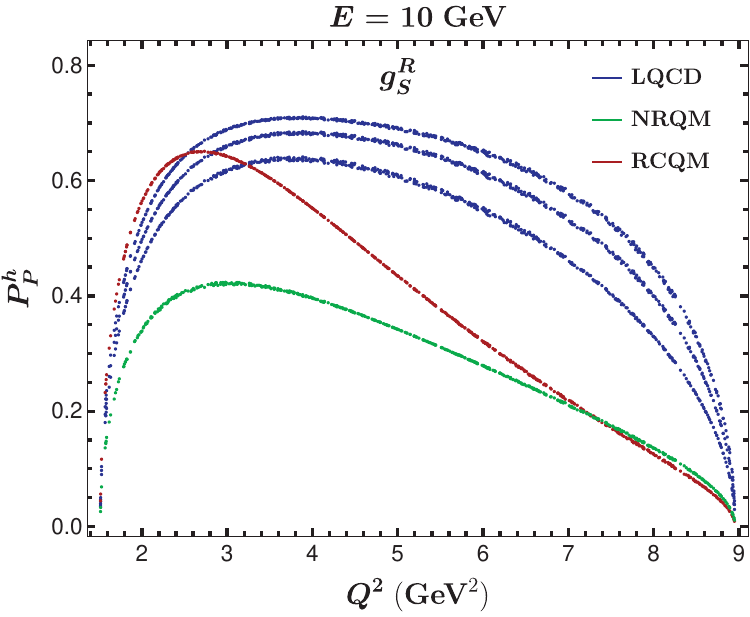} \;
	\includegraphics[width=4.38cm]{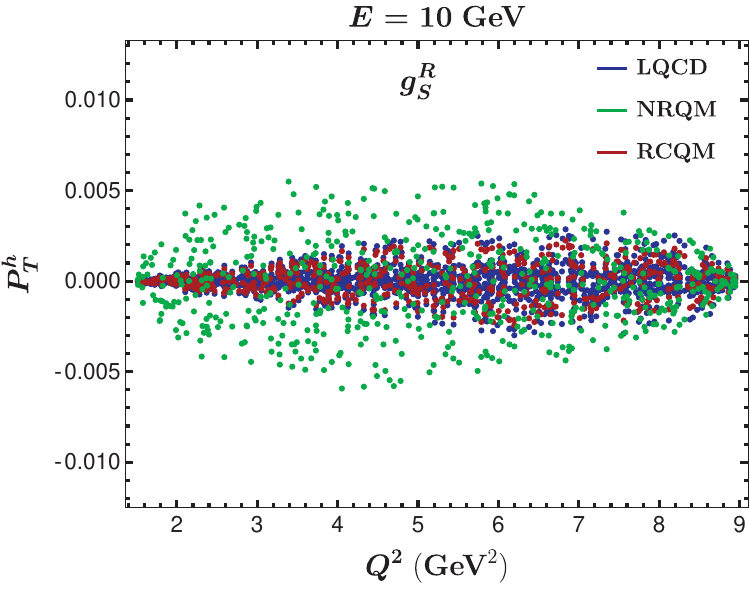} 
	\caption{The differential cross section as well as the polarizations $P^{l,h}_L$, $P^{l,h}_P$, and $P^{l,h}_T$ as a function of $Q^2$, predicted with the form factors calculated in LQCD (blue), NRQM (green), and RCQM (red), respectively. Here we focus only on the $g^R_S$ scenario. Note that the $1\sigma$-level statistical uncertainties of the form factors in LQCD have been propagated to all the observables, as denoted by the outer blue regions.}
	\label{fig:diff_FF}
\end{figure*}

As mentioned in subsection~\ref{subsec:Cross section}, one of the reasons that we adopt the LQCD calculations of the $\Lambda_c\to N$ transition form factors 
is that they provide us with an error estimation. Yet our calculation has only involved the central values of these inputs so far. In this subsection, we study how our predictions of the observables are affected by the uncertainties of these form factors. As a simple illustration, we focus on the NP scenarios in the presence of the WCs $g^R_V$ and $g^R_S$, and consider only the $Q^2$-dependent observables, i.e., the differential cross section and the polarizations $P^{l,h}_a$. To this end, we firstly scan randomly $g^R_V$ and $g^R_S$ within the available parameter space shown in Fig.~\ref{fig:constraints} and propagate the uncertainties of the form factors to each observable for all the allowed data points of $g^R_V$ and $g^R_S$. We then plot in Figs.~\ref{fig:ErrorL} and \ref{fig:ErrorH} the central, upper, and lower values of each observable in blue, green, and red accordingly, instead of presenting them in error bars. In this way, the combined regions of the green and red ones as well as the regions between them can be naively understood as the overall uncertainty of the observable considered. 

From Figs.~\ref{fig:ErrorL} and \ref{fig:ErrorH}, we see that there exist large overlaps among the three colored regions in the low-$Q^2$ region 
	for each observable in the $g^R_V$ scenario, indicating that the dominant factor determining the overall shape of these observables is still due to the vast available parameter space of $g^R_V$. But the impact from the uncertainties of the form factors becomes gradually distinct, particularly in the relatively high-$Q^2$ region where the uncertainty from the form factors for each observable can more than double. As for the $g^R_S$ scenario, the large blank spaces between the blue and green (red) regions represent the impacts on the observables from the uncertainties of the form factors, which clearly dwarf the effect of the WC $g^R_S$ due to the stringent experimental constraint on it. The only exceptions are $P^l_T$ and $P^h_T$ in both NP scenarios, on which the impacts from the uncertainties of the form factors and the available parameter space of the WCs seem comparable. These observations can be easily applied to other NP scenarios too.

Besides the above comparisons, it may be also interesting to explore how the uncertainties of the observables propagate along the kinematics $Q^2$. To this end, let us focus on the observables in the $g^R_S$ scenario as an illustration. Firstly, the green and red regions on the bottom panel of Figs.~\ref{fig:ErrorL} and \ref{fig:ErrorH} clearly indicate that the overall uncertainties of the differential cross section and the polarizations $P^{l,h}_L$ increase along with the increase of $Q^2$. Secondly, the uncertainties of $P^{l,h}_P$ and $P^{l,h}_T$ shrink at the $Q^2_{\text{min}}$ and $Q^2_{\text{max}}$ regions, mainly due to the characteristic behaviors of $P^{l,h}_P$ and $P^{l,h}_T$, but the general pattern is still consistent with what we have just observed. Such a pattern is closely related to the behaviors of the form factors with respect to $Q^2$. As can be seen from Fig.~\ref{fig:form_error}, the uncertainties of all the form factors follow the same pattern as the observables do---the total uncertainties in particular increase dramatically along with the increase of $Q^2$. Because of the relatively milder behaviors of the statistical uncertainties, we take them instead of the total uncertainties into account in Figs.~\ref{fig:ErrorL} and \ref{fig:ErrorH}, as well as in the rest of this work.
     
In short, although the LQCD calculation~\cite{Meinel:2017ggx} of the $\Lambda_c\to N$ transition form factors comes with an error estimation---one of its advantages over the model evaluations presented in Refs.~\cite{Perez-Marcial:1989sch,Avila-Aoki:1989arc,Gutsche:2014zna}, the persistently increasing uncertainties along with the increase of $Q^2$ have become one of the major obstacles to further probe or constrain the NP scenarios through the QE neutrino scattering process. This calls for either better control of the uncertainties of the form factors in future LQCD calculations or new model estimations of these form factors with a good error estimation within the relevant kinematic ranges.  

\begin{figure*}[t]
	\centering
	\includegraphics[width=0.98\textwidth]{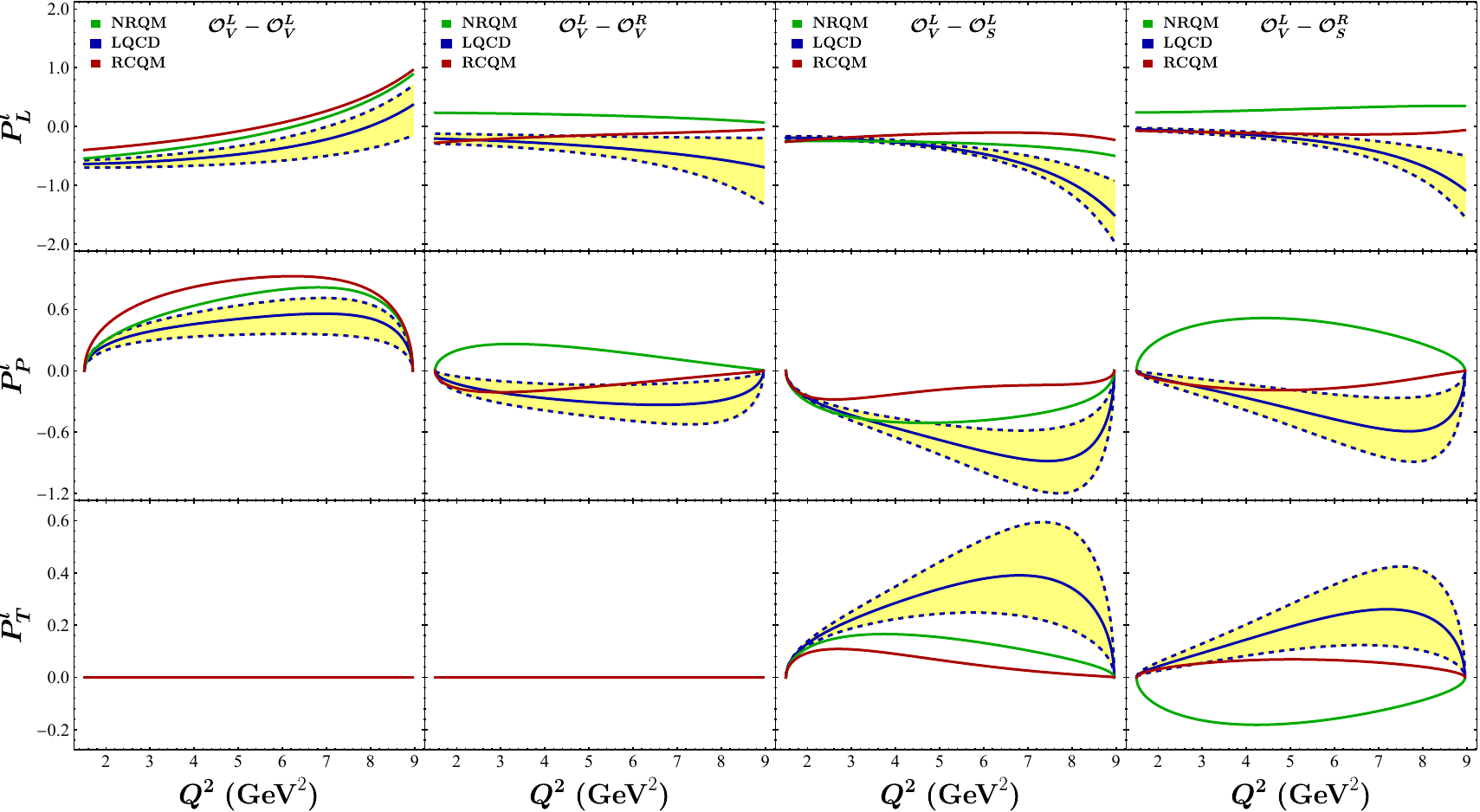}
	\caption{Variations of $(P_{\text{Int}})^l_a$ with respect to $Q^2$ in various NP scenarios, as predicted with the form factors calculated in LQCD (blue), NRQM (green), and RCQM (red), respectively. The $1\sigma$-level statistical uncertainties of the form factors in LQCD have been propagated to $(P_{\text{Int}})^l_a$, as denoted by the yellow region. The neutrino beam energy has been set to $E=10$~GeV.}
	\label{fig:CombindifFF}
\end{figure*}

\begin{figure*}[t]
	\centering
	\includegraphics[width=0.98\textwidth]{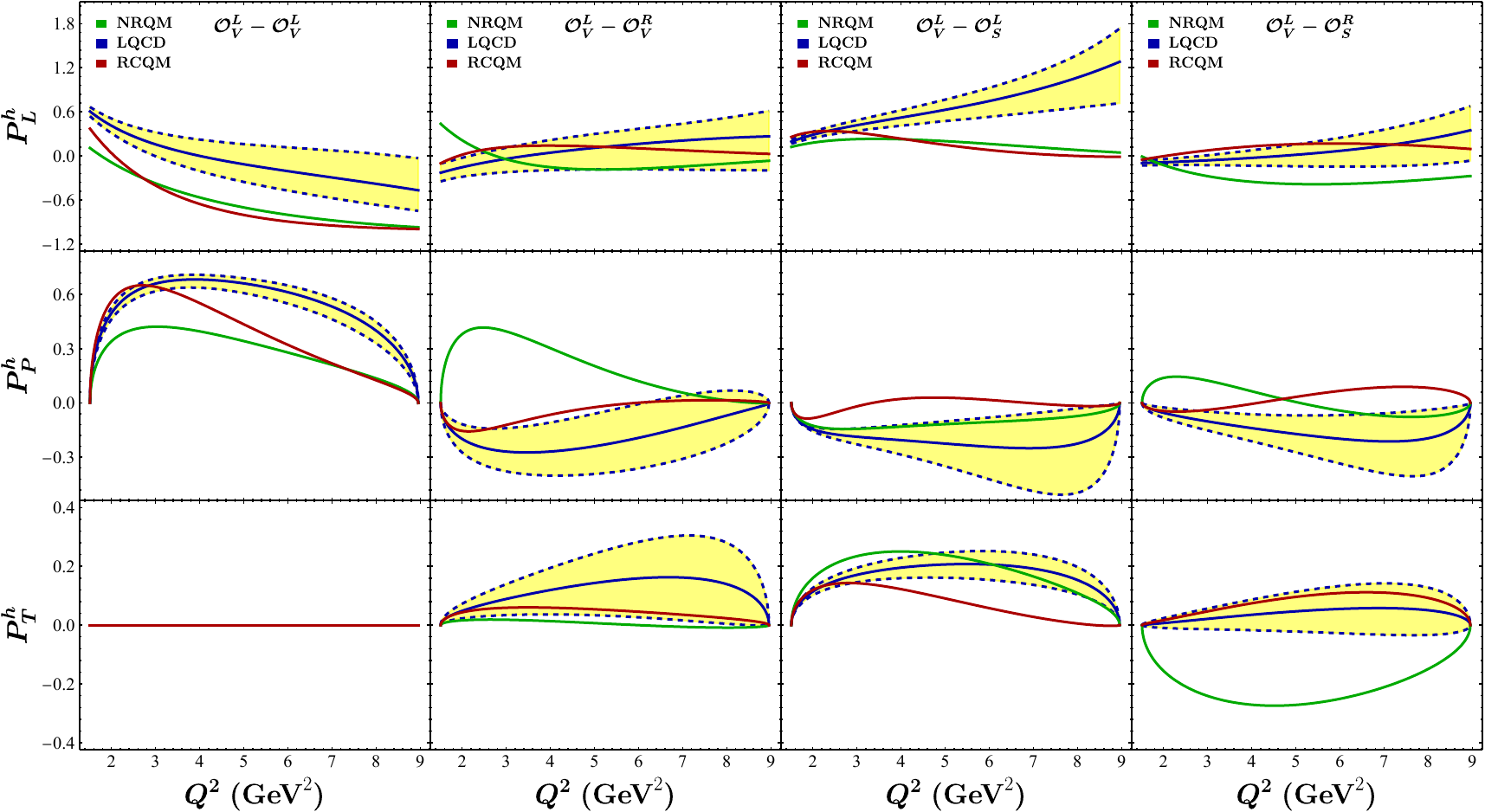}
	\caption{Variations of $(P_{\text{Int}})^h_a$ with respect to $Q^2$ in various NP scenarios, as predicted with the form factors calculated in LQCD (blue), NRQM (green), and RCQM (red), respectively. The other captions are the same as in Fig.~\ref{fig:CombindifFF}.}
	\label{fig:CombindifFFH}
\end{figure*}

\subsection{Observables with different form-factor parametrizations}
\label{sec:obsdiffform}

The parametrization scheme adopted in Ref.~\cite{Meinel:2017ggx} is not the only way to describe the $q^2$-dependence of the $\Lambda_c \to N$ transition form factors; nor is the LQCD the only method for evaluating the form factors. 
As discussed in subsection~\ref{subsec:Cross section} and detailed in Appendix~\ref{appendix:form factor}, there exist already three different parametrization schemes, which can be extended to the $q^2<0$ range, and have been employed by the MBM, NRQM, and RCQM models, 
as well as the LQCD calculations. Moreover, these parametrization schemes are validated against the experimental measurements of the $\Lambda_c$ semileptonic decays reported by the BESIII Collaboration~\cite{BESIII:2015ysy,BESIII:2016ffj}.\footnote{Note that the BESIII Collaboration has improved the measurement of the absolute branching fraction of $\Lambda^+_c\to \Lambda e^+\nu_e$ decay~\cite{BESIII:2022ysa}.} 
However, direct calculations of the QE weak production of the $\Lambda_c$ baryon through the $\nu_{\mu}$ scattering 
off nuclei reveal that large deviations arise by using the different schemes of the form factors, 
demonstrating a direct consequence of the ambiguities induced by extrapolating the form factors to the 
moderately large positive $Q^2$~\cite{Sobczyk:2019uej}.  
Given that our analysis is based on the same extrapolation, 
we examine in this subsection if the same observation applies to the observables considered here in various NP scenarios. 

In Fig.~\ref{fig:diff_FF}, we evaluate the differential cross section and the polarizations $P^{l,h}_a$ with the form factors calculated in LQCD (blue), NRQM (green), and RCQM (red), respectively.\footnote{We do not present the results with the form factors calculated in MBM, because both MBM and NRQM employ the dipole form for the $q^2$ dependence of the form factors~\cite{Perez-Marcial:1989sch,Avila-Aoki:1989arc} (see Appendix~\ref{appendix:form factor} for details).} To be thorough, we also take account of the $1\sigma$-level statistical uncertainties of the form factors in the LQCD case. As an illustration, we focus only on the NP scenario in the presence of $g^R_S$. From Fig.~\ref{fig:diff_FF}, it can be seen that there exists large disparity between the red (green) and blue regions, indicating that the resulting deviations of $d\sigma$, $P^{l,h}_L$, and $P^{l,h}_P$ due to the different parametrization schemes of the form factors dwarf that from the $1\sigma$-level statistical uncertainties of the form factors in LQCD. For the polarization $P^l_T$, on the other hand, the overall blue region prevails over the others, indicating a totally opposite situation. Finally, comparing the red region with the overall blue one in the $P^h_T$-$Q^2$ plot, one can see that the deviation of $P^h_T$ in RCQM from the LQCD predication can be comparable to that from the $1\sigma$-level statistical uncertainties of the form factors in LQCD. 

The SM predictions of $(P_{\text{Int}})^l_a$ and $(P_{\text{Int}})^h_a$ are presented in the first columns of Figs.~\ref{fig:CombindifFF} and \ref{fig:CombindifFFH}, respectively. One can see that among the three cases, the LQCD predicts the largest differential cross section of the QE scattering process in the SM, while the NRQM yields the smallest. Such a pattern is also consistent with that observed in the QE weak production of the $\Lambda_c$ baryon through the process $\nu_{\mu}+\ce{^16O}\to \mu^- +\Lambda_c+X$~\cite{Sobczyk:2019uej}. However, the situation becomes more complicated for other observables. For instance, the crossover behavior of $P^{l,h}_L$ makes the $P^{l,h}_L=0$ line a watershed: above it the RCQM (LQCD) predicts the largest $P^{l}_L$ ($P^{h}_L$), while below it the LQCD (RCQM) predicts the largest $P^{l}_L$ ($P^{h}_L$). In addition, the RCQM always seems to produce a larger $P^{l,h}_P$ than the NRQM does. 

The small width of each fuzzy colored region in Fig.~\ref{fig:diff_FF} results from the variation of the WC $g^R_S$ within the allowed parameter space shown in Fig.~\ref{fig:constraints}. To have a clearer view of this effect, we work in the small-$g_i$ limit and plot in Figs.~\ref{fig:CombindifFF} and \ref{fig:CombindifFFH} the variations of $(P_{\text{SM}})^{l,h}_{a}$ 
and $(P_{\text{Int}})^{l,h}_{a}$ with respect to $Q^2$ with the form factors calculated in LQCD (blue), NRQM (green), and RCQM (red), both within the SM and in the $g^R_V$, $g^L_S$, and $g^R_S$ scenarios. Note that the $g^L_T$ scenario is not considered here, because the relevant tensor form factors have not been calculated in NRQM and RCQM. Once again, the $1\sigma$-level statistical uncertainties of the form factors have been taken into account in the LQCD case (see the yellow regions shown in Figs.~\ref{fig:CombindifFF} and \ref{fig:CombindifFFH}).

Since the resulting $P^{l,h}_a$ due to the mixing $\mathcal{O}^L_V$-$\mathcal{O}^L_V$ correspond exactly to the SM case, which has been discussed above, let us now move on to the next three mixing scenarios. For the mixing $\mathcal{O}^L_V$-$\mathcal{O}^R_V$, it can be seen that, contrary to $(P_{\text{Int}})^{h}_{T}$, the resulting $(P_{\text{Int}})^{l,h}_{L,P}$ from NRQM and RCQM are opposite in sign. At the same time, the absolute values of all the $(P_{\text{Int}})^{l,h}_{a}$ in these two models are compatible with the LQCD results at $1\sigma$ level. These observations can be applied to the mixing $\mathcal{O}^L_V$-$\mathcal{O}^R_S$ as well, except that the NRQM forecasts the largest absolute value of $(P_{\text{Int}})^{l,h}_{T}$. For the mixing $\mathcal{O}^L_V$-$\mathcal{O}^L_S$, on the other hand, one can see that the RCQM always predicts the smallest absolute values of all the $(P_{\text{Int}})^{l,h}_{a}$, while the NRQM results are in general compatible with that of the LQCD at $1\sigma$ level. 
 
All in all, despite the complicated behaviors of each polarization observable calculated with various form-factor parametrization schemes in different scenarios, an overall observation is that the uncertainties of the polarization observables due to the different schemes even overwhelm that from the error propagation of the 
statistical uncertainties of the form factors.
 
\section{Conclusion}
\label{sec:con}

The absence of semitauonic decays of charmed hadrons makes the decay processes mediated by the quark-level $c\to d \tau^+ \nu_{\tau}$ transition inadequate for probing a generic NP with all kinds of Dirac structures. To fill in this gap, we have considered in this paper the QE neutrino scattering process $\nu_{\tau}+n\to \tau^-+\Lambda_c$, and proposed searching for NP through the polarizations of the $\tau$ lepton and the $\Lambda_c$ baryon. Working in the framework of a general low-energy effective Lagrangian given by Eq.~\eqref{eq:Leff} and using the combined constraints from the measured branching fraction of the purely leptonic $D^+\to\tau^+\nu_{\tau}$ decay 
and the analysis of the high-$p_T$ dilepton invariant mass tails in $pp\to \tau \nu_{\tau}$ processes, we have performed a comprehensive analysis of the (differential) cross sections and polarization vectors of the $\nu_{\tau}+n\to \tau^-+\Lambda_c$ process both within the SM and in various NP scenarios. 

For the SM, we have shown that the dominant polarization mode of the outgoing $\tau$ lepton is longitudinal and that of the $\Lambda_c$ baryon is perpendicular, whereas the transverse polarizations $\langle P_T\rangle$ of both the $\tau$ and $\Lambda_c$ remain zero in such a QE scattering process. We have also explored the variations of the polarization vectors with respect to the kinematics $Q^2$, and observed that both $P^l_L$ and $P^h_L$ experience a crossover, and the peaks of $P^l_P$ and $P^h_P$ are both reached within the available kinematic range, though happening at different $Q^2$ points. 

For the various NP scenarios, the overall observation we have made is that, due to the stringent experimental constraints on the WCs $g^L_S$, $g_S^R$, and $g^L_T$, there exist only small (of $\mathcal{O}(10^{-2})$) deviations between the SM and the $g^L_S$, $g_S^R$, and $g^L_T$ scenarios for the polarizations $P^{l,h}_a$. By contrast, the larger available parameter space of the WC $g_V^R$ makes all the deviations $\delta P^{l,h}_a$ much bigger, except for $\delta P^{l}_T$ which remains zero. As for the $g^L_V$ scenario, since it shares the same effective operator 
$\mathcal{O}^L_V$ with the SM, all the deviations $\delta P^{l,h}_a$ always remain zero, making the (differential) cross section the only avenue to probe $g^{L}_V$ through the QE scattering process. 

We have also explored the impacts of the uncertainties of the $\Lambda_c\to N$ transition form factors, and shown that they have become one of the major challenges to further probe or constrain the NP scenarios through the QE neutrino scattering process. Furthermore, we have considered three different form-factor parametrization schemes employed by NRQM, 
RCQM, and LQCD respectively, and discovered large differences among their predictions in the SM, which is also consistent with the observation made in the QE weak production of the $\Lambda_c$ baryon through the $\nu_{\mu}$ scattering off nuclei~\cite{Sobczyk:2019uej}. For the NP scenarios, although the deviations $\delta P^{l,h}_a$ predicted in NRQM and RCQM are still compatible with the LQCD results at $1\sigma$ level,  
the overall observation is that large uncertainties of the polarization observables arise from using 
the different schemes and dwarf that from the error propagation of the 
form factors, which demonstrates a direct consequence of the ambiguities induced by extrapolating the form factors to the large positive $Q^2$.

Finally, we would like to make a comment on the detection of the outgoing $\tau$ lepton. It is known that the $\tau$ lepton decays rapidly and its decay
products contain at least one undetected neutrino, making its identification very challenging and its polarization states hard to be measured. However, its kinematic and polarization information can be inferred from the visible final-state kinematics in its subsequent decays~\cite{Ivanov:2017mrj,Alonso:2017ktd,Asadi:2020fdo,Hu:2020axt,Penalva:2021gef,Hu:2021emb,Penalva:2021wye,Penalva:2022vxy,Li:2023gev,Hernandez:2022nmp,Isaacson:2023gwp}. In our upcoming work, we will incorporate this idea into our further analysis of the QE scattering process.

\section*{Acknowledgments}

This work is supported by the National Natural Science Foundation of China under Grant Nos. 12135006 and 12075097, the Fundamental Research Funds for the Central Universities under Grant Nos. CCNU22LJ004 and CCNU19TD012, as well as the 
Pingyuan Scholars Program under Grand No.5101029470306.  

\appendix

\section{\boldmath Definitions and parametrizations of the $\Lambda_c\to N$ transition form factors}
\label{appendix:form factor}

\begin{figure*}[t]
	\centering
	\includegraphics[width=4.1cm]{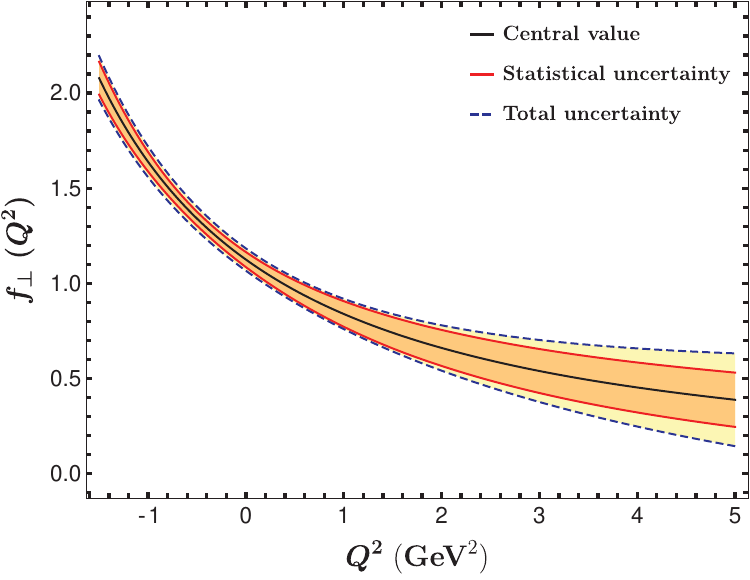} \quad
	\includegraphics[width=4.1cm]{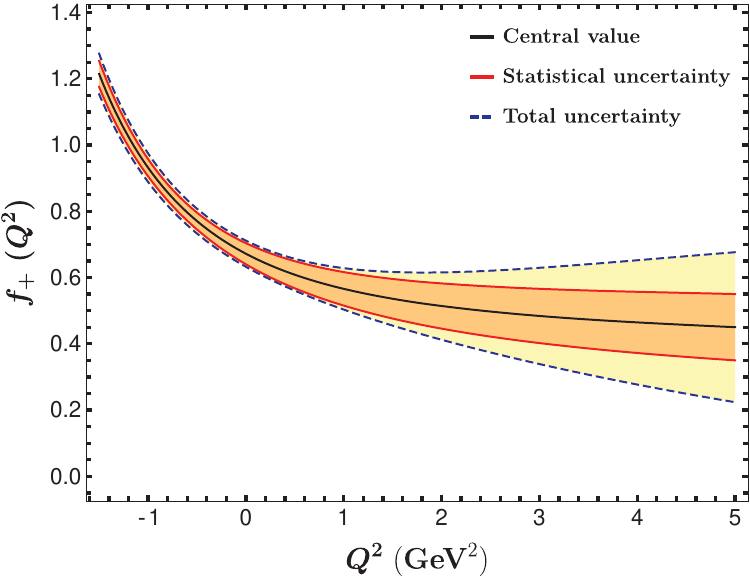} \quad
	\includegraphics[width=4.1cm]{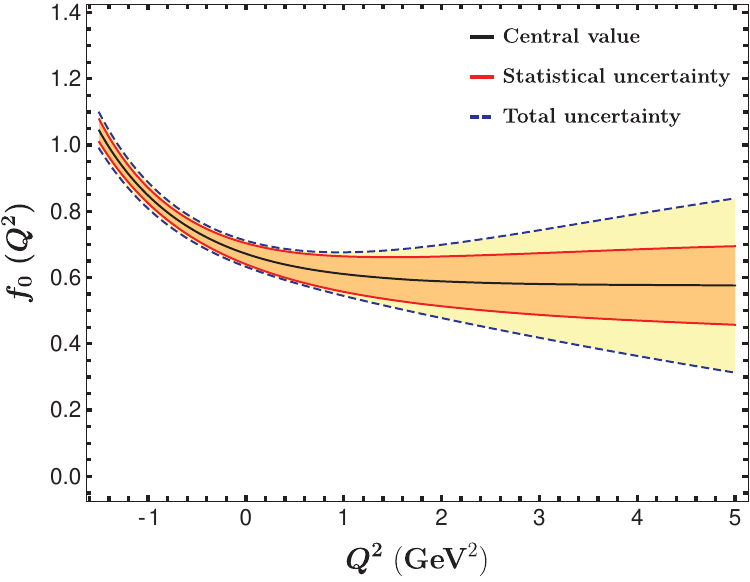} \quad
	\includegraphics[width=4.1cm]{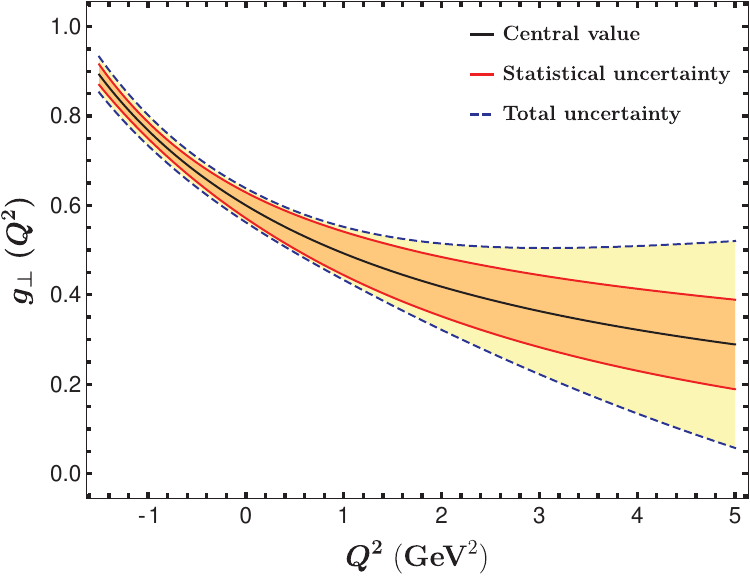} \\ 
	\vspace{0.2cm}
	\includegraphics[width=4.1cm]{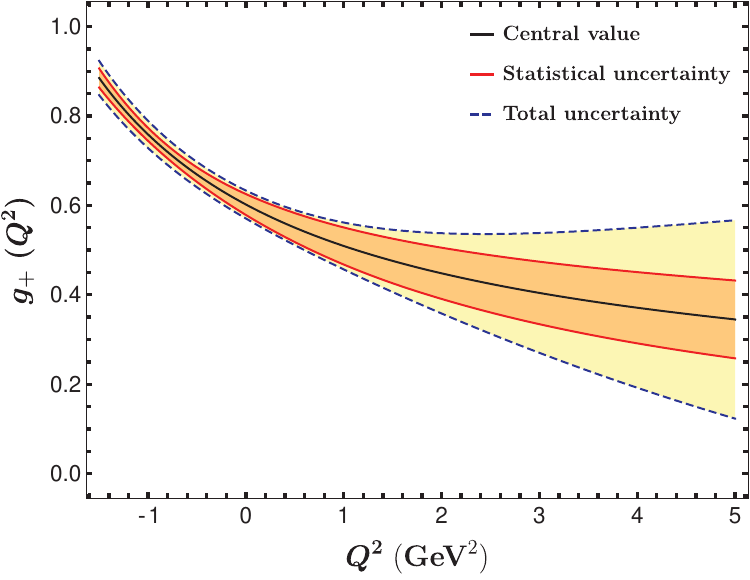} \quad
	\includegraphics[width=4.1cm]{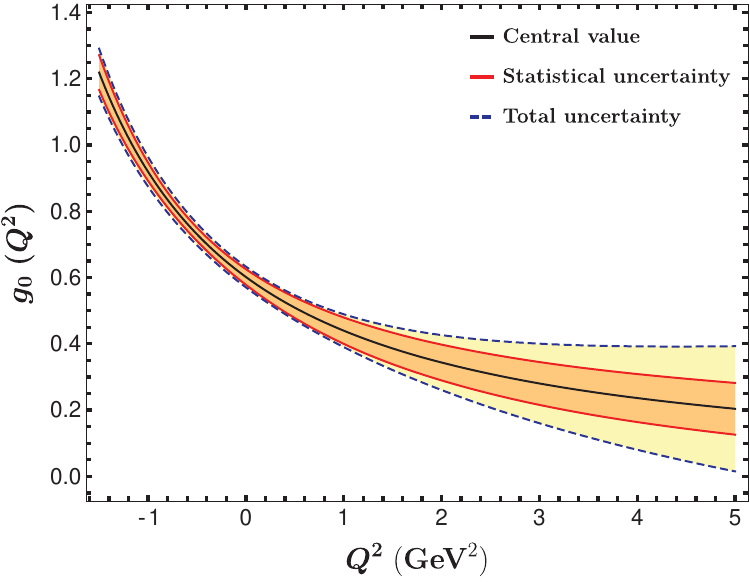} \quad
	\includegraphics[width=4.1cm]{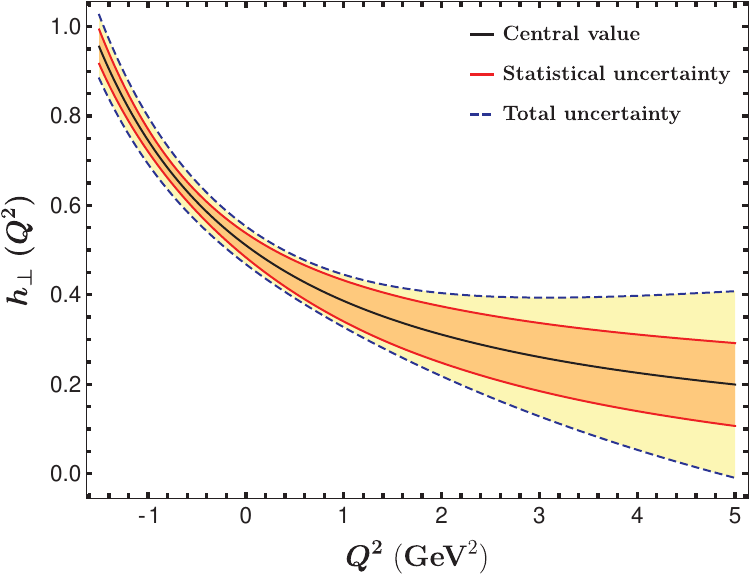} \quad
	\includegraphics[width=4.1cm]{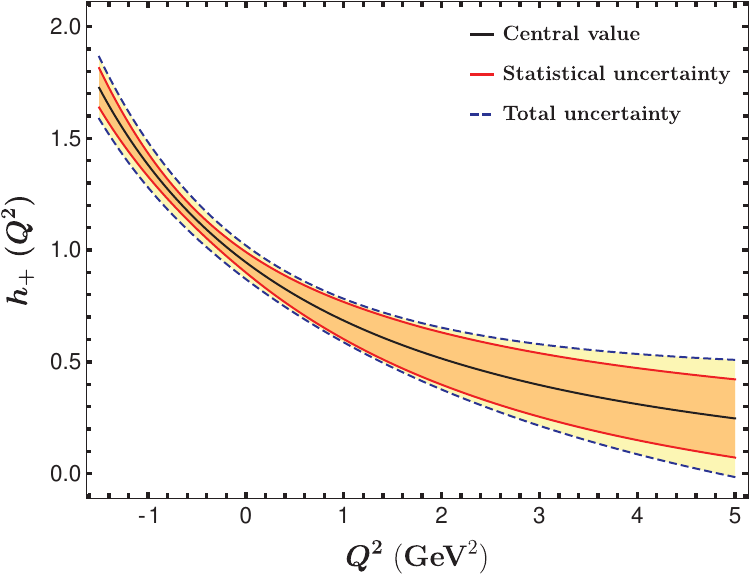}\\
	\vspace{0.2cm}
	\includegraphics[width=4.1cm]{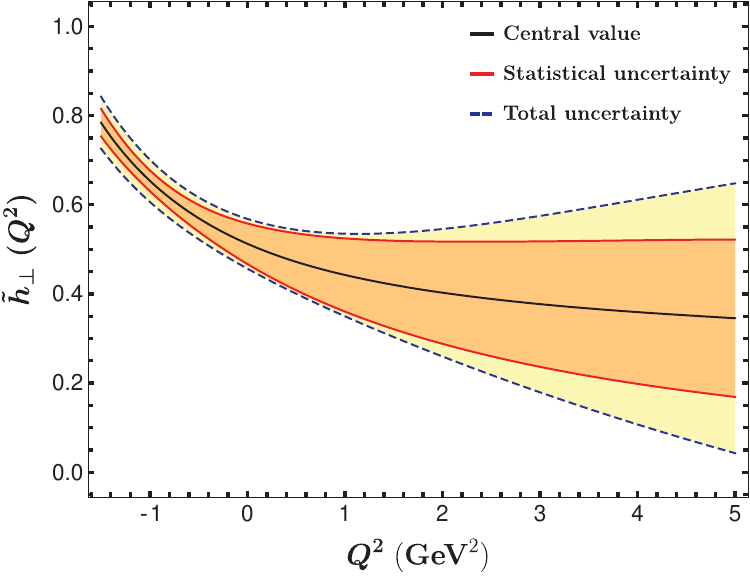} \quad
	\includegraphics[width=4.1cm]{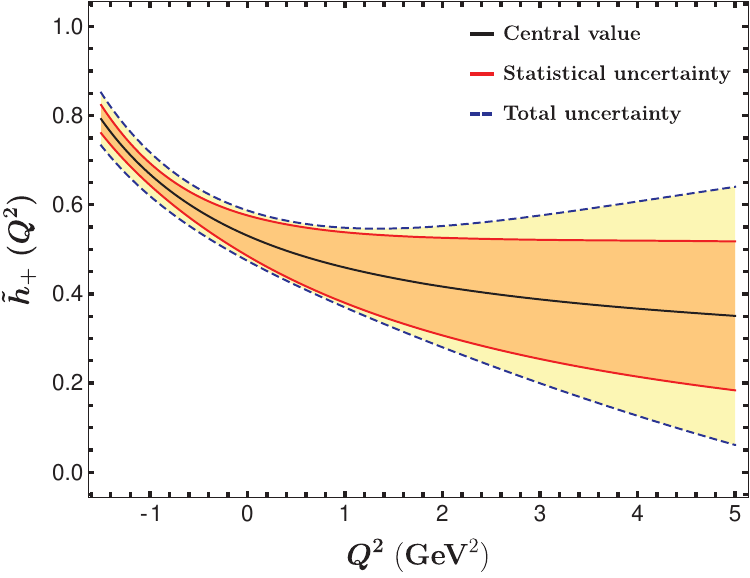} 
	\caption{The $Q^2$ dependence of the different form factors, where the red and blue dashed lines denote the statistical and total uncertainties of the form factors within $1\sigma$ error bars, respectively.}
	\label{fig:form_error}
\end{figure*}

\begin{figure}[t]
	\centering
	\includegraphics[width=4.132cm]{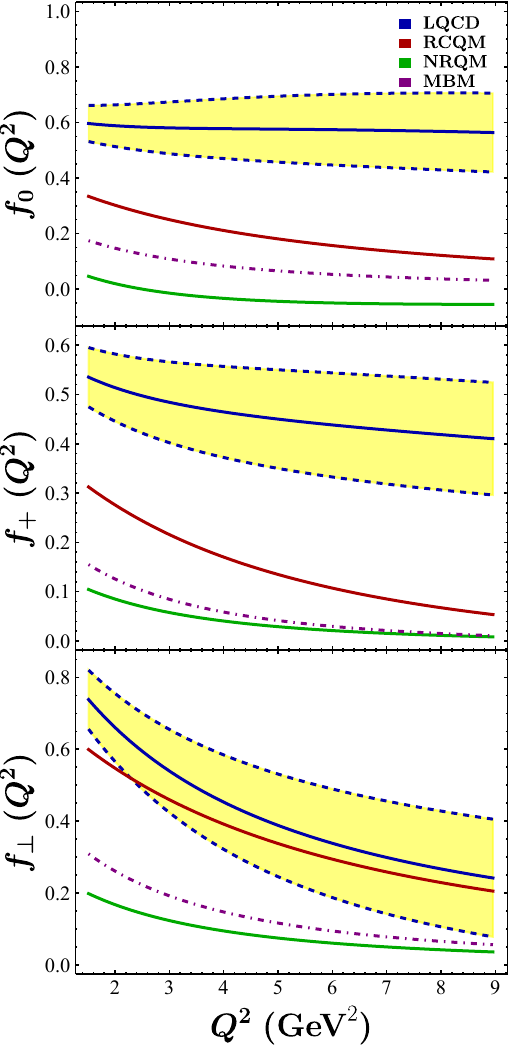} 
	\hspace{0.1cm}
	\includegraphics[width=4.25cm]{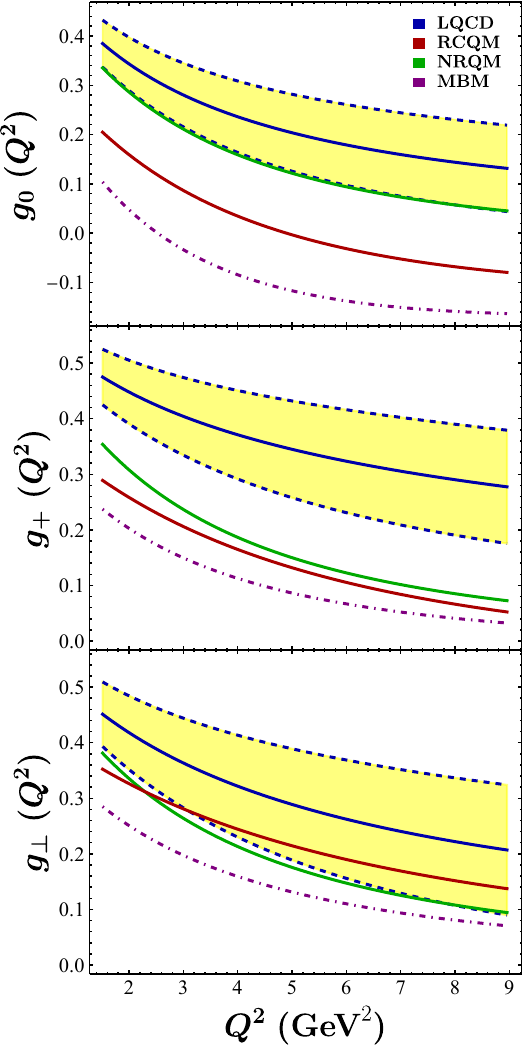}  
	\caption{The $Q^2$ dependence of the $N\to \Lambda_c$ transition form factors deduced from extrapolating to $Q^2>0$ the results of LQCD~\cite{Meinel:2017ggx}, RCQM~\cite{Gutsche:2014zna}, NRQM~\cite{Perez-Marcial:1989sch,Avila-Aoki:1989arc}, and MBM~\cite{Perez-Marcial:1989sch,Avila-Aoki:1989arc}, respectively.}
	\label{fig:form_comb}
\end{figure}

The $\Lambda_c\to N$ transition form factors used in this work are defined in 
the helicity basis~\cite{Feldmann:2011xf,Meinel:2017ggx,Das:2018sms}. For the vector and axial-vector currents, their hadronic matrix elements are defined, respectively, by 
\begin{align}
&\langle N(p,s)|\bar{d}\gamma^{\mu}c|\Lambda_c(p^\prime,s^\prime)\rangle \nonumber \\[0.12cm]
&\ \ \!=\!\bar{u}_N(p,s)\bigg[f_0(q^2)(m_{\Lambda_c}\!-\! m_N)\frac{q^{\mu}}{q^2}\nonumber \\[0.12cm]
&\ \  \ \ \ \!+\!f_+(q^2)\frac{m_{\Lambda_c}\!+\!m_N}{s_+}
\Big(p^{\prime\mu}\!+\!p^{\mu}\!-\!(m^2_{\Lambda_c}\!-\!m^2_N)\frac{q^{\mu}}{q^2}\Big) \nonumber \\[0.12cm]
&\ \ \ \ \  \!+\! f_{\perp}(q^2)\Big(\gamma^{\mu}\!-\!\frac{2m_N}{s_+}p^{\prime\mu}\!-\!\frac{2m_{\Lambda_c}}{s_+}p^{\mu}\Big)\bigg]u_{\Lambda_c}(p^\prime,s^\prime)\,, \label{eq:vect}
\end{align}
and 
\begin{align}
&\langle N(p,s)|\bar{d}\gamma^{\mu}\gamma^5c|\Lambda_c(p^\prime,s^\prime)\rangle \nonumber \\[0.12cm]
&\ \ \!=\!-\bar{u}_N(p,s)\gamma^5\bigg[g_0(q^2)(m_{\Lambda_c}\!+\! m_N)\frac{q^{\mu}}{q^2}\nonumber \\[0.12cm]
&\ \ \ \ \ \!+\! g_+(q^2)\frac{m_{\Lambda_c}\!-\!m_N}{s_-}
\Big(p^{\prime\mu}\!+\!p^{\mu}\!-\!(m^2_{\Lambda_c}\!-\! m^2_N)\frac{q^{\mu}}{q^2}\Big) \nonumber \\[0.12cm]
&\ \ \ \ \ \!+\!g_{\perp}(q^2)\Big(\gamma^{\mu}\!+\!\frac{2m_N}{s_-}p^{\prime\mu}\!-\!\frac{2m_{\Lambda_c}}{s_-}p^{\mu}\Big)\bigg]u_{\Lambda_c}(p^\prime,s^\prime)\,, \label{eq:psvect}
\end{align}
where $q=p^\prime-p$ and $s_{\pm}=(m_{\Lambda_c}\pm m_N)^2-q^2$. From Eqs.~\eqref{eq:vect} and \eqref{eq:psvect}, we can obtain the hadronic matrix elements of the scalar and pseudo-scalar currents through the equation of motion, which are given, respectively, by  
\begin{align}
&\langle N(p,s)|\bar{d}c|\Lambda_c(p^\prime,s^\prime)\rangle \nonumber \\[0.15cm]
&\ \ =\frac{(m_{\Lambda_c}- m_N)}{m_{c}- m_{d}}f_0(q^2)\bar{u}_N(p,s)u_{\Lambda_c}(p^\prime,s^\prime)\,, \\[0.2cm]
&\langle N(p,s)|\bar{d}\gamma^5	c|\Lambda_c(p^\prime,s^\prime)\rangle \nonumber \\[0.15cm]
&\ \  =\frac{(m_{\Lambda_c}+m_N)}{m_{c}+m_{d}}g_0(q^2)\bar{u}_N(p,s)\gamma^5 u_{\Lambda_c}(p^\prime,s^\prime)\,,\label{eq:scalar}
\end{align}
where $m_{d(c)}$ denotes the $d(c)$-quark running mass. Finally, the hadronic matrix element of the tensor current is given by 
\begin{align}
&\langle N(p,s)|\bar{d}i\sigma_{\mu\nu} c|\Lambda_c(p^\prime,s^\prime)\rangle \nonumber \\[0.15cm]
&\ \ \!=\!\bar{u}_N(p,s) \bigg[2h_+ \frac{p^\prime_{\mu}p_{\nu}\! -\! p^\prime_{\nu}p_{\mu}}{s_+}\!
+\! h_{\perp}\Big(\frac{m_{\Lambda_c}\! +\! m_N}{q^2}\nonumber \\[0.15cm]
&\ \ \ \ \ \! \times\! (q_{\mu}\gamma_{\nu}\!-\! q_{\nu}\gamma_{\mu})\!-\! 2\left(\frac{1}{q^2}\!
+\!\frac{1}{s_+}\right)(p^\prime_{\mu}p_{\nu}\!-\! p^\prime_{\nu}p_{\mu})\Big)\nonumber \\[0.15cm]
&\ \ \ \ \ \!+\! \widetilde{h}_+\Big(i\sigma_{\mu\nu}\!-\!\frac{2}{s_-}[m_{\Lambda_c}(p_{\mu}\gamma_{\nu}\!-\!p_{\nu}\gamma_{\mu})\nonumber \\[0.15cm]
&\ \ \ \ \ \!-\!m_N(p^\prime_{\mu}\gamma_{\nu}\!-\!p^\prime_{\nu}\gamma_{\mu})+\!p^\prime_{\mu}p_{\nu}\!-\!p^\prime_{\nu}p_{\mu}]\Big)\nonumber \\[0.15cm]
&\ \ \ \ \ \!+\widetilde{h}_{\perp}\frac{m_{\Lambda_c}-m_N}{q^2s_-}\Big((m_{\Lambda_c}^2-m_N^2-q^2)(\gamma_{\mu}p^\prime_{\nu}-\gamma_{\nu}p^\prime_{\mu})\nonumber \\[0.15cm]
&\ \ \ \ \ \!-(m_{\Lambda_c}^2-m_N^2+q^2)(\gamma_{\mu}p_{\nu}-\gamma_{\nu}p_{\mu})\nonumber \\[0.15cm]
&\ \ \ \ \ \!+\!2(m_{\Lambda_c}-m_N)(p^\prime_{\mu}p_{\nu}-p^\prime_{\nu}p_{\mu})\Big)\bigg]u_{\Lambda_c}(p^\prime,s^\prime)\,, \label{eq:tensor}
\end{align}
where $\sigma_{\mu\nu}=i[\gamma_{\mu},\gamma_{\nu}]/2$. 

The parametrization of these $\Lambda_c\to N$ transition form factors calculated in LQCD takes the form~\cite{Bourrely:2008za,Meinel:2017ggx}
\begin{align}\label{eq:formfactorpara}
	f(q^{2})=\frac{1}{1-q^{2} /(m_{\text{pole}}^{f})^{2}} \sum_{n=0}^{n_{\max}} a_{n}^{f}\left[z(q^{2})\right]^{n},
\end{align}
with the expansion variable defined by
\begin{align}
	z(q^{2})=\frac{\sqrt{t_{+}-q^{2}}-\sqrt{t_{+}-t_{0}}} {\sqrt{t_{+}-q^{2}}+\sqrt{t_{+}-t_{0}}},
\end{align}
where $t_{+}=(m_{D}+m_{\pi})^{2}$ is set equal to the threshold of $D\pi$ two-particle states, $t_{0}=(m_{\Lambda_c}-m_{N})^{2}$ determines which value of $q^2$ gets mapped to $z=0$, and the lowest poles are already factored out before the $z$ expansion, with their quantum numbers and masses listed in Table~IV of Ref.~\cite{Meinel:2017ggx} for the different form factors. The central values and the statistical uncertainties of $a^f_{0,1,2}$ in Eq.~\eqref{eq:formfactorpara} for different form factors $f(q^2)$ have been evaluated in Ref.~\cite{Meinel:2017ggx} by the nominal fit ($n_{\max}=2$), while their systematic uncertainties can be obtained by a combined analysis of both the nominal and higher-order ($n_{\max}=3$) fits; we refer the readers to Ref.~\cite{Meinel:2017ggx} for further details. 

In Fig.~\ref{fig:form_error}, we depict the central values as well as the statistical and total uncertainties of these form factors with respect to the kinematics $Q^2$. It can be seen that the yellow region in each plot increases dramatically along with the increase of $Q^2$, indicating a larger total uncertainty in the larger $Q^2$ range. Although the statistical uncertainties also increases along with the increase of $Q^2$, their behaviors are much milder. Therefore, we only take the statistical uncertainties into account throughout this work. 

Often, the hadronic matrix elements of the vector and axial-vector currents are expressed in terms of another set of form factors $f^{V,A}_{i}$ with $i=1,2,3$, which are related to the ones introduced in 
Eqs.~\eqref{eq:vect} and \eqref{eq:psvect} by
\begin{align}
	f_0&=\frac{q^2}{m_{\Lambda_c}(m_{\Lambda_c}-m_N)}f^V_3+f^V_1\,, \nonumber \\[0.15cm]
	f_{+}&=f^V_1+\frac{q^2}{m_{\Lambda_c}(m_{\Lambda_c}+m_N)}f^V_2\,, \nonumber \\[0.15cm]
	f_{\perp}&=f^V_1+f^V_2\frac{(m_N+m_{\Lambda_c})}{m_{\Lambda_c}}\,, \nonumber \\[0.15cm]
	g_0&=-\frac{q^2}{m_{\Lambda_c}(m_{\Lambda_c}-m_N)}f^A_3+f^A_1\,, \nonumber \\[0.15cm]
	g_{+}&=f^A_1-\frac{q^2}{m_{\Lambda_c}(m_{\Lambda_c}-m_N)}f^A_2\,, \nonumber \\[0.15cm]
	g_{\perp}&=f^A_1+f^A_2\frac{(m_N-m_{\Lambda_c})}{m_{\Lambda_c}}\,.
\end{align}
To parametrize the $q^2$ dependence of this set of form factors, the RCQM model adopts the following double-pole form~\cite{Gutsche:2014zna}: 
\begin{align}\label{eq:RCQM_Para}
	f(q^2)=\frac{f(0)}{1-a\hat{s}+b\hat{s}^2}\,,
\end{align}
with $\hat{s}=q^2/m_{\Lambda_c}^2$, where the values of the parameters $f(0)$, $a$, and $b$ are listed in Table~\ref{tab:RCQM}. On the other hand, the MBM and NRQM models employ both the monopole and dipole parametrizations for these form factors~\cite{Perez-Marcial:1989sch,Avila-Aoki:1989arc}. For simplicity, we only consider the later, which has the following form: 
\begin{align}\label{eq:NRQM_Para}
	f(q^2)=\frac{A}{\left(1-q^2/M_{R}^2\right)^2}\,,
\end{align}
where the values of the parameters $A$ and $M_R$ are reported in Table~\ref{tab:NRQM}. We refer the readers to Ref.~\cite{Sobczyk:2019uej} for more details about the form-factor parametrizations in different models. 

\begin{table}[t]	
	\renewcommand*{\arraystretch}{1.5}
	\tabcolsep=0.73cm
	\centering
	\caption{Values of the parameters employed in Eq.~\eqref{eq:RCQM_Para} to construct the $q^2$ dependence of the form factors associated with the vector and axial-vector currents (see Eqs.~\eqref{eq:vect} and \eqref{eq:psvect}) for the RCQM model~\cite{Gutsche:2014zna}.}
	\begin{tabular}{cccc}
		\hline \hline
		& $f(0)$ & $a$  & $b$ \\
		\hline
		$f_1^V$ &  0.470 & 1.111 & 0.303 \\
		$f_2^V$ & 0.247 & 1.240  & 0.390  \\
		$f_3^V$ &  0.038 &  0.308 & 1.998 \\
		$f_1^A$ & 0.414 & 0.978 & 0.235 \\
		$f_2^A$ & -0.073  & 0.781 & 0.225\\
		$f_3^A$ & -0.328 & 1.330  & 0.486\\
		\hline \hline
	\end{tabular}
	\label{tab:RCQM}
\end{table}

\begin{table}[t]
	\renewcommand\arraystretch{1.5} 
	\tabcolsep=0.36cm
	\centering
	\caption{Values of the parameters employed in Eq.~\eqref{eq:NRQM_Para} to construct the $q^2$ dependence of the form factors associated with the vector and axial-vector currents (see Eqs.~\eqref{eq:vect} and \eqref{eq:psvect}) for the MBM and NRQM models~\cite{Perez-Marcial:1989sch,Avila-Aoki:1989arc}.}
	\begin{tabular}{ccccc}
		\hline \hline
		& \multicolumn{2}{c}{NRQM}&
		\multicolumn{2}{c}{MBM}\\
		\cline{2-5}
		& A & $M_R$~(GeV)& A & $M_R$~(GeV)\\
		\hline
		$f_1^V$ & $0.22$ & $2.01$ & $0.33$ & $2.01$ \\
		$f_2^V$ & $0.11$ & $2.01$ & $0.18$ & $2.01$ \\
		$f_3^V$ & $0.27$ & $2.01$ & $0.00$ & $2.01$ \\
		$f_1^A$ & $0.58$ & $2.42$ & $0.41$ & $2.42$ \\
		$f_2^A$ & $-0.04$ & $2.42$ & $-0.07$ & $2.42$ \\
		$f_3^A$ & $-0.10$ & $2.42$ & $-0.50$ & $2.42$ \\
		\hline \hline
	\end{tabular}
	\label{tab:NRQM} 
\end{table}

In Fig.~\ref{fig:form_comb}, we show the $Q^2$ dependence of these six form factors associated with the matrix elements of the vector and axial-vector currents in the suitable kinematic range ($Q^2>0$) for the low-$E$ QE scattering process. It can be seen that the LQCD predicts the largest values for all these six form factors. Especially for $f_{0}$, $f_{+}$, and $g_{+}$, the central values provided by the three models lie outside the $1\sigma$ error bars of the LQCD calculations. Interestingly enough, the NRQM model produces the lowest values for $f_{0,+,\perp}$, while the MBM model provides the lowest values for $g_{0,+,\perp}$.

Finally, it should be mentioned that another set of form factors has also been employed to parametrize the transition matrix elements of the vector and axial-vector currents. They can be related to $f^{V,A}_{i}$ in a trivial way, and have been investigated in the (light-cone) QCD sum rule approach (see, e.g., Refs.~\cite{Khodjamirian:2011jp,Li:2016qai,Zhang:2023nxl}) and the light-front constituent quark model (see, e.g., Refs.~\cite{Geng:2020gjh,Geng:2020fng}). However, since the form factors $f^{V,A}_3$ were not calculated, the results presented in these references will not be considered in this work.

\section{Amplitude squared of the QE scattering process}
\label{app:Amplitude}

For the convenience of future discussions, we provide here the explicit expression of the amplitude squared $|\mathcal{M}|^2 $ of the QE scattering process $\nu_{\tau}(k)+n(p)\to \tau^-(k')+\Lambda_c(p')$ mediated by the general effective Lagrangian $\mathcal{L}_{\text{eff}}$ (see Eq.~\eqref{eq:Leff}). With all the operators of $\mathcal{L}_{\text{eff}}$ taken into account, the amplitude square $|\mathcal{M}|^2 $ is given explicitly by 
\begin{widetext}
	\begin{align}\label{eq:amplitude2}
		 |\mathcal{M}|^2=&|1+g_{V}^{L}|^2\mathcal{A}_{V_{L}-V_{L}}
		+|g_{V}^{R}|^2\mathcal{A}_{V_{R}-V_{R}}
		+(|g_{S}^{L}|^2+|g_{S}^{R}|^2)\mathcal{A}_{S_{L}-S_{L}}
		+|g_{T}^{L}|^2\mathcal{A}_{T_{L}-T_{L}}
		+2\text{Re}[g_{S}^{L}g_{S}^{R*}]\mathcal{A}_{S_{L}-S_{R}} \nonumber\\[0.15cm]
		&+2\text{Re}[g_{V}^{R}(1+g_{V}^{L*})]\mathcal{A}_{V_{R}-V_{L}}
		+2\text{Re}[g_{S}^{L}(1+g_{V}^{L*})+g_{S}^{R}g_{V}^{R*}]\mathcal{A}_{S_{L}-V_{L}}
		+2\text{Re}[g_{S}^{R}(1+g_{V}^{L*})+g_{S}^{L}g_{V}^{R*}]\mathcal{A}_{S_{R}-V_{L}}\nonumber\\[0.15cm]
		&+2\text{Re}[g_{T}^{L}(1+g_{V}^{L*})]\mathcal{A}_{T_{L}-V_{L}}
		+2\text{Re}[g_{T}^{L}g_{V}^{R*}]\mathcal{A}_{T_{L}-V_{R}}
		+2\text{Re}[g_{T}^{L}g_{S}^{LL*}]\mathcal{A}_{T_{L}-S_{L}}
		+2\text{Re}[g_{T}^{L}g_{S}^{R*}]\mathcal{A}_{T_{L}-S_{R}}\,,
	\end{align}
\end{widetext}
where the various subscripts attached to the different $\mathcal{A}$ on the right-hand side represent the possible interference between the two operators (see Ref.~\cite{Lai:2021sww} for more details). Note that, because of the chiral structures of the lepton and quark currents involved, $\mathcal{A}$ with different subscripts can be identical to each other, e.g., 
$\mathcal{A}_{S_{L}-V_{L}} = \mathcal{A}_{S_{R}-V_{R}}$ and thus only one of them is kept in Eq.~\eqref{eq:amplitude2}. The amplitudes associated with other interference terms that are not shown in Eq.~\eqref{eq:amplitude2} are all zero. For convenience, we provide here the explicit expressions of the $\mathcal{A}$ on the right-hand side of Eq.~\eqref{eq:amplitude2} as 
\begin{widetext}
	\begin{align}
			\mathcal{A}_{V_{L}-V_{L}}=&\frac{m_\tau^2 \left(m_\tau^2\!-\!q^2\right)}{2q^4}\left[f_0^2 (m_{\Lambda_c}\!-\!m_n)^2 s_++g_0^2 (m_{\Lambda_c}+m_n)^2s_-\right]\!-\!\frac{m_\tau^2\left(m_{\Lambda_c}^2-m_n^2\right) }{q^4}		\nonumber\\[0.15cm]
		&\times\!(f_0 f_+\!+\!g_0 g_+)\left[4 E m_n q^2\!+\!\left(m_\tau^2\!-\!q^2\right) \left(m_{\Lambda_c}^2\!-\!m_n^2-q^2\right)\right]
		+\left[\frac{f_+^2 (m_{\Lambda_c}\!+\!m_n)^2}{2q^4 s_+}
		\right.\nonumber\\[0.15cm]
		&\left.+\frac{ g_+^2 (m_{\Lambda_c}\!\!-\!m_n)^2}{2q^4 s_-}\right]\Big{\{}4 m_n^2 q^4 \left(4 E^2\!-\!m_\tau^2+q^2\right)+\left(m_\tau^2\!-\!q^2\right) \left(m_{\Lambda_c}^2-m_n^2-q^2\right) 
		\nonumber\\[0.15cm]
		&\times\!\left[8 E m_n q^2\!+\!m_\tau^2 \left(m_{\Lambda_c}^2\!\!-\!m_n^2\!-\!q^2\right)\right]\Big{\}}
		\!+\!\left(\frac{f_\perp^2}{s_+}\!+\!\frac{ g_\perp^2}{s_-}\right)\Bigg\{8 E^2 m_n^2 q^2\!+\!\left(m_\tau^2\!-\!q^2\right) \nonumber\\[0.12cm]
		&\times\left[2 m_{\Lambda_c}^2 q^2-4 E m_n \left(m_n^2-m_{\Lambda_c}^2+q^2\right)-\left(m_{\Lambda_c}^2-m_n^2\right)^2+2 m_n^2 m_\tau^2-q^4\right]\Bigg\}
		\nonumber\\[0.15cm]
		&-2f_\perp g_\perp \left[4 E m_n q^2+\left(m_\tau^2-q^2\right) \left(m_{\Lambda_c}^2-m_n^2-q^2\right)\right]
		\,,
		\\[0.2cm]
		\mathcal{A}_{V_{R}-V_{R}}=&\frac{m_\tau^2 \left(m_\tau^2\!-\!q^2\right)}{2q^4}\left[f_0^2 (m_{\Lambda_c}\!-\!m_n)^2 s_++g_0^2 (m_{\Lambda_c}+m_p)^2s_-\right]\!-\!\frac{m_\tau^2\left(m_{\Lambda_c}^2-m_n^2\right) }{q^4}		\nonumber\\[0.15cm]
		&\times\!(f_0 f_+\!+\!g_0 g_+)\left[4 E m_n q^2\!+\!\left(m_\tau^2\!-\!q^2\right) \left(m_{\Lambda_c}^2\!-\!m_n^2-q^2\right)\right]
		+\left[\frac{f_+^2 (m_{\Lambda_c}\!+\!m_n)^2}{2q^4 s_+}
		\right.\nonumber\\[0.15cm]
		&\left.+\frac{ g_+^2 (m_{\Lambda_c}\!\!-\!m_n)^2}{2q^4 s_-}\right]\Big{\{}4 m_n^2 q^4 \left(4 E^2\!-\!m_\tau^2+q^2\right)+\left(m_\tau^2\!-\!q^2\right) \left(m_{\Lambda_c}^2-m_p^2-q^2\right) 
		\nonumber\\[0.15cm]
		&\times\!\left[8 E m_n q^2\!+\!m_\tau^2 \left(m_{\Lambda_c}^2\!\!-\!m_n^2\!-\!q^2\right)\right]\Big{\}}
		\!+\!\left(\frac{f_\perp^2}{s_+}\!+\!\frac{ g_\perp^2}{s_-}\right)\Bigg\{8 E^2 m_n^2 q^2\!+\!\left(m_\tau^2\!-\!q^2\right) 	\nonumber\\[0.15cm]
		&\times\left[2 m_{\Lambda_c}^2 q^2-4 E m_n \left(m_n^2-m_{\Lambda_c}^2+q^2\right)-\left(m_{\Lambda_c}^2-m_n^2\right)^2+2 m_n^2 m_\tau^2-q^4\right]\Bigg\}
		\nonumber\\[0.15cm]
		&+2f_\perp g_\perp \left[4 E m_n q^2+\left(m_\tau^2-q^2\right) \left(m_{\Lambda_c}^2-m_n^2-q^2\right)\right]
		\,,
		\\[0.2cm]
	   \mathcal{A}_{S_{L}-S_{L}}=&\dfrac{m_\tau^2-q^2 }{2m_c^2}\left[f_0^2 (m_{\Lambda_c}-m_n)^2 s_++g_0^2 (m_{\Lambda_c}+m_n)^2 s_-\right]\,,
		\\[0.2cm]		
       \mathcal{A}_{T_{L}-T_{L}}=&-8\left(\dfrac{h_+^2}{s_+}\!+\!\dfrac{\tilde{h}_+^2}{s_-}\right)\Big\{4 m_\tau^4 m_n^2 \!+ \!m_{\Lambda_c}^4 (q^2\!-\!m_\tau^2)\!+\!2 m_{\Lambda_c}^2 (m_\tau^2\!-\!q^2) (4 E m_n\!+\!m_n^2\nonumber\\[0.15cm]
		&+\!q^2)\!+\!q^2 (4 E m_n\!+\!m_n^2 + q^2)^2\!-\! m_\tau^2 \big[m_n^4\!+\!6 m_n^2 q^2\!+\!q^4\!+\!8 E m_n (m_n^2\!+\! q^2)\big]\Big\}\nonumber\\[0.15cm]
		&+\!16\left[\dfrac{h_\perp^2(m_{\Lambda_c}\!+\!m_n)^2}{s_+q^4}\!+\!\dfrac{\tilde{h}_\perp^2(m_{\Lambda_c}\!-\!m_n)^2}{s_-q^4}\right]
		\Big\{2 m_n (2 E\!+\!m_n) q^4 (2 E m_n\!+\!q^2)\nonumber\\[0.15cm]
		& - m_\tau^2 q^2 (m_n^2 + q^2) (4 E m_n + m_n^2 + q^2) + m_{\Lambda_c}^4 m_\tau^2 (m_\tau^2 - q^2) +m_\tau^4 (m_n^4 + q^4)\nonumber\\[0.15cm]
		& - 2 m_{\Lambda_c}^2 (m_\tau^2 - q^2) \big[m_\tau^2 (m_n^2 + q^2)-2 E m_n q^2\big]\Big\}-\frac{32 m_\tau^2 (m_{\Lambda_c}^2 - m_n^2)}{q^4}\nonumber\\[0.15cm]
		&\times \big[m_{\Lambda_c}^2 (m_\tau^2 - q^2) - m_\tau^2 (m_n^2 + q^2) + 
		q^2 (4 E m_n + m_n^2 + q^2)\big]h_\perp\,\tilde{h}_\perp\,, 
		\\[0.2cm]	
		\mathcal{A}_{V_{R}-V_{L}}=&\frac{m_\tau^2 \left(m_\tau^2\!-\!q^2\right)}{2q^4}\left[f_0^2 (m_{\Lambda_c}\!-\!m_n)^2 s_+-g_0^2 (m_{\Lambda_c}+m_n)^2s_-\right]\!-\!\frac{m_\tau^2\left(m_{\Lambda_c}^2-m_n^2\right) }{q^4}		\nonumber\\[0.15cm]
		&\times\!(f_0 f_+\!-\!g_0 g_+)\left[4 E m_n q^2\!+\!\left(m_\tau^2\!-\!q^2\right) \left(m_{\Lambda_c}^2\!-\!m_n^2-q^2\right)\right]
		+\left[\frac{f_+^2 (m_{\Lambda_c}\!+\!m_n)^2}{2q^4 s_+}
		\right.\nonumber\\[0.15cm]
		&\left.-\frac{ g_+^2 (m_{\Lambda_c}\!\!-\!m_n)^2}{2q^4 s_-}\right]\Big{\{}4 m_n^2 q^4 \left(4 E^2\!-\!m_\tau^2+q^2\right)+\left(m_\tau^2\!-\!q^2\right) \left(m_{\Lambda_c}^2-m_n^2-q^2\right) 
		\nonumber\\[0.15cm]
		&\times\!\left[8 E m_n q^2\!+\!m_\tau^2 \left(m_{\Lambda_c}^2\!\!-\!m_n^2\!-\!q^2\right)\right]\Big{\}}
		\!+\!\left(\frac{f_\perp^2}{s_+}\!-\!\frac{ g_\perp^2}{s_-}\right)\left\{8 E^2 m_n^2 q^2\!+\!\left(m_\tau^2\!-\!q^2\right) 	\right.\nonumber\\[0.15cm]
		&\left.\times\left[2 m_{\Lambda_c}^2 q^2-4 E m_n \left(m_n^2-m_{\Lambda_c}^2+q^2\right)-\left(m_{\Lambda_c}^2-m_n^2\right)^2+2 m_n^2 m_\tau^2-q^4\right]\right\}
		\,,
		\\[0.2cm]	
		\mathcal{A}_{S_{L}-V_{L}}=&\frac{m_\tau \left(q^2\!-\!m_\tau^2\right)}{2m_c q^2} \left[f_0^2 (m_{\Lambda_c}\!-\!m_n)^2 s_+\!+\!g_0^2 (m_{\Lambda_c}\!+\!m_n)^2s_-\right]\!+\!\frac{m_\tau\left(m_{\Lambda_c}^2\!-\!m_n^2\right) }{2m_c q^2}	\nonumber\\[0.15cm]
		&\times(f_0 f_++g_0 g_+)\left[4 E m_n q^2+\left(m_\tau^2-q^2\right) \left(m_{\Lambda_c}^2-m_n^2-q^2\right)\right]\,,
		\\[0.2cm]
		\mathcal{A}_{S_{R}-V_{L}}=&\frac{m_\tau \left(q^2\!-\!m_\tau^2\right)}{2m_c q^2} \left[f_0^2 (m_{\Lambda_c}\!-\!m_n)^2 s_+\!-\!g_0^2(m_{\Lambda_c}+m_n)^2s_-\right]
		+\frac{m_\tau\left(m_{\Lambda_c}^2-m_n^2\right) }{2m_c q^2} 	\nonumber\\[0.15cm]
		&\times(f_0 f_+-g_0 g_+)\left[4 E m_n q^2+\left(m_\tau^2-q^2\right) \left(m_{\Lambda_c}^2-m_n^2-q^2\right)\right]\,,
		\\[0.2cm]
		\mathcal{A}_{T_{L}-V_{L}}=&-\frac{2m_\tau}{q^2}\Big\{\big[m_{\Lambda_c}^2 (m_\tau^2 - q^2) - m_\tau^2 (m_n^2 + q^2) + q^2 (4 E m_n + m_n^2 + q^2)\big]\big[(m_{\Lambda_c}- m_n)  \nonumber\\[0.15cm]
		&\times(f_0\,h_+ +2f_\perp\,\tilde{h}_\perp)+(m_{\Lambda_c} + m_n) (g_0 \,\tilde{h}_+ + 2g_\perp\,h_\perp)\big]-(m_\tau^2 - q^2)\big[(m_{\Lambda_c}+ m_n) \nonumber\\[0.15cm]
		&\times s_-(f_+\, h_+ + 2f_\perp\, h_\perp) +(m_{\Lambda_c}- m_n)s_+(g_+\,\tilde{h}_+ + 2g_\perp\, \tilde{h}_\perp)\big]\Big\}\,,
		\\[0.2cm]
		\mathcal{A}_{T_{L}-V_{R}}=&-\frac{2m_\tau}{q^2}\Big\{\big[m_{\Lambda_c}^2 (m_\tau^2 - q^2) - m_\tau^2 (m_n^2 + q^2) + q^2 (4 E m_n + m_n^2 + q^2)\big]\big[(m_{\Lambda_c}- m_n)  \nonumber\\[0.15cm]
		&\times (f_0\,h_+ +2f_\perp\,\tilde{h}_\perp)-(m_{\Lambda_c} + m_n) (g_0 \,\tilde{h}_+ + 2g_\perp\,h_\perp)\big]-(m_\tau^2 - q^2)\big[(m_{\Lambda_c}+ m_n) \nonumber\\[0.15cm]
		&\times s_-(f_+\, h_+ + 2f_\perp\, h_\perp)-(m_{\Lambda_c}- m_n)s_+(g_+\,\tilde{h}_+ + 2g_\perp\, \tilde{h}_\perp)\big]\Big\}\,,
 	\\[0.2cm]
 		\mathcal{A}_{T_{L}-S_{L}}=&\frac{2}{m_c}\left[4 E m_n q^2+\left(m_\tau^2-q^2\right) \left(m_{\Lambda_c}^2-m_p^2-q^2\right)\right]\left[f_0 h_+ (m_{\Lambda_c}-m_p)+g_0 \tilde{h}_+ (m_{\Lambda_c}+m_n)\right]\,,	\\[0.2cm]
	\mathcal{A}_{T_{L}-S_{R}}=&\frac{2}{m_c} \left[4 E m_n q^2+\left(m_\tau^2-q^2\right) \left(m_{\Lambda_c}^2-m_n^2-q^2\right)\right]\left[f_0 h_+ (m_{\Lambda_c}-m_p)-g_0 \tilde{h}_+ (m_{\Lambda_c}+m_p)\right]\,,
\\[0.2cm] 		
	\mathcal{A}_{S_{L}-S_{R}}=&\dfrac{m_\tau^2-q^2 }{2m_c^2}\Big{[}f_0^2 (m_{\Lambda_c}-m_n)^2 s_+-g_0^2 (m_{\Lambda_c}+m_n)^2 s_-\Big{]}\,.
 \end{align}
\end{widetext}

\section{\boldmath Details of the polarization vectors of $\tau$ and $\Lambda_c$}
\label{app:numerator}

We now present the explicit expressions of $P_L^{l,h}$, $P_P^{l,h}$, and $P_T^{l,h}$ of the outgoing $\tau$ and $\Lambda_c$. These components of the polarization vectors are defined in Eq.~\eqref{eq:polcomp} and read 
\begin{align} 
	P_a^{l,h}&=-(\mathcal{P}\cdot N_a)^{l,h}\nonumber \\
	&=\frac{\text{Tr}[\rho_{l,h} \gamma_5\,\slashed{N}_a]}{\text{Tr}[\rho_{l,h}]}
	=\frac{\mathcal{A}_a^{(l,h)}}{2m_{(\tau,\Lambda_c)}|\mathcal{M}|^2}\,.
\end{align}
Note that the trace over the spin density matrices $\rho_{l,h}$ has been replaced in the last step by 
\begin{align}
	\text{Tr}[\rho_{l,h}]=2m_{(\tau,\Lambda_c)}|\mathcal{M}|^2\,,
\end{align}
which can be inferred from Eqs.~\eqref{eq:rho_tau} and \eqref{eq:rho_lambda}, and the amplitude squared $|\mathcal{M}|^2$ has been given in Eq.~\eqref{eq:amplitude2}. In addition, the trace in the numerator has been redefined as $\mathcal{A}_a^{(l,h)}$, which are given, respectively, by 
\begin{widetext}
	\begin{align}
		\mathcal{A}_{a}^{l}=&|1+g_{V}^{LL}|^2\mathcal{A}^{l}_{V_{L}-V_{L}}
		+|g_{V}^{R}|^2\mathcal{A}^{l}_{V_{R}-V_{R}}
		+(|g_{S}^{L}|^2+|g_{S}^{R}|^2)\mathcal{A}^{l}_{S_{L}-S_{L}}\nonumber\\[0.15cm]
		&+|g_{T}^{L}|^2\mathcal{A}^{l}_{T_{L}-T_{L}}+2\text{Re}[g_{V}^{R}(1+g_{V}^{L*})\mathcal{A}^{l}_{V_{R}-V_{L}}
		+2\text{Re}[g_{T}^{L}(1+g_{V}^{L*})\mathcal{A}^{l}_{T_{L}-V_{L}}]\nonumber\\[0.15cm]
		&+2\text{Re}[(g_{S}^{L}(1+g_{V}^{L*})+g_{S}^{R}g_{V}^{R*})\mathcal{A}^{l}_{S_{L}-V_{L}}]
		+2\text{Re}[g_{S}^{L}g_{S}^{R*}\mathcal{A}^{l}_{S_{L}-S_{R}}]\nonumber\\[0.15cm]
		&+2\text{Re}[(g_{S}^{R}(1+g_{V}^{L*})+g_{S}^{L}g_{V}^{R*})\mathcal{A}^{l}_{S_{R}-V_{L}}]
		+2\text{Re}[g_{T}^{L}g_{V}^{R*}\mathcal{A}^{l}_{T_{L}-V_{R}}]\nonumber\\[0.15cm]
		&+2\text{Re}[g_{T}^{L}g_{S}^{L*}\mathcal{A}^{l}_{T_{L}-S_{L}}]
		+2\text{Re}[g_{T}^{L}g_{S}^{R*}\mathcal{A}^{l}_{T_{L}-S_{R}}]\,, \label{eq:Aal}\\[0.2cm]
		\mathcal{A}_a^{h}=&|1+g_{V}^{LL}|^2\mathcal{A}^{h}_{V_{L}-V_{L}}
		+|g_{V}^{R}|^2\mathcal{A}^{h}_{V_{R}-V_{R}}
		+(|g_{S}^{L}|^2-|g_{S}^{R}|^2)\mathcal{A}^{h}_{S_{L}-S_{L}}\nonumber\\[0.15cm]
		&+|g_{T}^{L}|^2\mathcal{A}^{h}_{T_{L}-T_{L}}+2\text{Re}[g_{V}^{R}(1+g_{V}^{L*})\mathcal{A}^{h}_{V_{R}-V_{L}}]
		+2\text{Re}[g_{T}^{L}(1+g_{V}^{L*})\mathcal{A}^{h}_{T_{L}-V_{L}}]\nonumber\\[0.15cm]
		&+2\text{Re}[g_{S}^{L}(1+g_{V}^{L*})\mathcal{A}^{h}_{S_{L}-V_{L}}]
		+2\text{Re}[g_{S}^{R}(1+g_{V}^{L*})\mathcal{A}^{h}_{S_{R}-V_{L}}]\nonumber\\[0.15cm]
		&+2\text{Re}[g_{S}^{R}g_{V}^{R*}\mathcal{A}^{h}_{S_{R}-V_{R}}]+2\text{Re}[g_{S}^{L}g_{V}^{R*}\mathcal{A}^{h}_{S_{L}-V_{R}}]
		+2\text{Re}[g_{T}^{L}g_{V}^{R*}\mathcal{A}^{h}_{T_{L}-V_{R}}]\nonumber\\[0.15cm]
		&+2\text{Re}[g_{T}^{L}g_{S}^{L*}\mathcal{A}^{h}_{T_{L}-S_{L}}]
		+2\text{Re}[g_{T}^{L}g_{S}^{R*}\mathcal{A}^{h}_{T_{L}-S_{R}}]+2\text{Re}[g_{S}^{L}g_{S}^{R*}\mathcal{A}^{h}_{S_{L}-S_{R}}]\,. \label{eq:Aah}
	\end{align}
\end{widetext}
The explicit expressions of all the $\mathcal{A}^{l,h}$ on the right-hand side of Eqs.~\eqref{eq:Aal} and \eqref{eq:Aah} are presented as follows:
\begin{widetext}
\begin{align}
	\mathcal{A}^{l}_{V_{L}-V_{L}}=&\frac{2m_\tau^4 (N_a\cdot k)}{q^4}\left[f_0^2 (m_{\Lambda_c}-m_n)^2 s_++g_0^2 (m_{\Lambda_c}+m_n)^2s_-\right]-\frac{2 m_\tau^2 \left(m_{\Lambda_c}^2-m_n^2\right)}{q^4} \nonumber\\[0.15cm]
	& \times (f_0 f_++g_0 g_+)\left\{(N_a\cdot k) \left[4 E m_n q^2+\left(m_\tau^2-q^2\right) \left(2 m_{\Lambda_c}^2-2 m_n^2-q^2\right)\right] 
	\right.\nonumber\\[0.15cm]
	&\left.-q^2\left(m_\tau^2-q^2\right) (N_a\cdot p+N_a\cdot p^\prime)\right\}+2m_\tau^2\left[\frac{ f_+^2 (m_{\Lambda_c}+m_n)^2}{q^4 s_+}+\frac{g_+^2 (m_{\Lambda_c}-m_n)^2}{q^4 s_-}\right]
	\nonumber\\[0.15cm]
	&\times\Big{\{}(N_a\cdot k) \left[\left(m_{\Lambda_c}^2\!-\!m_n^2\right)\Big(m_\tau^2 (m_{\Lambda_c}^2-m_n^2-q^2)\!-\!q^2 (2 m_{\Lambda_c}^2\!-\!4m_n E \!-\!2m_n^2+q^2)\Big) 	\right.\nonumber\\[0.15cm]
	&\left.+q^4\left(4 m_{\Lambda_c}^2-q^2\right)\right]-q^2 (N_a\cdot p+N_a\cdot p^\prime)\left[4 E m_n q^2+\left(m_\tau^2-q^2\right) (m_{\Lambda_c}^2-m_n^2-q^2)\right]\Big{\}} \nonumber\\[0.15cm]
	&-8m_\tau^2 f_\perp g_\perp\left[(N_a\cdot p) \left(m_\tau^2-q^2-2 E m_n\right)+2 E m_n (N_a\cdot p^\prime)\right]
	\nonumber\\[0.15cm]
	&+4 m_\tau^2\left(\frac{ f_\perp^2}{s_+}+\frac{ g_\perp^2}{s_-}\right)\Big{\{}(N_a\cdot p) \left[2 E m_n \left(m_{\Lambda_c}^2-m_p^2+q^2\right)+\left(m_\tau^2-q^2\right) (m_{\Lambda_c}^2+m_p^2-q^2)\right]\nonumber\\[0.15cm]
	&+2 m_n (N_a\cdot p^\prime) \left[E \left(m_n^2-m_{\Lambda_c}^2+q^2\right)+m_n \left(q^2-m_\tau^2\right)\right]\Big{\}}\,,
	\\[0.2cm]
    \mathcal{A}^{l}_{V_{R}-V_{R}}=&\frac{2m_\tau^4 (N_a\cdot k)}{q^4}\left[f_0^2 (m_{\Lambda_c}-m_p)^2 s_++g_0^2 (m_{\Lambda_c}+m_n)^2s_-\right]-\frac{2 m_\tau^2 \left(m_{\Lambda_c}^2-m_n^2\right)}{q^4} \nonumber\\[0.15cm]
    &\times  (f_0 f_++g_0 g_+)\left\{(N_a\cdot k) \left[4 E m_n q^2+\left(m_\tau^2-q^2\right) \left(2 m_{\Lambda_c}^2-2 m_n^2-q^2\right)\right] 
    \right.\nonumber\\[0.15cm]
    &\left.-q^2\left(m_\tau^2-q^2\right) (N_a\cdot p+N_a\cdot p^\prime)\right\}+2m_\tau^2\left[\frac{ f_+^2 (m_{\Lambda_c}+m_n)^2}{q^4 s_+}+\frac{g_+^2 (m_{\Lambda_c}-m_n)^2}{q^4 s_-}\right]
    \nonumber\\[0.15cm]
    &\times \Big{\{}(N_a\cdot k) \left[\left(m_{\Lambda_c}^2\!-\!m_n^2\right)\Big(m_\tau^2 (m_{\Lambda_c}^2-m_n^2-q^2)\!-\!q^2 (2 m_{\Lambda_c}^2\!-\!4m_n E\!-\!2m_n^2+q^2)\Big) 	\right.\nonumber\\[0.15cm]
    &\left.+q^4\left(4 m_{\Lambda_c}^2-q^2\right)\right]-q^2 (N_a\cdot p+N_a\cdot p^\prime)\left[4 E m_n q^2+\left(m_\tau^2-q^2\right) (m_{\Lambda_c}^2-m_n^2-q^2)\right]\Big{\}}
    \nonumber\\[0.15cm]
    &+8m_\tau^2 f_\perp g_\perp\left[(N_a\cdot p) \left(m_\tau^2-q^2-2 E m_n\right)+2 E m_n (N_a\cdot p^\prime)\right]
    \nonumber\\[0.15cm]
    &+4 m_\tau^2\left(\frac{ f_\perp^2}{s_+}+\frac{ g_\perp^2}{s_-}\right)\Big{\{}(N_a\cdot p) \left[2 E m_n \left(m_{\Lambda_c}^2-m_n^2+q^2\right)+\left(m_\tau^2-q^2\right) (m_{\Lambda_c}^2+m_n^2-q^2)\right]
    \nonumber\\[0.12cm]
    &+2 m_n (N_a\cdot p^\prime) \left[E \left(m_n^2-m_{\Lambda_c}^2+q^2\right)+m_n \left(q^2-m_\tau^2\right)\right]\Big{\}}\,,
    \\[0.2cm]
	\mathcal{A}^{l}_{S_{L}-S_{L}}=&\mathcal{A}_{S_{L}-S_{L}}\frac{4m_\tau^2( k\cdot N_a)}{m_\tau^2-q^2}\,,
    \\[0.2cm]
    \mathcal{A}^{l}_{T_{L}-T_{L}}=&32 m_\tau^2\left(\frac{h_+^2}{s_+}+\frac{\tilde{h}_+^2}{s_-}\right)\Big{\{}2 (N_a\cdot p) 
	\big[2 E m_n \left(m_{\Lambda_c}^2\!-\!m_n^2+q^2\right)+\left(m_\tau^2\!-\!q^2\right) (m_{\Lambda_c}^2+m_n^2\!-\!q^2)\big]
    \nonumber\\[0.15cm]
    &+\!4 m_n (N_a\cdot p^\prime) \left[E \left(m_n^2\!-\!m_{\Lambda_c}^2\!+\!q^2\right)\!+\!m_n \left(q^2\!-\!m_\tau^2\right)\right]\!
    +\!(N_i\cdot k) \,s_+ s_- \Big{\}}
    \nonumber\\[0.15cm]
    &-\!32m_\tau^2\Big\{(N_a\cdot k) \left(m_\tau^2\!-\!q^2\right) s_-s_+ \!-\!\left[4 E m_n q^2\!+\!\left(m_\tau^2\!\!-\!q^2\right) \left(m_{\Lambda_c}^2\!-\!m_n^2\!-\!q^2\right)\right]  
    \nonumber\\[0.15cm]
    &\times\big[(N_a\cdot p)(m_{\Lambda_c}^2-m_n^2+q^2)+(N_a\cdot p^\prime) \left(m_n^2-m_{\Lambda_c}^2+q^2\right)\big]\Big\}
    \left[\frac{(m_{\Lambda_c}-m_n)^2\tilde{h}_\perp^2}{q^4 s_-}
    \right.\nonumber\\[0.15cm]
    &\left.+\frac{(m_{\Lambda_c}+m_n)^2h_\perp^2}{q^4 s_+}\right]-\frac{128m_\tau^2 (m_{\Lambda_c}^2-m_n^2)h_\perp\tilde{h}_\perp}{q^4}\Big\{ (N_a\cdot p) \left[2 E m_n q^2+(m_{\Lambda_c}^2-m_p^2) (m_\tau^2-q^2)\right]
    \nonumber\\[0.15cm]
    &-(N_a\cdot p^\prime) \left[2 E m_n q^2+\left(m_\tau^2-q^2\right) \left(m_{\Lambda_c}^2-m_n^2-q^2\right)\right]\Big\}\,,
    \\[0.2cm]
    \mathcal{A}^{l}_{V_{R}-V_{L}}=&\frac{2m_\tau^4 (N_a\cdot k)}{q^4}\left[f_0^2 (m_{\Lambda_c}-m_n)^2 s_+-g_0^2 (m_{\Lambda_c}+m_n)^2s_-\right]-\frac{2 m_\tau^2 \left(m_{\Lambda_c}^2-m_n^2\right)}{q^4} \nonumber\\[0.15cm]
    &\times (f_0 f_+-g_0 g_+)\left\{(N_a\cdot k) \left[4 E m_n q^2+\left(m_\tau^2-q^2\right) \left(2 m_{\Lambda_c}^2-2 m_n^2-q^2\right)\right]
    \right.\nonumber\\[0.15cm]
    &\left.-q^2 \left(m_\tau^2-q^2\right) (N_a\cdot p+N_a\cdot p^\prime)\right\}+2m_\tau^2\left[\frac{ f_+^2 (m_{\Lambda_c}+m_n)^2}{q^4 s_+}-\frac{g_+^2 (m_{\Lambda_c}-m_n)^2}{q^4 s_-}\right]
    \nonumber\\[0.15cm]
    &\times\Big{\{}(N_a\cdot k) \left[\left(m_{\Lambda_c}^2\!-\!m_n^2\right)\Big(m_\tau^2 (m_{\Lambda_c}^2-m_n^2-q^2)\!-\!q^2 (2 m_{\Lambda_c}^2\!-\!4m_n E\!-\!2m_n^2+q^2)\Big) 	\right.\nonumber\\[0.15cm]
    &\left.+q^4\left(4 m_{\Lambda_c}^2-q^2\right)\right]-q^2 (N_a\cdot p+N_a\cdot p^\prime)\left[4 E m_n q^2+\left(m_\tau^2-q^2\right) (m_{\Lambda_c}^2-m_n^2-q^2)\right]\Big{\}} \nonumber\\[0.15cm]
    &+4 m_\tau^2\left(\frac{ f_\perp^2}{s_+}-\frac{ g_\perp^2}{s_-}\right)\Big{\{}(N_a\cdot p) \left[2 E m_n \left(m_{\Lambda_c}^2-m_n^2+q^2\right)+\left(m_\tau^2-q^2\right)
    \right.\nonumber\\[0.15cm]
    &\left. \times (m_{\Lambda_c}^2+m_n^2-q^2)\right]+2 m_n (N_a\cdot p^\prime) \left[E \left(m_n^2-m_{\Lambda_c}^2+q^2\right)+m_n \left(q^2-m_\tau^2\right)\right]\Big{\}}\,,
    \\[0.2cm]
    \mathcal{A}^{l}_{S_{L}-V_{L}}=&-\!\frac{2 m_\tau^3 (N_a\cdot k)}{m_c q^2}\Big{\{}f_0^2 (m_{\Lambda_c}\!-\!m_n)^2 s_+\!+\!g_0^2 (m_{\Lambda_c}\!+\!m_n)^2 s_-\Big{\}}\!-\!\frac{2 m_\tau \left(m_{\Lambda_c}^2\!-\!m_n^2\right) }{m_c q^2}\nonumber\\[0.15cm]
    &\times(f_0 f_++g_0 g_+)\Big{\{}q^2 \left(m_\tau^2-q^2\right) (N_a\cdot p^\prime)+(N_a\cdot p-N_a\cdot p^\prime) \nonumber\\[0.15cm]
    &\times \left[2 E m_n q^2+\left(m_{\Lambda_c}^2-m_n^2\right) (m_\tau^2-q^2)\right]-2 i q^2 \varepsilon _{\{k\},\{k^\prime\},\{N_a\},\{p\}}\Big{\}}\,,	
    \\[0.2cm]
    \mathcal{A}^{l}_{S_{R}-V_{L}}=&-\!\frac{2 m_\tau^3 (N_i\cdot k)}{m_c q^2}\Big{\{}f_0^2 (m_{\Lambda_c}\!-\!m_n)^2 s_+\!-\!g_0^2 (m_{\Lambda_c}\!+\!m_n)^2 s_-\Big{\}}\!-\!\frac{2 m_\tau \left(m_{\Lambda_c}^2\!-\!m_n^2\right) }{m_c q^2}\nonumber\\[0.15cm]
    &\times(f_0 f_+-g_0 g_+)\Big{\{}q^2 \left(m_\tau^2-q^2\right) (N_a\cdot p^\prime)+(N_a\cdot p-N_a\cdot p^\prime)\nonumber\\[0.15cm]
    &\times \left[2 E m_n q^2+\left(m_{\Lambda_c}^2-m_p^2\right) (m_\tau^2-q^2)\right]-2 i q^2 \varepsilon _{\{k\},\{k^\prime\},\{N_a\},\{p\}}\Big{\}}\,,	
   \\[0.2cm]
   \mathcal{A}^{l}_{T_{L}-V_{L}}=&-8 m_\tau^3\left[\frac{(m_{\Lambda_c}\!-\!m_n)f_0h_+}{q^2}\!+\!\frac{(m_{\Lambda_c}\!+\!m_n)g_0\tilde{h}_+}{q^2}\right]
   \left[ (N_a\cdot p) \left(2 E m_n\!-\!m_\tau^2+q^2\right)
   \right.	\nonumber\\[0.15cm]	
   &\left.-\!2 \left(E m_n (N_a\cdot p^\prime)\!+\!i \varepsilon _{\{k\}\{k^\prime\}\{N_a\}\{p\}}\right)\right]\!\!+\!\!\left[4 E m_n q^2\!+\!\left(m_\tau^2\!-\!q^2\right) \left(m_{\Lambda_c}^2\!\!-\!m_n^2\!-\!q^2\right)\right]
   \nonumber\\[0.15cm]	
   &\times\!\big[2 i \varepsilon_{\{k\}\{k^\prime\}\{N_a\}\{p\}}\!+\!(N_a\cdot p) \left(m_{\Lambda_c}^2\!\!-\!m_n^2\!+\!m_\tau^2\!-\!2 E m_n\right)\!+\!(N_a\cdot p^\prime)(2 E m_n\!-\!m_{\Lambda_c}^2\!+\!m_n^2\!+\!q^2)\big]
   \nonumber\\[0.15cm]	
   &\times \left[\frac{8m_\tau(m_{\Lambda_c}\!+\!m_n)f_\perp h_\perp}{s_+q^2}\!+\!\frac{8m_\tau(m_{\Lambda_c}\!-\!m_n)g_\perp \tilde{h}_\perp}{s_-q^2}\right]\!+\!8m_\tau \left(q^2\!-\!m_\tau^2\right)
   \nonumber\\[0.15cm]	
   &\times\left[\frac{(m_{\Lambda_c}\!-\!m_n)f_\perp\tilde{h}_\perp}{q^2}\!+\!\frac{(m_{\Lambda_c}\!+\!m_n)g_\perp h_\perp}{q^2}\right]\left[2 i \varepsilon _{\{k\}\{k^\prime\}\{N_a\}\{p\}}\!+\!(N_a\cdot p) (m_{\Lambda_c}^2\!-\!m_n^2\!+\!m_\tau^2\!\!-\!2 E m_n)
   \right.	\nonumber\\[0.15cm]	
   &\left.	\!+\!(N_a\cdot p^\prime) \left(2 E  m_n\!-\!m_{\Lambda_c}^2\!+\!m_n^2\!+\!q^2\right)\right]\!\!+\!8m_\tau\Big{\{}\!\!\left[4 m_n^2 q^2 \left(m_\tau^2\!\!-\!2 E^2\right)
   \right.	\nonumber\\[0.15cm]	
   &\left.+2 E m_n \left(m_\tau^2\!-\!3 q^2\right) \left(m_n^2\!-\!m_{\Lambda_c}^2\!+\!q^2\right)\!-\!q^2 s_- s_+\right](N_a\cdot p\!+\!N_a\cdot p^\prime)\!-\!\left(m_\tau^2\!+\!q^2\right)
   \nonumber\\[0.15cm]	
   &\times\!\left[\left(m_\tau^2\!-\!q^2\right) \left(m_n^2\!-\!m_{\Lambda_c}^2\!+\!q^2\right)\!-\!4 E m_n q^2\right]  (N_a\cdot p)\!-\!2 i \left[4 E m_n q^2\!+\!\left(m_\tau^2\!-\!q^2\right)
   \right.\nonumber\\[0.15cm]	
   & \left.\times\!\left(m_{\Lambda_c}^2\!\!-m_n^2\!-\!q^2\right)\right] \varepsilon _{\{k\}\{k^\prime\}\{N_a\}\{p\}}\Big{\}}
   \left[\frac{ (m_{\Lambda_c}\!+\!m_n)f_+h_+}{s_+q^2}\!+\!\frac{ (m_{\Lambda_c}\!\!-\!m_n)g_+\tilde{h}_+}{s_-q^2}\right]
    \,,
    \\[0.2cm]
   \mathcal{A}^{l}_{T_{L}-V_{R}}&=-8 m_\tau^3\left[\frac{(m_{\Lambda_c}\!-\!m_n)f_0h_+}{q^2}\!-\!\frac{(m_{\Lambda_c}\!+\!m_n)g_0\tilde{h}_+}{q^2}\right]\left[ (N_a\cdot p) \left(2 E m_n\!-\!m_\tau^2+q^2\right)
   \right.	\nonumber\\[0.15cm]	
   &\left.-\!2 \left(E m_n (N_a\cdot p^\prime)\!+\!i \varepsilon _{\{k\}\{k^\prime\}\{N_a\}\{p\}}\right)\right]\!\!+\!\!\left[4 E m_n q^2\!+\!\left(m_\tau^2\!-\!q^2\right) \left(m_{\Lambda_c}^2\!\!-\!m_n^2\!-\!q^2\right)\right]
   \nonumber\\[0.15cm]	
   &\times\!\big[2 i \varepsilon_{\{k\}\{k^\prime\}\{N_a\}\{p\}}\!+\!(N_a\cdot p) \left(m_{\Lambda_c}^2\!\!-\!m_n^2\!+\!m_\tau^2\!-\!2 E  m_n\right)\!+\!(N_a\cdot p^\prime)(2 E m_n\!-\!m_{\Lambda_c}^2
   \nonumber\\[0.15cm]	
   &+\!m_n^2\!+\!q^2)\big]\left[\frac{8m_\tau(m_{\Lambda_c}\!+\!m_n)f_\perp h_\perp}{s_+q^2}\!-\!\frac{8m_\tau(m_{\Lambda_c}\!-\!m_n)g_\perp \tilde{h}_\perp}{s_-q^2}\right]\!+\!8m_\tau \left(q^2\!-\!m_\tau^2\right)
\nonumber\\[0.15cm]	
&\times\left[\frac{(m_{\Lambda_c}\!-\!m_n)f_\perp\tilde{h}_\perp}{q^2}\!-\!\frac{(m_{\Lambda_c}\!+\!m_n)g_\perp h_\perp}{q^2}\right]\left[2 i \varepsilon _{\{k\}\{k^\prime\}\{N_a\}\{p\}}\!+\!(N_a\cdot p) (m_{\Lambda_c}^2\!-\!m_n^2
\right.	\nonumber\\[0.15cm]	
&\left.	+m_\tau^2\!\!-\!2 E m_n)\!+\!(N_a\cdot p^\prime) \left(2 E m_n\!-\!m_{\Lambda_c}^2\!+\!m_n^2\!+\!q^2\right)\right]\!\!+\!8m_\tau\Big{\{}\!\!\left[4 m_n^2 q^2 \left(m_\tau^2\!\!-\!2 E^2\right)
\right.	\nonumber\\[0.15cm]	
&\left.+2 E m_n \left(m_\tau^2\!-\!3 q^2\right) \left(m_n^2\!-\!m_{\Lambda_c}^2\!+\!q^2\right)\!-\!q^2 s_- s_+\right](N_a\cdot p\!+\!N_a\cdot p^\prime)\!-\!\left(m_\tau^2\!+\!q^2\right)
\nonumber\\[0.15cm]	
& \times\!\left[\left(m_\tau^2\!-\!q^2\right) \left(m_n^2\!-\!m_{\Lambda_c}^2\!+\!q^2\right)\!-\!4 E m_n q^2\right]  (N_a\cdot p)\!-\!2 i \left[4 E m_n q^2\!+\!\left(m_\tau^2\!-\!q^2\right)
\right.\nonumber\\[0.15cm]	
& \left.\times\!\left(m_{\Lambda_c}^2\!\!-m_n^2\!-\!q^2\right)\right] \varepsilon _{\{k\}\{k^\prime\}\{N_a\}\{p\}}\Big{\}}
\left[\frac{ (m_{\Lambda_c}\!+\!m_n)f_+h_+}{s_+q^2}\!-\!\frac{ (m_{\Lambda_c}\!\!-\!m_n)g_+\tilde{h}_+}{s_-q^2}\right]
\,,
\\[0.2cm]
\mathcal{A}^{l}_{T_{L}-S_{L}}=&\frac{8 m_\tau^2}{m_c} \left[2 i \varepsilon _{\{k\},\{N_a\},\{p\},\{p^\prime\}}+(N_a\cdot p) \left(2 E m_n-m_\tau^2+q^2\right)-2 E m_n (N_a\cdot p^\prime)\right]
\nonumber\\[0.15cm]
&\times \left[f_0 h_+ (m_{\Lambda_c}-m_n)+g_0 \tilde{h}_+ (m_{\Lambda_c}+m_n)\right]\,,
\\[0.2cm]
\mathcal{A}^{l}_{T_{L}-S_{R}}=&\frac{8 m_\tau^2}{m_c} \left[2 i \varepsilon _{\{k\},\{N_a\},\{p\},\{p^\prime\}}+(N_a\cdot p) \left(2 E  m_n-m_\tau^2+q^2\right)-2 E m_n (N_a\cdot p^\prime)\right]
\nonumber\\[0.15cm]
&\times \left[f_0 h_+ (m_{\Lambda_c}-m_n)-g_0 \tilde{h}_+ (m_{\Lambda_c}+m_n)\right]\,,
\\[0.2cm]
\mathcal{A}^{l}_{S_{L}-S_{R}}=&\mathcal{A}_{S_{L}-S_{R}}\frac{2m_\tau (k\cdot N_a)}{m_\tau^2-q^2}\,,\\[0.2cm]
	\mathcal{A}^{h}_{V_{L}-V_{L}}=&-2m_{\Lambda_c} \left\{(N_a\cdot k) \left[4 E m_n \left(m_n^2-m_{\Lambda_c}^2\right)+\left(q^2-m_\tau^2\right) \left(m_{\Lambda_c}^2+3 m_n^2-q^2\right)+2 s_+ s_-\right]
	\right.\nonumber\\[0.12cm]
	&\left.+(N_a\cdot p+N_a\cdot p^\prime) \left[4 E m_n q^2+\left(m_\tau^2-q^2\right) \left(m_{\Lambda_c}^2-m_n^2-q^2\right)\right]\right\} 
	\nonumber\\[0.12cm]
	&\times\left[\frac{(m_{\Lambda_c}+m_n)f_+ f_\perp}{s_+}+\frac{(m_{\Lambda_c}\!-\!m_n)g_+ g_\perp}{s_-}\right]+\frac{2f_0 g_0m_\tau^2(m_{\Lambda_c}^2\!-\!m_n^2)}{q^4}\left(m_\tau^2\!-\!q^2\right) 
	\nonumber\\[0.12cm]
	&\times\left[2 m_{\Lambda_c}^2 (N_a\cdot p)-(N_a\cdot p^\prime) (m_{\Lambda_c}^2
	+m_n^2-q^2)\right]+4 m_{\Lambda_c}\big[4 E m_n q^2+\left(m_\tau^2-q^2\right)
	\nonumber\\[0.12cm]
	&\times (m_{\Lambda_c}^2\!\!-\!m_n^2\!-\!q^2)\big]\left\{\left[s_+s_-\!-\!2 E m_n \left(m_{\Lambda_c}^2\!-\!m_n^2\!+\!q^2\right)\!+\!\left(q^2\!-\!m_\tau^2\right) (m_{\Lambda_c}^2\!+\!m_n^2\!-\!q^2)\right] (N_a\cdot p)
	\right.\nonumber\\[0.12cm]
	&\left.\!-\!\left[s_+s_-\!+\!2 E m_n \left(m_n^2-m_{\Lambda_c}^2+q^2\right)+2 m_n^2 \left(q^2-m_\tau^2\right)\right](N_a\cdot p^\prime)\right\}
	\nonumber\\[0.12cm]
	&\times\!\frac{\left[(m_{\Lambda_c}\!\!+\!m_n)f_+ g_\perp\!+\!(m_{\Lambda_c}\!\!-\!m_n)f_\perp g_+\right]}{q^2s_- s_+}\!+\!\left[\frac{(m_{\Lambda_c}\!\!-\!m_n)f_0 g_\perp}{q^2 s_-}\!+\!\frac{(m_{\Lambda_c}\!\!+\!m_n)f_\perp g_0}{q^2 s_+}\right]
	\nonumber\\[0.12cm]
	&\times\!4 m_{\Lambda_c} m_\tau^2\left\{(N_a\cdot p )\left[2 E m_n \left(m_{\Lambda_c}^2\!\!-\!m_n^2\!+\!q^2\right)\!\!+\!\!\left(m_\tau^2\!\!-\!q^2\right) \left(m_{\Lambda_c}^2\!\!+\!m_n^2\!-\!q^2\right)\right]
	\right.\nonumber\\[0.12cm]
	&\left.\!+\!2 m_n\! (N_a\cdot p^\prime)\left[E \left(m_n^2\!-\!m_{\Lambda_c}^2\!+\!q^2\right)\!+\!m_n \left(q^2\!-\!m_\tau^2\right)\right]\!
	+\! (N_a\cdot k) s_+s_-\right\}\!+\!2\left(\frac{f_\perp^2}{s_+}\!+\!\frac{g_\perp^2}{s_-}\right)
	\nonumber\\[0.12cm]
	&\times\!\left[4 E m_n q^2\!+\!\left(m_\tau^2\!-\!q^2\right) \left(m_{\Lambda_c}^2\!\!-\!m_n^2\!-\!q^2\right)\right]\left[2 m_{\Lambda_c}^2 (N_a\cdot p)\!-\!(N_a\cdot p^\prime) (m_{\Lambda_c}^2\!+\!m_n^2\!-\!q^2)\right] \nonumber\\[0.12cm]
	&\!+\!\frac{4f_\perp g_\perp}{s_- s_+} \left\{8 E^2 m_n^2 q^2\!+\!\left(m_\tau^2\!-\!q^2\right) \left[2 m_{\Lambda_c}^2 \left(m_n^2\!+\!q^2\right)\!-\!4 E m_n (m_p^2\!-\!m_{\Lambda_c}^2+\!q^2)
	\right.\right.\nonumber\\[0.12cm]
	&\left.\left.-\!m_n^4+2 m_n^2 m_\tau^2-q^4\right]\right\}\left[2 m_{\Lambda_c}^2 (N_a\cdot p)-(N_a\cdot p^\prime) \left(m_{\Lambda_c}^2+m_n^2-q^2\right)\right]
	\nonumber\\[0.12cm]
	&-\!2m_\tau^2\left\{\left[4 E m_n q^2\!+\!\left(m_\tau^2\!-\!q^2\right) \left(m_{\Lambda_c}^2\!\!-\!m_n^2\!-\!q^2\right)\right] \left[2 m_{\Lambda_c}^2( N_a\cdot p)\!-\!(N_a\cdot p^\prime )
	\right.\right.\nonumber\\[0.12cm]
	&\left.\left.\times (m_{\Lambda_c}^2\!+\!m_n^2\!-\!q^2)\right]\right\}\left[\frac{f_0 g_+  (m_{\Lambda_c}\!\!-\!m_n)^2}{q^4 s_-}\!+\!\frac{f_+g_0  (m_{\Lambda_c}\!+\!m_n)^2}{q^4 s_+}\right]
	\!\!-\!\!\frac{2(m_{\Lambda_c}^2\!\!-\!m_n^2)f_+ g_+}{q^4 s_- s_+}
	\nonumber\\[0.12cm]
	&\times\!\left\{16 E^2 m_n^2 q^4\!+\!\left(m_\tau^2\!-\!q^2\right) \left[m_\tau^2 \left(m_n^2\!-\!m_{\Lambda_c}^2\!+\!q^2\right)^2\!
	-\!4 m_n q^2 \big(2 E(m_n^2\!-\!m_{\Lambda_c}^2+q^2)+m_n q^2\big)\right]\right\}
      \nonumber\\[0.12cm]
	&\times \left[2 m_{\Lambda_c}^2 (N_a\cdot p)-(N_a\cdot p^\prime) \left(m_{\Lambda_c}^2+m_n^2-q^2\right)\right]\,,
	\\[0.2cm]
	\mathcal{A}^{h}_{V_{R}-V_{R}}=&-2m_{\Lambda_c} \left\{(N_a\cdot k) \left[4 E m_n \left(m_n^2-m_{\Lambda_c}^2\right)+\left(q^2-m_\tau^2\right) \left(m_{\Lambda_c}^2+3 m_n^2-q^2\right)+2 s_+ s_-\right]
	\right.\nonumber\\[0.12cm]
	&\left.+(N_a\cdot p+N_a\cdot p^\prime) \left[4 E m_n q^2+\left(m_\tau^2-q^2\right) \left(m_{\Lambda_c}^2-m_n^2-q^2\right)\right]\right\} 
	\nonumber\\[0.12cm]
	&\times\left[\frac{(m_{\Lambda_c}+m_n)f_+ f_\perp}{s_+}+\frac{(m_{\Lambda_c}\!-\!m_n)g_+ g_\perp}{s_-}\right]+\frac{2f_0 g_0m_\tau^2(m_{\Lambda_c}^2\!-\!m_n^2)}{q^4}\left(m_\tau^2\!-\!q^2\right) 
	\nonumber\\[0.12cm]
	&\times\left[2 m_{\Lambda_c}^2 (N_a\cdot p)-(N_a\cdot p^\prime) (m_{\Lambda_c}^2
	+m_n^2-q^2)\right]+4 m_{\Lambda_c}\big[4 E m_n q^2+\left(m_\tau^2-q^2\right)
	\nonumber\\[0.12cm]
	&\times (m_{\Lambda_c}^2\!\!-\!m_n^2\!-\!q^2)\big]\left\{\left[s_+s_-\!-\!2 E m_n \left(m_{\Lambda_c}^2\!-\!m_n^2\!+\!q^2\right)\!+\!\left(q^2\!-\!m_\tau^2\right) (m_{\Lambda_c}^2\!+\!m_n^2\!-\!q^2)\right] (N_a\cdot p)
	\right.\nonumber\\[0.12cm]
	&\left.-\!\left[s_+s_-\!+\!2 E m_n \left(m_n^2-m_{\Lambda_c}^2+q^2\right)+2 m_n^2 \left(q^2-m_\tau^2\right)\right](N_a\cdot p^\prime)\right\}
	\nonumber\\[0.12cm]
	&\times\!\frac{\left[(m_{\Lambda_c}\!\!+\!m_n)f_+ g_\perp\!+\!(m_{\Lambda_c}\!\!-\!m_n)f_\perp g_+\right]}{q^2s_- s_+}\!+\!\left[\frac{(m_{\Lambda_c}\!\!-\!m_n)f_0 g_\perp}{q^2 s_-}\!+\!\frac{(m_{\Lambda_c}\!\!+\!m_n)f_\perp g_0}{q^2 s_+}\right]
	\nonumber\\[0.12cm]
	&\times\!4 m_{\Lambda_c} m_\tau^2\left\{(N_a\cdot p) \left[2 E m_n \left(m_{\Lambda_c}^2\!\!-\!m_n^2\!+\!q^2\right)\!\!+\!\!\left(m_\tau^2\!\!-\!q^2\right) \left(m_{\Lambda_c}^2\!\!+\!m_n^2\!-\!q^2\right)\right]
	\right.\nonumber\\[0.12cm]
	&\left.+\!2 m_n \! (N_a\cdot p^\prime)\left[E \left(m_n^2\!-\!m_{\Lambda_c}^2\!+\!q^2\right)\!+\!m_n \left(q^2\!-\!m_\tau^2\right)\right]\!+\! (N_a\cdot k) s_+s_-\right\}\!+\!2\left(\frac{f_\perp^2}{s_+}\!+\!\frac{g_\perp^2}{s_-}\right)
	\nonumber\\[0.12cm]
	&\times\!\left[4 E m_n q^2\!+\!\left(m_\tau^2\!-\!q^2\right) \left(m_{\Lambda_c}^2\!\!-\!m_n^2\!-\!q^2\right)\right]\left[2 m_{\Lambda_c}^2 (N_a\cdot p)\!-\!(N_a\cdot p^\prime) (m_{\Lambda_c}^2\!+\!m_p^2-\!q^2)\right]
	 \nonumber\\[0.12cm]
	&-\!\frac{4f_\perp g_\perp}{s_- s_+} \left\{8 E^2 m_n^2 q^2\!+\!\left(m_\tau^2\!-\!q^2\right) \left[2 m_{\Lambda_c}^2 \left(m_n^2\!+\!q^2\right)\!-\!4 E m_n (m_n^2\!-\!m_{\Lambda_c}^2\!+\!q^2)
	\right.\right.\nonumber\\[0.12cm]
	&\left.\left.-\!m_n^4\!+\!2 m_n^2 m_\tau^2-q^4\right]\right\}\left[2 m_{\Lambda_c}^2 (N_a\cdot p)-(N_a\cdot p^\prime) \left(m_{\Lambda_c}^2+m_n^2-q^2\right)\right]
	\nonumber\\[0.12cm]
	&-\!2m_\tau^2\left\{\left[4 E  m_n q^2\!+\!\left(m_\tau^2\!-\!q^2\right) \left(m_{\Lambda_c}^2\!\!-\!m_n^2\!-\!q^2\right)\right] \left[2 m_{\Lambda_c}^2 (N_a\cdot p)\!-\!(N_a\cdot p^\prime)
	\right.\right.\nonumber\\[0.12cm]
	&\left.\left.\times (m_{\Lambda_c}^2+\!m_n^2\!-\!q^2)\right]\right\}\left[\frac{f_0 g_+  (m_{\Lambda_c}\!\!-\!m_n)^2}{q^4 s_-}\!+\!\frac{f_+g_0  (m_{\Lambda_c}\!+\!m_n)^2}{q^4 s_+}\right]
	\!\!+\!\!\frac{2(m_{\Lambda_c}^2\!\!-\!m_n^2)f_+ g_+}{q^4 s_- s_+}
	\nonumber\\[0.12cm]
	&\times\!\left\{16 E^2 m_n^2 q^4\!+\!\left(m_\tau^2\!-\!q^2\right) \left[m_\tau^2 \left(m_n^2\!-\!m_{\Lambda_c}^2\!+\!q^2\right)^2\!
	-\!4 m_n q^2 \big(2 E (m_n^2\!-\!m_{\Lambda_c}^2\!+\!q^2)
	\right.\right.\nonumber\\[0.12cm]
	&\left.\left.+m_n q^2\big)\right]\right\}\left[2 m_{\Lambda_c}^2 (N_a\cdot p)-(N_a\cdot p^\prime) \left(m_{\Lambda_c}^2+m_n^2-q^2\right)\right]\,,
	\\[0.2cm]
	\mathcal{A}^{h}_{S_{L}-S_{L}}=&\frac{2 f_0 g_0 \left(m_{\Lambda_c}^2-m_n^2\right) \left(m_\tau^2-q^2\right)}{m_{c}^2}\left[2 m_{\Lambda_c}^2 (N_a\cdot p)-(N_a\cdot p^\prime) \left(m_{\Lambda_c}^2+m_n^2-q^2\right)\right]\,,
	\\[0.2cm]
	\mathcal{A}^{h}_{T_{L}-T_{L}}=&32m_\tau^2\left[\frac{\tilde{h}_\perp^2 (m_{\Lambda_c}\!-\!m_n)^2}{s_-q^4}+\frac{h_\perp^2 (m_{\Lambda_c}+m_n)^2}{s_+q^4}\right]\left[\left(m_\tau^2\!-\!q^2\right) \left(m_{\Lambda_c}^2\!-\!m_n^2\!-\!q^2\right)\!+\!4 E m_n q^2\right]\nonumber\\[0.12cm]
	&\times\left[2 m_{\Lambda_c}^2 (N_a\cdot p)\!-\!(N_a\cdot p^\prime) \left(m_{\Lambda_c}^2\!\!+\!m_n^2\!-\!q^2\right)\right]\!-\!64m_{\Lambda_c}\left[\left(m_\tau^2\!-\!q^2\right) (m_{\Lambda_c}^2\!-\!m_n^2\!-\!q^2)\!
	\right.\nonumber\\[0.12cm]
	&\left.+\!4 E m_n q^2\right]\left\{\left[s_-s_+\!-\!2 E m_n \left(m_{\Lambda_c}^2\!\!-\!m_n^2\!+\!q^2\right)\!+\!\left(q^2\!-\!m_\tau^2\right)(m_{\Lambda_c}^2+m_n^2\!-\!q^2)\right](N_a\cdot p)
	\right.\nonumber\\[0.12cm]
	&\left.-\!\left[s_-s_+\!+\!2 m_n E_{\nu_\tau} \left(m_n^2\!-\!m_{\Lambda_c}^2\!\!+\!q^2\right)\!+\!2m_n^2\left(q^2\!\!-\!m_\tau^2\right)\right](N_a\cdot p^\prime)\right\}
	\nonumber\\[0.12cm]
	&\times\!\frac{h_+ \tilde{h}_\perp (m_{\Lambda_c}\!\!-\!m_n)\!+\!\tilde{h}_+ h_\perp (m_{\Lambda_c}\!\!+\!m_n)}{s_-s_+q^2}\!\!+\!\!\left[\frac{h_+ h_\perp (m_{\Lambda_c}\!+\!m_n)}{s_+q^2}\!+\!\frac{\tilde{h}_+ \tilde{h}_\perp (m_{\Lambda_c}\!\!-\!m_n)}{s_-q^2}\right]
	\nonumber\\[0.12cm]
	&\times\!64m_{\Lambda_c}m_\tau^2\left\{\left[s_-s_+\!-\!2 E m_n \left(m_{\Lambda_c}^2\!\!-\!m_n^2\!+\!q^2\right)\!+\!\left(q^2\!\!-\!m_\tau^2\right) \left(m_{\Lambda_c}^2\!+\!m_n^2\!-\!q^2\right)\right] (N_a\cdot p)
	\right.\nonumber\\[0.12cm]
	&\left. -\left[s_-s_++2 m_n E \left(m_n^2-m_{\Lambda_c}^2+q^2\right)+2m_n^2 \left(q^2-m_\tau^2\right)\right](N_a\cdot p^\prime)\right\}
	\nonumber\\[0.12cm]
	&-\frac{64 h_\perp \tilde{h}_\perp \left(m_{\Lambda_c}^2-m_n^2\right)}{s_-s_+q^4}\left\{(m_\tau^2-q^2)\left[m_\tau^2\left(s_-s_++2 m_n^2 q^2\right)-4 m_n q^2 E (m_n^2
	\right.\right.\nonumber\\[0.12cm]
	&\left.\left.-m_{\Lambda_c}^2+q^2)\!-\!2 m_n^2 q^4\right]+8 E^2 m_n^2 q^4\right\}\left[2 m_{\Lambda_c}^2 (N_a\cdot p)\!-\!(N_a\cdot p^\prime) \left(m_{\Lambda_c}^2+m_n^2-q^2\right)\right]
	\nonumber\\[0.12cm]
	&-\frac{32 h_+ \tilde{h}_+}{s_-s_+}\left\{(m_\tau^2\!-\!q^2)\left[-4m_n^2 \left(q^2\!-\!m_\tau^2\right)\!-\!8m_n E \left(m_n^2\!-\! m_{\Lambda_c}^2+q^2\right)\!-s_-s_+\right] \!+\!16 E^2 m_n^2 q^2\right\}
	\nonumber\\[0.12cm]
	&\times \left[2 m_{\Lambda_c}^2 (N_a\cdot p)-(N_a\cdot p^\prime) \left(m_{\Lambda_c}^2+m_n^2-q^2\right)\right]\,,
	\\[0.2cm]
	\mathcal{A}^{h}_{V_{R}-V_{L}}=&8 i m_{\Lambda_c} \big[f_+ g_\perp (m_{\Lambda_c}\!+\!m_n)\!-\!f_\perp g_+ (m_{\Lambda_c}\!-\!m_n)\big] \varepsilon_{\{k\}\{k^\prime\}\{N_a\}\{p\}}\!+\!2 \left[4 E m_n q^2
	\right.\nonumber\\[0.12cm]
	&\left.	+\left(m_\tau^2-q^2\right) \left(m_{\Lambda_c}^2-m_n^2-q^2\right)\right]\left[2 m_{\Lambda_c}^2 (N_a\cdot p)\!-\!(N_a\cdot p^\prime) \left(m_{\Lambda_c}^2\!+\!m_n^2\!-\!q^2\right)\right]\left(\frac{f_\perp^2}{s_+}\!-\!\frac{g_\perp^2}{s_-}\right)
	\nonumber\\[0.12cm]
	&-\!2 m_{\Lambda_c}\left\{\left[2s_+s_-\!+\!4 E m_n \left(m_n^2\!-\!m_{\Lambda_c}^2\right)\!
	+\!\left(q^2\!-\!m_\tau^2\right) (m_{\Lambda_c}^2\!+\!3 m_n^2-q^2)\right](N_a\cdot k)
	\right.\nonumber\\[0.12cm]
	&\left.+ \left[4 E m_n q^2+\left(m_\tau^2-q^2\right) \left(m_{\Lambda_c}^2-m_n^2-q^2\right)\right]\right\}(N_a\cdot p\!+\!N_a\cdot p^\prime) 
	\nonumber\\[0.12cm]
	&\times\left[\frac{f_+ f_\perp (m_{\Lambda_c}+m_n)}{s_+}-\frac{g_+ g_\perp (m_{\Lambda_c}-m_n)}{s_-}\right]\,,
	\\[0.2cm]
	\mathcal{A}^{h}_{S_{L}-V_{L}}=&-\frac{4 i m_{\Lambda_c} m_\tau  \varepsilon _{\{k\}\{N_a\}\{p\}\{p^\prime\}}}{m_{c}}\big[(m_{\Lambda_c}\!+\!m_n)g_0 g_\perp\!+\!(m_{\Lambda_c}-m_n)f_0 f_\perp\big]\!
	\nonumber\\[0.12cm]
	&-\!\frac{2 m_{\Lambda_c} m_\tau }{m_{c} }\left\{\frac{(m_{\Lambda_c}+m_p)f_\perp g_0}{s_+}+\frac{(m_{\Lambda_c}\!-\!m_n)f_0 g_\perp}{s_-}\right\}\left\{(N_a\cdot p) \left[2 E m_n \left(m_{\Lambda_c}^2\!-\!m_n^2+q^2\right)
	\right.\right.\nonumber\\[0.12cm]
	&\left.+\!\left(m_\tau^2\!-\!q^2\right) \left(m_{\Lambda_c}^2\!\!\!+\!m_n^2\!-\!q^2\right)\right]\!+
	\!2 m_n (N_a\cdot p^\prime) \left[E \left(m_n^2\!\!-\!m_{\Lambda_c}^2\!+\!q^2\right)\!\!+\!m_n \left(q^2\!\!-\!m_\tau^2\right)\right]
	\nonumber\\[0.12cm]
	&\left.+(N_a\cdot k )s_+s_-\right\}+m_\tau\big[4 E m_n q^2+\left(m_\tau^2\!-\!q^2\right)
	\left(m_{\Lambda_c}^2\!-\!m_n^2-q^2\right)\big]\left[2 m_{\Lambda_c}^2( N_a\cdot p)
	\right.\nonumber\\[0.12cm]
	&\left. -(N_a\cdot p^\prime )\left(m_{\Lambda_c}^2+m_n^2-q^2\right)\right]
	\left[\frac{(m_{\Lambda_c}+m_n)^2f_+ g_0}{m_{c} q^2s_+}+\frac{(m_{\Lambda_c}-m_n)^2f_0 g_+}{m_{c} q^2s_-}\right]
	\nonumber\\[0.12cm]
	&+\frac{2f_0 g_0m_\tau\left(m_{\Lambda_c}^2\!-\!m_n^2\right)}{m_{c} q^2}\left(q^2\!-\!m_\tau^2\right) 
	\left[2 m_{\Lambda_c}^2 (N_a\cdot p)\!-\!(N_a\cdot p^\prime) \left(m_{\Lambda_c}^2+m_n^2\!-q^2\right)\right]
	\,,
	\\[0.2cm]
	\mathcal{A}^{h}_{S_{R}-V_{L}}&=\frac{4 i m_{\Lambda_c} m_\tau  \varepsilon _{\{k\}\{N_a\}\{p\}\{p^\prime\}}}{m_{c}}\big[(m_{\Lambda_c}\!+\!m_n)g_0 g_\perp\!-\!(m_{\Lambda_c}-m_n)f_0 f_\perp\big]
	\nonumber\\[0.12cm]
	&-\!\frac{2 m_{\Lambda_c} m_\tau }{m_{c} }\left\{\frac{(m_{\Lambda_c}\!-\!m_n)f_0 g_\perp}{s_-}\!-\!\frac{(m_{\Lambda_c}+m_n)f_\perp g_0}{s_+}\right\}
	\left\{(N_a\cdot p) \left[2 E m_n \left(m_{\Lambda_c}^2\!-\!m_p^2+q^2\right)
	\right.\right.\nonumber\\[0.12cm]
	&\left.+\!\left(m_\tau^2\!\!-\!q^2\right) \left(m_{\Lambda_c}^2\!\!+\!m_n^2\!\!-\!q^2\right)\right]\!\!\!+\!2 m_n (N_a\cdot p^\prime) 
	\left[E \left(m_n^2\!\!-\!m_{\Lambda_c}^2\!+\!q^2\right)\!+\!m_n \left(q^2\!\!-\!m_\tau^2\right)\right]
	\nonumber\\[0.12cm]
	&\left.+(N_a\cdot k )s_+s_-\right\}\!+m_\tau\big[4 E m_n q^2+\left(m_\tau^2\!-\!q^2\right) \left(m_{\Lambda_c}^2\!-\!m_n^2\!-\!q^2\right)\big]\left[2 m_{\Lambda_c}^2 (N_a\cdot p)
	\right.\nonumber\\[0.12cm]
	&\left.-(N_a\cdot p^\prime) \left(m_{\Lambda_c}^2+m_n^2-q^2\right)\right]
	\left[\frac{(m_{\Lambda_c}-m_n)^2f_0 g_+}{m_{c} q^2s_-}-\frac{(m_{\Lambda_c}+m_n)^2f_+ g_0}{m_{c} q^2s_+}\right]
	\,,
	\\[0.2cm]
	\mathcal{A}^{h}_{S_{L}-V_{R}}=&\frac{4 i m_{\Lambda_c} m_\tau  \varepsilon _{\{k\}\{N_a\}\{p\}\{p^\prime\}}}{m_{c}}\big[(m_{\Lambda_c}\!+\!m_n)g_0 g_\perp\!-\!(m_{\Lambda_c}-m_n)f_0 f_\perp\big]
	\nonumber\\[0.12cm]
	&+\!\frac{2 m_{\Lambda_c} m_\tau }{m_{c} }\left\{\frac{(m_{\Lambda_c}\!-\!m_n)f_0 g_\perp}{s_-}\!-\!\frac{(m_{\Lambda_c}+m_n)f_\perp g_0}{s_+}\right\}\left\{(N_a\cdot p) \left[2 E m_n \left(m_{\Lambda_c}^2\!-\!m_n^2+q^2\right)
	\right.\right.\nonumber\\[0.12cm]
	&\left.+\!\left(m_\tau^2\!\!-\!q^2\right) \left(m_{\Lambda_c}^2\!\!+\!m_n^2\!\!-\!q^2\right)\right]\!\!\!+\!2 m_n (N_a\cdot p^\prime) 
	\left[E \left(m_n^2\!\!-\!m_{\Lambda_c}^2\!+\!q^2\right)\!+\!m_n \left(q^2\!\!-\!m_\tau^2\right)\right]
	\nonumber\\[0.12cm]
	&\left.+(N_a\cdot k) s_+s_-\right\}\!-m_\tau\big[4 E m_n q^2+\left(m_\tau^2\!-\!q^2\right) \left(m_{\Lambda_c}^2\!-\!m_n^2\!
	-\!q^2\right)\big]\left[2 m_{\Lambda_c}^2 (N_a\cdot p)
	\right.\nonumber\\[0.12cm]
	&\left.-(N_a\cdot p^\prime )\left(m_{\Lambda_c}^2+m_n^2-q^2\right)\right]
	\left[\frac{(m_{\Lambda_c}-m_n)^2f_0 g_+}{m_{c} q^2s_-}-\frac{(m_{\Lambda_c}+m_n)^2f_+ g_0}{m_{c} q^2s_+}\right]
	\,,
	\\[0.2cm]
	\mathcal{A}^{h}_{S_{R}-V_{R}}&=\frac{4 i m_{\Lambda_c} m_\tau  \varepsilon _{\{k\}\{N_a\}\{p\}\{p^\prime\}}}{m_{c}}\big[(m_{\Lambda_c}\!+\!m_n)g_0 g_\perp\!+\!(m_{\Lambda_c}-m_n)f_0 f_\perp\big]
	\nonumber\\[0.12cm]
	&+\!\frac{2 m_{\Lambda_c} m_\tau }{m_{c}}\left\{\frac{(m_{\Lambda_c}\!-\!m_n)f_0 g_\perp}{s_-}\!
	+\!\frac{(m_{\Lambda_c}+m_n)f_\perp g_0}{s_+}\right\}\left\{(N_a\cdot p) \left[2 E m_n \left(m_{\Lambda_c}^2\!-\!m_n^2+q^2\right)
	\right.\right.\nonumber\\[0.12cm]
	&\left.+\!\left(m_\tau^2\!\!-\!q^2\right) \left(m_{\Lambda_c}^2\!\!+\!m_n^2\!\!-\!q^2\right)\right]\!\!\!+\!2 m_n (N_a\cdot p^\prime) 
	\left[E \left(m_n^2\!\!-\!m_{\Lambda_c}^2\!+\!q^2\right)\!+\!m_n \left(q^2\!\!-\!m_\tau^2\right)\right]
	\nonumber\\[0.12cm]
	&\left.+(N_a\cdot k) s_+s_-\right\}\!-m_\tau\big[4 Em_n q^2+\left(m_\tau^2\!-\!q^2\right) \left(m_{\Lambda_c}^2\!-\!m_n^2\!-\!q^2\right)\big]\left[2 m_{\Lambda_c}^2 (N_a\cdot p)
	\right.\nonumber\\[0.12cm]
	&\left.-(N_a\cdot p^\prime) \left(m_{\Lambda_c}^2+m_n^2-q^2\right)\right]
	\left[\frac{(m_{\Lambda_c}-m_n)^2f_0 g_+}{m_{c} q^2s_-}+\frac{(m_{\Lambda_c}+m_n)^2f_+ g_0}{m_{c} q^2s_+}\right]
	\nonumber\\[0.12cm]
	&-\frac{2f_0 g_0m_\tau\left(m_{\Lambda_c}^2\!-\!m_n^2\right)}{m_{c} q^2}\left(q^2\!-\!m_\tau^2\right) \left[2 m_{\Lambda_c}^2
	(N_a\cdot p)\!-\!(N_a\cdot p^\prime) \left(m_{\Lambda_c}^2+m_n^2\!-q^2\right)\right]\,,
	\\[0.2cm]
	\mathcal{A}^{h}_{T_{L}-V_{L}}=&16 i m_{\Lambda_c} m_\tau  \left[\frac{\left(m_{\Lambda_c}^2\!-\!m_n^2\right) (f_0 h_\perp+g_0 \tilde{h}_\perp+f_+ \tilde{h}_\perp	+g_+ h_\perp)}{q^2}\!+\!f_\perp \tilde{h}_+\!+\!g_\perp h_+\right]
	\varepsilon _{\{k\}\{k^\prime\}\{N_a\}\{p\}}
	\nonumber\\[0.12cm]
	&+\!8m_{\Lambda_c}m_\tau\left[\frac{g_+ \tilde{h}_\perp (m_{\Lambda_c}\!-\!m_n)^2}{s_-q^2}\!+\!\frac{f_+ h_\perp (m_{\Lambda_c}\!+\!m_n)^2}{s_+q^2}\right]\left\{\left[s_-s_+-2 E m_n \left(m_{\Lambda_c}^2\!-\!m_n^2+q^2\right)
	\right.\right.\nonumber\\[0.12cm]
	&\left.+\left(q^2-m_\tau^2\right) \left(m_{\Lambda_c}^2+m_n^2-q^2\right)\right]
	(N_a\cdot p)-\left[s_-s_++2 E m_n \left(m_n^2\!-\!m_{\Lambda_c}^2+q^2\right)
	\right.	\nonumber\\[0.12cm]
	&\left.\left.+2 m_n^2 \left(q^2\!-\!m_\tau^2\right)\right](N_a\cdot p^\prime)\right\}\!-\!\left(\frac{g_\perp \tilde{h}_+}{s_-}+\frac{f_\perp h_+}{s_+}\right)
	\nonumber\\[0.12cm]
	&\times8m_{\Lambda_c}m_\tau\Big{\{}s_-s_+(N_a\cdot k)\!+\!(N_a\cdot p) \left[2 E m_n \left(m_{\Lambda_c}^2\!\!-\!m_n^2\!+\!q^2\right)\!+\!\left(m_\tau^2\!-\!q^2\right) (m_{\Lambda_c}^2
	\right.	\nonumber\\[0.12cm]
	&\left.+m_n^2-q^2)\right]+2 m_n (N_a\cdot p^\prime) \left[E \left(m_n^2-m_{\Lambda_c}^2\!+\!q^2\right)+m_n \left(q^2-m_\tau^2\right)\right]\Big{\}}
	\nonumber\\[0.12cm]
	&\!+\!4m_\tau\left[\frac{(m_{\Lambda_c}\!\!-\!m_n) (2 g_\perp \tilde{h}_\perp\!\!-\!f_0 \tilde{h}_+)}{s_-q^2}\!+\!\frac{(m_{\Lambda_c}\!+\!m_p) (2 f_\perp h_\perp\!\!-\!g_0 h_+)}{s_+q^2}\right]\left[4 E m_n q^2
	\right.	\nonumber\\[0.12cm]
	&\left.+\left(m_\tau^2-q^2\right) \left(m_{\Lambda_c}^2-m_n^2-q^2\right)\right]\left[2 m_{\Lambda_c}^2 (N_a\cdot p)\!-\!(N_a\cdot p^\prime) \left(m_{\Lambda_c}^2\!+\!m_n^2\!-\!q^2\right)\right]
	\nonumber\\[0.12cm]
	&\!-\!4 m_\tau\left[\frac{(m_{\Lambda_c}\!+\!m_n) (2 g_\perp h_\perp\!\!-\!f_+ \tilde{h}_+)}{q^2}\!+\!\frac{(m_{\Lambda_c}\!\!-\!m_n) (2 f_\perp \tilde{h}_\perp\!\!-\!g_+ h_+)}{q^2}\right] \left(m_\tau^2\!-\!q^2\right)
	\nonumber\\[0.12cm]
	& \times\left[2 m_{\Lambda_c}^2 (N_a\cdot p)\!-\!(N_a\cdot p^\prime) \left(m_{\Lambda_c}^2\!+\!m_n^2\!-\!q^2\right)\right]\!+\!
	4m_{\Lambda_c}m_\tau\Big{\{}(m_\tau^2+q^2)s_-s_+(N_a\cdot k)
	\nonumber\\[0.12cm]
	&	+\left[4 E m_n q^2+\left(m_\tau^2-q^2\right) \left(m_{\Lambda_c}^2-m_n^2-q^2\right)\right] \left[(N_a\cdot p) \left(m_{\Lambda_c}^2-m_n^2+q^2\right)
	\right.	\nonumber\\[0.12cm]
	&\left.+(N_a\cdot p^\prime) \left(m_n^2-m_{\Lambda_c}^2+q^2\right)\right]\Big{\}}
	\left[\frac{f_0 \tilde{h}_\perp (m_{\Lambda_c}-m_n)^2}{s_-q^4}+\frac{g_0 h_\perp (m_{\Lambda_c}+m_n)^2}{s_+q^4}\right]
	\,,
	\\[0.2cm]
	\mathcal{A}^{h}_{T_{L}-V_{R}}=&16 i m_{\Lambda_c} m_\tau  \left[\frac{\left(m_{\Lambda_c}^2\!-\!m_n^2\right) (f_0 h_\perp\!-\!g_0 \tilde{h}_\perp+f_+ \tilde{h}_\perp	-g_+ h_\perp)}{q^2}\!+\!f_\perp \tilde{h}_+\!-\!g_\perp h_+\right]\!\varepsilon _{\{k\}\{k^\prime\}\{N_a\}\{p\}}
	\nonumber\\[0.12cm]
	&+\!8m_{\Lambda_c}m_\tau\left[\frac{f_+ h_\perp (m_{\Lambda_c}\!+\!m_n)^2}{s_+q^2}\!-\!\frac{g_+ \tilde{h}_\perp (m_{\Lambda_c}\!-\!m_n)^2}{s_-q^2}\right]\left\{\left[s_-s_+\!-\!2 E m_n \left(m_{\Lambda_c}^2\!-\!m_n^2\!+\!q^2\right)
	\right.\right.\nonumber\\[0.12cm]
	&\left.+\left(q^2\!-m_\tau^2\right) \left(m_{\Lambda_c}^2+m_n^2-q^2\right)\right](N_a\cdot p)-\left[s_-s_+\!+\!2 E m_n \left(m_n^2\!-\!m_{\Lambda_c}^2\!+\!q^2\right)
	\right.	\nonumber\\[0.12cm]
	&\left.\left.+2 m_n^2 \left(q^2\!-\!m_\tau^2\right)\right](N_a\cdot p^\prime)\right\}\!-\!\left(\frac{f_\perp h_+}{s_+}\!-\!\frac{g_\perp \tilde{h}_+}{s_-}\right)
	\nonumber\\[0.12cm]
	&\times\!8m_{\Lambda_c}m_\tau\left\{s_-s_+(N_a\cdot k)\!+\!(N_a\cdot p) \left[2 E m_n \left(m_{\Lambda_c}^2\!\!-\!m_n^2\!+\!q^2\right)\!+\!\left(m_\tau^2\!-\!q^2\right) (m_{\Lambda_c}^2
	\right.\right.	\nonumber\\[0.12cm]
	&\left.\left.+m_n^2-q^2)\right]+2 m_n (N_a\cdot p^\prime) \left[E \left(m_n^2-m_{\Lambda_c}^2+q^2\right)+m_n \left(q^2-m_\tau^2\right)\right]\right\}
	\nonumber\\[0.12cm]
	&\!+\!4m_\tau\left[\frac{(m_{\Lambda_c}\!\!+\!m_n) (2 f_\perp h_\perp\!\!+\!g_0 h_+)}{s_+q^2}\!-
	\!\frac{(m_{\Lambda_c}\!\!-\!m_n) (2 g_\perp \tilde{h}_\perp\!\!+\!f_0 \tilde{h}_+)}{s_-q^2}\right]\left[4 E m_n q^2
	\right.	\nonumber\\[0.12cm]
	&\left.+\left(m_\tau^2-q^2\right) \left(m_{\Lambda_c}^2-m_n^2-q^2\right)\right]\left[2 m_{\Lambda_c}^2 (N_a\cdot p)-(N_a\cdot p^\prime) \left(m_{\Lambda_c}^2+m_n^2-q^2\right)\right]
	\nonumber\\[0.12cm]
	&\!-\!4 m_\tau\left[\frac{(m_{\Lambda_c}\!\!-\!m_n) (2 f_\perp \tilde{h}_\perp\!\!+\!g_+ h_+)}{q^2}\!-\!\frac{(m_{\Lambda_c}\!+\!m_n) (2 g_\perp h_\perp\!\!+\!f_+ \tilde{h}_+)}{q^2}\right] \left(m_\tau^2\!-\!q^2\right)
	\nonumber\\[0.12cm]
	& \times\left[2 m_{\Lambda_c}^2 (N_a\cdot p)\!-\!(N_a\cdot p^\prime) \left(m_{\Lambda_c}^2\!+\!m_n^2\!-\!q^2\right)\right]\!+\!
	4m_{\Lambda_c}m_\tau\Big{\{}(m_\tau^2\!+\!q^2)s_-s_+(N_a\cdot k)
	\nonumber\\[0.12cm]
	&	+\left[4 E m_n q^2+\left(m_\tau^2-q^2\right) \left(m_{\Lambda_c}^2-m_n^2-q^2\right)\right] \left[(N_a\cdot p) \left(m_{\Lambda_c}^2-m_n^2+q^2\right)
	\right.	\nonumber\\[0.12cm]
	&\left.+(N_a\cdot p^\prime) \left(m_p^2-m_{\Lambda_c}^2+q^2\right)\right]\Big{\}}
	\left[\frac{f_0 \tilde{h}_\perp (m_{\Lambda_c}-m_n)^2}{s_-q^4}-\frac{g_0 h_\perp (m_{\Lambda_c}+m_n)^2}{s_+q^4}\right]
	\,,
	\\[0.2cm]
	\mathcal{A}^{h}_{T_{L}-S_{L}}=&4 \left[4 E m_n q^2\!+\!\left(m_\tau^2\!-\!q^2\right) \left(m_{\Lambda_c}^2\!\!-\!m_n^2\!-\!q^2\right)\right] \left[2 m_{\Lambda_c}^2 (N_a\cdot p)\!-\!(N_a\cdot p^\prime) (m_{\Lambda_c}^2\!+\!m_n^2\!-\!q^2)\right]
	\nonumber\\[0.12cm]
	&\times \left[\frac{(m_{\Lambda_c}\!\!-\!m_n)f_0 \tilde{h}_+}{m_c s_-}\!+\!\frac{(m_{\Lambda_c}\!+\!m_n)g_0 h_+}{m_c s_+}\right]
	\!\!-\!\frac{16 i m_{\Lambda_c} (m_{\Lambda_c}^2\!\!-\!m_n^2)\varepsilon _{\{k\}\{k^\prime\}\{N_a\}\{p\}}}{m_c}
	\nonumber\\[0.12cm]
	&\times\!(f_0 h_\perp\!+\!g_0 \tilde{h}_\perp)
	\!-\!4\left\{\left[4 E m_n q^2\!+\!\left(m_\tau^2\!-\!q^2\right) \left(m_{\Lambda_c}^2\!\!-\!m_n^2\!-\!q^2\right)\right] \left[(N_a\cdot p) (m_{\Lambda_c}^2\!\!-\!m_n^2+q^2)
	\right.\right.\nonumber\\[0.12cm]
	&\left.+(N_a\cdot p^\prime) \left(m_n^2-m_{\Lambda_c}^2+q^2\right)\right]+(N_a\cdot k) \left(m_\tau^2+q^2\right) \left[m_{\Lambda_c}^4-2 m_{\Lambda_c}^2 \left(m_n^2+q^2\right)
	\right.\nonumber\\[0.12cm]
	&\left.\left.+\!\left(m_n^2\!-\!q^2\right)^2\right]\right\}\left[\frac{m_{\Lambda_c} (m_{\Lambda_c}\!\!-\!m_n)^2f_0 \tilde{h}_\perp}{m_c q^2 s_-}\!+\!\frac{m_{\Lambda_c} (m_{\Lambda_c}\!+\!m_n)^2g_0 h_\perp}{m_c q^2 s_+}\right]\,,
	\\[0.2cm]
	\mathcal{A}^{h}_{T_{L}-S_{R}}=&4 \left[4 E m_n q^2\!+\!\left(m_\tau^2\!-\!q^2\right) \left(m_{\Lambda_c}^2\!\!-\!m_n^2\!-\!q^2\right)\right] \left[2 m_{\Lambda_c}^2 (N_a\cdot p)\!-\!(N_a\cdot p^\prime )(m_{\Lambda_c}^2\!+\!m_n^2\!-\!q^2)\right]
	\nonumber\\[0.12cm]
	&\times\left[\frac{(m_{\Lambda_c}\!\!-\!m_n)f_0 \tilde{h}_+}{m_c s_-}\!-\!\frac{(m_{\Lambda_c}\!+\!m_n)g_0 h_+}{m_c s_+}\right]
	\!\!-\!\frac{16 i m_{\Lambda_c} (m_{\Lambda_c}^2\!\!-\!m_n^2)\varepsilon _{\{k\}\{k^\prime\}\{N_a\}\{p\}}}{m_c}
	\nonumber\\[0.12cm]
	&\times\!(f_0 h_\perp\!-\!g_0 \tilde{h}_\perp)
	\!-\!4\left\{\left[4 E m_n q^2\!+\!\left(m_\tau^2\!-\!q^2\right) \left(m_{\Lambda_c}^2\!\!-\!m_p^2\!-\!q^2\right)\right] \left[(N_a\cdot p) (m_{\Lambda_c}^2\!\!-\!m_p^2+q^2)
	\right.\right.\nonumber\\[0.12cm]
	&\left.+(N_a\cdot p^\prime) \left(m_n^2-m_{\Lambda_c}^2+q^2\right)\right]+(N_a\cdot k) \left(m_\tau^2+q^2\right) \left[m_{\Lambda_c}^4-2 m_{\Lambda_c}^2 \left(m_n^2+q^2\right)
	\right.\nonumber\\[0.12cm]
	&\left.\left.+\!\left(m_n^2\!-\!q^2\right)^2\right]\right\}\left[\frac{m_{\Lambda_c} (m_{\Lambda_c}\!\!-\!m_n)^2f_0 \tilde{h}_\perp}{m_c q^2 s_-}\!-\!\frac{m_{\Lambda_c} (m_{\Lambda_c}\!+\!m_p)^2g_0 h_\perp}{m_c q^2 s_+}\right]\,,
	\\[0.2cm]
	\mathcal{A}^{h}_{S_{L}-S_{R}}&=0\,,
	\end{align}
\end{widetext}		
where $\varepsilon_{\{k\}\{k^\prime\}\{N_a\}\{p\}}\equiv\varepsilon _{\mu\nu\alpha\beta}k^{\mu}k^{\prime\nu}N_a^{\alpha}p^{\beta}$, with $\varepsilon$ being a totally antisymmetric tensor. From the equations above, it is clear that $\mathcal{A}^{l,h}$ with the same subscripts are always real. 
		
\bibliographystyle{apsrev4-1}
\bibliography{reference}

\end{document}